\documentclass[11pt]{article}
\newcommand{\Comment}[1]{{}}
\usepackage{amsfonts,amsthm,amsmath,amssymb,slashed}
\usepackage[textwidth = 430 pt, textheight = 630 pt]{geometry}
\usepackage{color}

\Comment{\usepackage{color}
\definecolor{MyDarkBlue}{rgb}{0.15,0.15,0.45}
\usepackage[linktocpage=true]{hyperref}
\hypersetup{
colorlinks=true,
citecolor=MyDarkBlue,
linkcolor=MyDarkBlue,
urlcolor=MyDarkBlue,
pdfauthor={Horatiu Nastase and Kostas Skenderis},
pdftitle={Holographic cosmology solutions of problems with pre-inflationary cosmology},
pdfsubject={hep-th}
}

\usepackage[numbers,sort&compress]{natbib}
\usepackage{hypernat}}
\usepackage{graphicx}
\usepackage{cite}

\newcommand\ignore[1]{}
\def\one{{\,\hbox{1\kern-.8mm l}}}

\def\Tr{{\rm Tr\, }}

\def\a{\alpha}\def\b{\beta}

\def\d{\partial}

\def\Tr{\mathop{\rm Tr}\nolimits}

\newcommand{\Cset}{{\,\,{{{^{_{\pmb{\mid}}}}\kern-.45em{\mathrm C}}}}}

\newcommand{\be}{\begin{equation}}
\newcommand{\bea}{\begin{eqnarray}}

\newcommand{\ee}{\end{equation}}
\newcommand{\eea}{\end{eqnarray}}

\providecommand{\gsim}{\gtrsim}

\begin{document}

\renewcommand{\thefootnote}{\fnsymbol{footnote}}

\makeatletter
\@addtoreset{equation}{section}
\makeatother
\renewcommand{\theequation}{\thesection.\arabic{equation}}

\rightline{}
\rightline{}
%   \vspace{1.8truecm}

%\begin{flushright}
% preprint nrs.
%\end{flushright}

%\vspace{10pt}

%%%%%%%%%%%%%%%%%

\begin{center}
{\LARGE \bf{\sc Holographic cosmology solutions of problems with pre-inflationary cosmology}}
\end{center} 
 \vspace{1truecm}
\thispagestyle{empty} \centerline{
{\large \bf {\sc Horatiu Nastase${}^{a}$}}\footnote{E-mail address: \Comment{\href{mailto:horatiu.nastase@unesp.br}}{\tt horatiu.nastase@unesp.br}}
%{\bf{\sc and}}
%{\large \bf {\sc Kostas Skenderis${}^{b}$}}\footnote{E-mail address: \Comment{\href{mailto:K.Skenderis@soton.ac.uk}}{\tt 
%K.Skenderis@soton.ac.uk}}
                                                        }

\vspace{.5cm}

%\vspace{.3cm}

\centerline{{\it ${}^a$Instituto de F\'{i}sica Te\'{o}rica, UNESP-Universidade Estadual Paulista}} 
\centerline{{\it R. Dr. Bento T. Ferraz 271, Bl. II, Sao Paulo 01140-070, SP, Brazil}}
%\vspace{.3cm}
%\centerline{{\it ${}^b$STAG Research Center and Mathematical Sciences, University of Southampton,}}
%\centerline{{\it Highfield, Southampton SO17 1BJ, United Kingdom}}

\vspace{1truecm}

%%%%%%%%%%%%%%%%%
\thispagestyle{empty}

\centerline{\sc Abstract}

\vspace{.4truecm}

\begin{center}
\begin{minipage}[c]{380pt}
{\noindent In this paper we describe in detail how to solve the problems of pre-inflationary cosmology within the holographic cosmology
model of McFadden and Skenderis \cite{McFadden:2009fg}. The solutions of the smoothness and horizon problems, the 
flatness problem, the entropy and perturbation problems and the baryon asymmetry problem are shown, and the mechanisms
for them complement the inflationary solutions. Most of the paper is devoted to the solution of the monopole relic problem, 
through a detailed calculation of 2-loop correlators of currents in a toy model which we perform in $d$ dimensions, in order
to extract its leading dependence on $g^2_{\rm eff}=g^2_{YM}N/q$ and find a dilution effect.
Taken together with the fact that holographic cosmology gives as good a fit to CMBR as the standard 
paradigm of $\Lambda$ CDM with inflation, it means holographic cosmology extends the inflationary paradigm into 
new corners, explorable only through a dual perturbative field theory in 3 dimensions. 

}
\end{minipage}
\end{center}

\vspace{.5cm}

\setcounter{page}{0}
\setcounter{tocdepth}{2}

\newpage

%\tableofcontents
\renewcommand{\thefootnote}{\arabic{footnote}}
\setcounter{footnote}{0}

\linespread{1.1}
\parskip 4pt

%{}~
%{}~

%---------------------------------------------------------

%%%%%%%%%%%%%%%%%%%%%%%%%%%%%%%%%%%%%%%%%%%%%%%%%%%%%%%%%%%%%%%%%%%%%%%%%%%%%%%%%%%%%%%%
\section{Introduction}
%%%%%%%%%%%%%%%%%%%%%%%%%%%%%%%%%%%%%%%%%%%%%%%%%%%%%%%%%%%%%%%%%%%%%%%%%%%%%%%%%%%%%%%%

Inflationary cosmology \cite{Brout:1977ix, Starobinsky:1979ty, Starobinsky:1980te, Sato:1980yn, Guth:1980zm, Linde:1981mu, Albrecht:1982wi} 
is currently considered the leading paradigm for the physics of the very early universe.  
Inflation  together with cold dark matter and a late time cosmological constant form the theoretical underpinnings
of the $\Lambda$CDM model, the concordance model of cosmology, which has been very successful 
in explaining astronomical observations. 
This framework is often assumed to be the only paradigm capable of describing cosmology as a whole, including explaining the cosmic 
microwave background radiation (CMBR) and solving the cosmological problems
of pre-inflationary cosmology. In this paper we will argue that the holographic models of \cite{McFadden:2009fg}, which
describe a strongly interacting non-geometric very early Universe, are also capable of explaining both the CMBR observations and 
also resolving the classic puzzles of the hot Big Bang puzzles, elaborating on our discussion in \cite{Nastase:2019rsn},

In inflation quantum fluctuations of the graviton plus the inflaton are treated perturbatively around an accelerating 
Friedmann-Lema\^itre-Robertson-Walker (FLRW) background. One may justify this 
treatment  by viewing inflation an effective field theory, if the scale of inflation is sufficiently smaller than the Planck scale. 
Nevertheless, it has been known for a long time that if we have sufficiently many e-folds of inflation we eventually go back 
to a time where gravity was strongly coupled, the so-called trans-Planckian problem see for example \cite{Brandenberger:2012aj} 
and references therein (but see also \cite{Starobinsky:2001kn} or the more recent \cite{Dvali:2020cgt} for the counter point of view). 
Even taking this issue aside, one needs {\em assumptions} about the initial conditions for the quantum fluctuations, usually taken to 
be the Bunch-Davies vacuum, which may also be afflicted by strong coupling issues. These issues together with the difficulty to find 
(quasi)-de Sitter solutions in string theory led to questions about the very existence of de Sitter in quantum gravity \cite{Obied:2018sgi} 
and led to swampland conjectures related to the trans-Planckian issues \cite{Bedroya_2020,Bedroya:2019tba} (see also
\cite{Brahma:2019vpl}); in particular, the 
Trans-Planckian Censorship Conjecture (TCC) was also found to constrain holographic cosmology in \cite{Bernardo:2019bbi}. 

It is thus imperative to both understand how to embed conventional inflation in a UV complete theory and also further develop 
alternatives to conventional inflation. Holographic cosmology offers the possibility to address both: conventional inflation is 
associated with dual QFTs which are strongly coupled while we obtain qualitatively new models for the very early Universe 
when the dual QFT is weakly interacting. The focus in this paper will be on the new models, which describe a non-geometric 
very early Universe, but we will also discuss universal features of the holographic framework.

The holographic models are defined by giving the three-dimensional QFT and the holographic dictionary.
The models of \cite{McFadden:2009fg} describing the non-geometric very early Universe are based on a set of super-renormalizable, 
large $N$ gauge theories in 3 dimensions, 
with a generalized conformal structure. Remarkably, these models fit the CMBR data as well as $\Lambda$CDM plus inflation
\cite{Afshordi:2016dvb,Afshordi:2017ihr}, despite 
the fact that the form of the power spectra is qualitatively different than that of inflationary models. One takes a 
phenomenological point of view, 
and fixes the parameters of a general action by fitting against the CMBR data, and in the comparison of the best fit to $\Lambda$ CDM 
plus inflation, one obtains a $\chi^2$ 
of $824.0$ for holographic cosmology vs. $823.5$ for $\Lambda$ CDM (table V in \cite{Afshordi:2017ihr}), so within half a 
sigma.

%In these models 
%the  the initial conditions 
%issue and the strong gravity possibility are solved by mapping to a (perturbative) 3 dimensional field theory. 
%This is the approach of the holographic cosmology, in the version 
%of McFadden and Skenderis ,
%which offers the possibility of extending the inflationary paradigm to domains previously unexplored, of strong gravity 
%but weak (perturbative) field theory, thus allowing us to calculate cosmological quantities using standard 3 dimensional 
%quantum field theory. One takes a phenomenological approach for the field theory dual to the cosmology, singling out a 
%set of super-renormalizable, large $N$ gauge theories in 3 dimensions, with a generalized conformal structure, and fixing 
%(part of) its parameters from observations.
%Moreover, in 2 remarkable papers , it was shown that 
%the (new, but smoothly connected to the inflationary one) paradigm of perturbative phenomenological 
%field theory dual to cosmology  
%matches the CMBR data as well as the $\Lambda$ CDM with inflation, namely that one obtains a $\chi^2$ 
%of $824.0$ for holographic cosmology vs. $823.5$ for $\Lambda$ CDM (table V in \cite{Afshordi:2017ihr}), so within half a 
%sigma. 

Given this fact, a natural question to ask is what about the cosmological problems (Hot Big Bang puzzles)
of pre-inflationary cosmology solved
by inflation in \cite{Guth:1980zm, Linde:1981mu, Albrecht:1982wi}? We have found in \cite{Nastase:2019rsn} 
that the answer is, they are also solved, 
either in similar way, or in 
new ways. In this article we explain in detail the analysis of \cite{Nastase:2019rsn}. In particular, 
the bulk of the paper is devoted to 
the detailed calculation of the 2-point function of global currents  in a toy model for the phenomenological field theory, 
which shows that there is a certain dilution 
effect as time evolves, allowing for a created monopole-type  perturbation in the cosmology to be scaled away, just like 
it is in inflation. 

During the calculation of the Feynman diagrams, we will also calculate several integrals in dimensional regularization, 
as well as find an algorithmic way to obtain the relevant divergences of the integrals, which can be interesting in their own way.

The paper is organized as follows. In section 2 we review holographic cosmology, and in section 3 we
review the puzzles of Hot Big Bang cosmology that led to inflation, and their solutions in
inflation. In section 4 we present a toy model for the resolution of the monopole problem
in holographic cosmology, and calculate the two-point function of currents in it, the main
aim of the paper. In section 5 we present the solutions of the problems of Hot Big Bang
cosmology within holographic cosmology, and in section 6 we conclude. In Appendix A we
present a set-up for an $x$-space calculation of the same two-point correlator of currents,
in Appendix B we calculate the integrals appearing in the Feynman diagrams in momentum space, in Appendix C
we present the details of the calculation of the same integrals, and in Appendix D we review particle-vortex duality.

\section{Holographic cosmology}

The idea that quantum gravity is holographic, initially developed in \cite{tHooft:1993dmi, Susskind:1994vu, Maldacena:1997re},
is now widely accepted, and is at the basis of the AdS/CFT correspondence \cite{Maldacena:1997re}
(see the books  \cite{Nastase:2015wjb,Ammon:2015wua} for more information). Specifically, it means that a (quantum)
gravitational theory must be described by a theory without gravity and with one dimension less.

It is then a natural question to ask, what happens for a cosmological theory? While still conjectural, there is a lot of 
evidence that cosmology is also holographic, and is described by a three-dimensional Euclidean QFT. Work on this was 
initiated in \cite{Witten:2001kn, Strominger:2001pn, Strominger:2001gp, Maldacena:2002vr}, and it was shown that 
standard, weakly coupled inflation corresponds to a strongly coupled QFT  (see for example \cite{Maldacena:2011nz, Hartle:2012qb, Hartle:2012tv,Schalm:2012pi, Bzowski:2012ih, Mata:2012bx, Garriga:2013rpa, McFadden:2013ria, Ghosh:2014kba, Garriga:2014ema, Kundu:2014gxa, Garriga:2014fda, McFadden:2014nta, Arkani-Hamed:2015bza, Kundu:2015xta, Hertog:2015nia,Garriga:2015tea,  Garriga:2016poh,Hawking:2017wrd,Arkani-Hamed:2018kmz}).

What we are interested in here, however, is the case of a strongly coupled cosmology, corresponding to a weakly 
coupled quantum field theory, for which  the holographic cosmology model was developed 
in \cite{McFadden:2009fg,McFadden:2010na}. The phenomenology of these models has been worked out in 
\cite{McFadden:2009fg, McFadden:2010na, McFadden:2010vh, McFadden:2011kk, Bzowski:2011ab, Coriano:2012hd, Kawai:2014vxa,McFadden:2010jw}, using methods from 
\cite{Skenderis:2002wp,Papadimitriou:2004ap, Papadimitriou:2004rz, Maldacena:2002vr}.

The cosmology that we want to describe is a 3+1 dimensional FLRW metric with scale factor $a(t)$, coupled to a 
scalar with background $\phi(t)$, and with fluctuations for both the spatial metric components, $h_{ij}(t,\vec{x})$, 
and the scalar field, $\delta \phi(t,\vec{x})$, combining into the usual transverse traceless tensor perturbations $\gamma_{ij}
(t,\vec{x})$ and gauge invariant scalar $\zeta(t,\vec{x})$. 

Therefore the cosmology has metric and scalar 
\bea
ds^2&=&-dt^2+a^2(t)[\delta_{ij}+h_{ij}(t,\vec{x})]dx^i dx^j\;,\cr
\Phi(t,\vec{x})&=&\phi(t)+\delta\phi(t,\vec{x})\;,
\eea
and the action
\be
S=-\frac{1}{2\kappa^2}\int d^4x \sqrt{-g}[-R+(\d_\mu\phi)^2+2\kappa^2V(\Phi)].
\ee

In order to construct a holographic model, one first needs to a certain Wick rotation, i.e., a ``domain wall/cosmology 
correspondence'' \cite{Skenderis:2006jq},
by writing $t=-iz$, leading to a solution 
\bea
ds^2&=&+dz^2+a^2(z)[\delta_{ij}+h_{ij}(z,\vec{x})]dx^i dx^j\;,\cr
\Phi(z,\vec{x})&=&\phi(z)+\delta\phi(z,\vec{x})\;,
\eea
for the action
\be
S=+\frac{1}{2\bar \kappa^2}\int d^4x \sqrt{g}[-R+(\d_\mu\phi)^2+2\kappa^2V(\Phi)].
\ee

In this formulation, we can consider $z$ as a holographic coordinate (energy in field theory) in  a gravity dual 
solution, and the solution as a kind of a domain wall. 

We consider that this background, for $a(z)$ that is either exponential (AdS space) or power-law (domain wall) 
corresponds phenomenologically to a certain $SU(\bar N) $ gauge theory at large $\bar N$. In order to go back to the 
original Lorentzian signature space, i.e., to the cosmology, we must perform the Wick rotation
\be
\bar \kappa^2=-\kappa^2\;,\;\;\; \bar q =-iq\;,
\ee
corresponding in field theory to 
\be
\bar q=-iq\;,\;\;\;\; \bar N=-iN.
\ee

The cosmological quantities of interest for the CMBR observations are the scalar and tensor power spectra, coming 
from the momentum space two-point functions of the perturbations $\gamma_{ij}$ and $\zeta$, 
\bea
 \Delta_S^2(q)&\equiv& \frac{q^3}{2\pi^3}\langle \zeta(q)\zeta(-q)\rangle \cr
 \Delta_T^2(q)&\equiv & \frac{q^3}{2\pi^3}\langle \gamma_{ij}(q)\gamma_{ij}(-q)\rangle.\label{DeltaTS}
 \eea

A holographic computation, based either on the formalism
 developed in \cite{Skenderis:2000in,Papadimitriou:2004ap,Papadimitriou:2004rz}, or alternatively 
 (see for instance \cite{Afshordi:2017ihr}) on the holographic relation proposed by Maldacena between the 
 wave function of the Universe in cosmology $\psi(\Phi)$ and the partition function of the field theory $Z[\Phi]$, 
 namely $Z[\Phi]=\psi(\Phi)$, for the case of inflation \cite{Maldacena:2002vr} and extended to this case, gives 
 a result for the power spectra in terms of the energy-momentum two-point functions in the field theory, 
 \bea
 \Delta_S^2(q)&=& -\frac{q^3}{16\pi^2{\rm Im}B(-iq)} \cr
 \Delta_T^2(q)&= &-\frac{2q^3}{\pi^2{\rm Im}\label{DeltaTSAB}
 A(-iq)}\;,
 \eea
where we have already performed the analytical continuation to Lorentzian signature through $\bar q=-iq$ and $\bar N=-iN$.
Here $A$ and $B$ are the coefficients of the expansion of the energy-momentum tensor two-point function into 
given Lorentz structures, 
\be
\langle T_{ij}(\bar q)T_{kl}(-\bar q)\rangle =A(\bar q) \Pi_{ijkl}+B(\bar q)\pi_{ij}\pi_{kl}\;,\label{TTAB}
\ee
where 
\be
\Pi_{ijkl}=\pi_{i(k}\pi_{l)j}-\frac{1}{2}\pi_{ij}\pi_{kl}\;,\;\;\;
\pi_{ij}=\delta_{ij}-\frac{\bar q_i\bar q_j}{\bar q^2}
\ee
are the 4-index transverse traceless projection operator ($\Pi_{ijkl}$), and the 2-index transverse projection operator
($\pi_{ij}$).

The Euclidean field theory corresponding to the domain wall gravity dual is a super-renormalizable $SU(N)$ gauge 
theory with gauge field $A_i=A_i^aT_a$, scalars $\phi^M =\phi^{aM}T_a$ and fermions $\psi^L=\psi^{aL}T_a$, 
all in the adjoint representation ($T_a$ are the generators of $SU(N)$), and with flavor indices $M,L$. In 3 dimensions, 
Yukawa couplings and $\phi^4$ couplings are dimensional, therefore super-renormalizable, and there are no higher powers 
of fields with dimensional couplings. Therefore the phenomenological action considered is
\bea
S_{\rm QFT}&=&\int d^3x \Tr\left[\frac{1}{2}F_{ij}F^{ij}+\delta_{M_1M_2}D_i\Phi^{M_1}
D^i\Phi^{M_2}+2\delta_{L_1L_2}\bar \psi^{L_1}\gamma^i D_i
\psi^{L_2}\right.\cr
&&\left.+\sqrt{2}g_{YM}\mu_{ML_1L_2}\Phi^M \bar \psi^{L_1}\psi^{L_2}
+\frac{1}{6}g^2_{YM}\lambda_{M_1...M_4}\Phi^{M_1}...\Phi^{M_4}\right]\cr
&=&\frac{1}{g^2_{YM}}\int d^3x \Tr\left[\frac{1}{2}F_{ij}F^{ij}+\delta_{M_1M_2}D_i\Phi^{M_1}D^i\Phi^{M_2}
+2\delta_{L_1L_2}\bar \psi^{L_1}\gamma^i D_i
\psi^{L_2}\right.\cr
&&\left.+\sqrt{2}\mu_{ML_1L_2}\Phi^M \bar \psi^{L_1}\psi^{L_2}+\frac{1}{6}
\lambda_{M_1...M_4}\Phi^{M_1}...\Phi^{M_4}\right]\;,\label{phenoaction}
\eea
plus a nonminimal coupling of gravity to the scalar $1/(2g^2_{YM})\int \xi_M R(\Phi^M)^2$,
where in the second form we have rescaled the fields by $g_{YM}$, in order to have $g^2_{YM}$ as a common factor. 
The coupling constants $\lambda_{M_1...M_4}$ and $\mu_{ML_1L_2}$ are dimensionless, $\Tr[T_aT_b]=\frac{1}{2}
\delta_{ab}$ here, and in the first line the dimensions of the fields are $[A_i]=1/2= [\Phi^M]$ and $[\psi^L]=1$, whereas in 
the second they are as in 4 dimensions, $[A_i]=1=[\Phi^M]$ and $[\psi^L]=3/2$. 

This last form of the action shows the 
property of ``generalized conformal structure'', since this theory has the same properties as the dimensional reduction of a 4-dimensional 
conformal field theory (the dimensional reduction of a 4-dimensional conformal theory has generalized conformal structure, but 
a theory with generalized conformal structure is not necessarily the dimensional reduction of a conformal theory): 
the dimensions are contained in powers of the momenta only, and they appear through the 
effective coupling 
\be
g^2_{\rm eff}=\frac{g^2N}{q}.
\ee
Another way of saying it is that if one promotes $g^2_{YM}$ to a field with appropriate conformal 
transformations the theory becomes conformal  \cite{Jevicki:1998yr, Kanitscheider:2008kd}.

For the CMBR, we are interested in the two-point functions of energy-momentum tensors, specifically the 
coefficients $A$ and $B$, that appear in the holographic calculation of the power spectra. Considering their classical 
dimensions, and moreover the fact that they scale as $N^2$ in the large $N$ limit, from the 
generalized conformal structure we find the general scaling forms (the 1/4 in $B$ is conventional)
\be
A(q,N)=q^3N^2f_T(g^2_{\rm eff})\;,\;\;\;
B(q,N)=\frac{1}{4}q^3 N^2 f(g^2_{\rm eff})\;,
\ee
and an explicit calculation in the phenomenological action (\ref{phenoaction}) finds the two-loop form
\bea
f(g^2_{\rm eff})&=& f_0\left[1-f_1 g^2_{\rm eff}\ln g^2_{\rm eff}+f_2 g^2_{\rm eff}+{\cal O}(g^4_{\rm eff})\right]\cr
f_T(g^2_{\rm eff})&=& f_{T0}\left[1-f_{T1} g^2_{\rm eff}\ln g^2_{\rm eff}+f_{T2} g^2_{\rm eff}
+{\cal O}(g^4_{\rm eff})\right].
\eea
Here $f_0$ and $f_{T0}$ are obtained from a 1-loop computation, and $f_1, f_{T1}$ and $f_2, f_{T2}$ are 
obtained from a two-loop computation. 
In this formula (coming from a quantum field theory calculation), 
we can set the RG scale $\mu$ equal to the pivot scale $q_*$ of the observational CMBR spectrum.

Another quantity of interest for this paper is the two-point function of (nonabelian) global symmetry currents $j_i^A$, 
where $A$ belongs to the adjoint representation of some global symmetry group $G$. By a similar reasoning, 
the two-point function should take the form 
\be
\langle j_i^A(q)j_k^B(-q)\rangle =N^2q\delta^{AB}\pi_{ik} f_J(g^2_{\rm eff})\;,
\ee
where again 
\be
f_J(g^2_{\rm eff})= f_{J0}\left[1-f_{J1} g^2_{\rm eff}\ln g^2_{\rm eff}+f_{J2} g^2_{\rm eff}+{\cal O}(g^4_{\rm eff})\right].
\ee

The cosmological power spectra in terms of the coefficients $A$ and $B$ are then given by 
\be
\Delta^2_S(q)=\frac{q^3}{4\pi^2N^2 f(g^2_{\rm eff})}\;,\;\;\;
\Delta^2_T(q)=\frac{2q^3}{\pi^2N^2 f_T(g^2_{\rm eff})}\;,
\ee
which means one can parametrize the power spectra as 
\bea
\Delta^2_S(q)&=& \frac{\Delta_0^2}{1+\frac{gq_*}{q}\ln \left|\frac{q}{\b g q_*}\right|+{\cal O}\left(\frac{gq_*}{q}
\right)^2}\cr
\Delta^2_T(q)&=& \frac{\Delta_{0T}^2}{1+\frac{g_Tq_*}{q}\ln \left|\frac{q}{\b_T g q_*}\right|+{\cal O}
\left(\frac{g_Tq_*}{q}\right)^2}\;,\label{spectrumholo}
\eea
where 
\be
\Delta_0^2=\frac{1}{4\pi^2 N^2 f_0}\;,\;\;\;\;
\Delta_{T0}^2=\frac{2}{\pi^2N^2 f_{T0}}.
\ee

This parametrization, differing from the one coming from ($\Lambda$ CDM plus)  inflationary cosmology, was found 
in \cite{Afshordi:2016dvb,Afshordi:2017ihr} to be as good a fit to the CMBR data
(these models were previously compared against WMAP data in \cite{Dias:2011in, Easther:2011wh}), and to 
fix the parameters of the phenomenological model in a simplified version, as we already mentioned. 
The model also predicts non-Gaussianity of exactly factorisable equilateral shape 
with $f_{NL}^{\rm equil}=5/36$ \cite{McFadden:2010vh}.
Progress towards deriving (as a ``top-down'' model) the phenomenological set-up described here
from a modification of the usual $AdS_5\times S^5$ vs. ${\cal N}=4$ SYM gravity dual pair was made in 
\cite{Bernardo:2018cow}, but it doesn't seem to be in the region matching the CMBR data. 
The holographic cosmology set-up also maps the cosmological constant problem in gravity to a 
solved one in field theory \cite{Nastase:2018cbf}.

\section{Questions and their solutions in inflation}

We now want to see that the holographic cosmology  does as well as inflation also for the pre-inflationary problems 
that it solved. In order to do that, in this section we will first review the problems and their solutions in inflation. 

1. {\bf Smoothness and horizon problems}

The question that needed to be answered was, why is the Universe uniform and isotropic?

When we look at the sky on the largest scales, we see a remarkably uniform and isotropic Universe, up to small fluctuations.
In particular, the CMBR is uniform up to the order $10^{-5}$ fluctuations, and even those are correlated on the sky. 
All of this points to causal correlation, but the light rays that we see come from different parts of the sky, and from the 
moment of decoupling, shortly after the Big Bang, when they shouldn't have been in causal contact. 

One can be quantitative about this issue. Denote by $d_H(t_0)$ the horizon distance at the time of last scattering, when 
the CMBR was emitted, translated into today's scales, 
\be
d_H(t_0)=a(t_0)\int_0^{t_{\rm ls}}\frac{dt'}{a(t')}\;,
\ee
and by $r_H(t_0)$ the distance travelled by light from the time of last scattering until today, when we detect it, 
\be
r_H(t_0)=a(t_0)\int_{t_{\rm ls}}^{t_0}\frac{dt'}{a(t')}.
\ee

Assuming that the Universe was radiation dominated all of the time (actually, most of the time, but that is enough, 
since the contribution of the really early times, when we don't know what happens, is assumed to be small here) {\em 
before} the moment of last scattering, we can calculate the ratio of $2r_H(t_0)$, the size we observe to be causally 
connected and correlated, to $d_H(t_0)$, which should be correlated if nothing new appears, and we find
\bea
N&=&\frac{2r_H(t_0)}{d_H(t_0)}\simeq 2\left(\frac{t_0}{t_{\rm ls}}\right)^{1/3}=2\left(\frac{a_0}{a_{\rm ls}}\right)^{1/2}\cr
&=&2(1+z_{\rm ls})^{1/2}\simeq 72.
\eea

That means that we need to increase the size of the horizon, or more precisely the {\em size of a 
causally connected patch} by at least 72-fold at the time of last scattering, when 
the CMBR was created, if we are to match observations. 

In usual cosmology, the size of a patch increases with the expansion of the Universe as $\propto a(t)\propto t^n$, with 
$n<1$, corresponding to an equation of state $w>-1/3$ (true both for radiation domination and for matter domination), 
whereas the horizon size grows linearly with time, both the Hubble horizon $H^{-1}=(\dot a /a)^{-1}\propto t$ and 
the particle horizon $d_H(t_{\rm ls})\propto 
t_{\rm ls}$, thus horizons grow faster than scales. 
That means that the particle horizon size is the right measure of causal connection 
of points in the sky.

But inflation's answer to how it is possible to increase the size of the causally connected patch is to exponentially (or in any case, 
at least polynomially with $n>1$) blow up a small patch that will create the whole Universe, even outside the boundary of the
current horizon. The patch will get outside the (almost constant) Hubble horizon $H^{-1}$ during (exponential)
inflation but, more 
relevant for us, also the particle horizon $d_H$ will grow larger relative to the distance travelled by light $r_H$ with the 
needed amount. If inflation starts at $t_{\rm bi}$ and ends at $t_I$, with $N_e=H_I(t_I-t_{\rm bi})$ number of e-folds, 
since the early times dominate the integral due to the exponential inflation, we find the particle horizon
\be
d_H(t_{\rm ls})\simeq \frac{a(t_{\rm ls})}{a(t_I)}\int_{t_{\rm bi}}^{t_I}dt\; e^{H_I(t_I-t_{\rm bi})}\simeq 
\frac{a(t_{\rm ls})}{a(t_I)H_I}e^{N_e}\;,\label{dh}
\ee
whereas the distance travelled from last scattering to now, but measured at last scattering $t_{\rm ls}$ is
(note that $H_0=2/(3t_0)$ now, during matter domination)
\be
r_H(t_{\rm ls})=a(t_{\rm ls})\int_{t_{\rm ls}}^{t_0}\frac{dt'}{a(t')}\simeq \frac{2a(t_{\rm ls})}{a_0 H_0}.
\ee

Then the condition that the light from the CMBR is causally correlated in the sky today is 
\be
\frac{d_H(t_{\rm ls})}{2r_H(t_{\rm ls})}>1\;,
\ee
leading to 
\be
e^{N_e}> \frac{a(t_I)H_I}{a_0 H_0}\;,\label{condition}
\ee

A standard analysis leads then to a bound on the number of e-folds of inflation, 
\be
e^{N_e}\gsim e^{56}\frac{\rho_{\rm begRD}^{1/4}}{5\times 10^{13}GeV}\;,\label{condition2}
\ee
where $\rho_{\rm begRD}$ is the energy density at the beginning of the radiation dominated era. 

We see that the result of inflation is that solving the smoothness and horizon problems is turned into a quantitative bound 
on the number of e-folds of inflation.

2. {\bf Flatness problem}

The question was, why do we have $\Omega \simeq 1$ in the past?

Experimentally, we know that $\Omega\simeq 1$ today with only an approximate precision, so we can assume that there 
is some deviation. But the time evolution of this deviation is 
\be
\Omega(t)-1=\frac{k}{a(t)^2H(t)^2}\propto \left(\frac{t}{a(t)}\right)^2\propto t^{2(1-p)}\;,
\ee
for $a(t)\propto t^p$, which means that during the matter dominated ($p=2/3$) and radiation dominated ($p=1/2$)
eras, with $p<1$, $\Omega(t)-1$ actually grows with time, thus it was even smaller in the past, giving an 
unacceptable fine-tuning. 

That means that the simplest way to get rid of this fine-tuning is to consider a period of inflation, with $p>1$ or exponential, 
during which time $\Omega(t)-1$ decreases drastically, to then increase back until today. For exponential inflation,
we obtain the time evolution 
\be
\Omega (t)-1=\frac{k}{a(t)^2H^2}\propto e^{-2 H_I  t}.
\ee

That means that we can relate the value of $\Omega-1$ today, $\Omega_0-1$, to its value at the beginning of inflation, 
$\Omega(t_{\rm bi})-1$, as 
\bea
\Omega_0-1&=&\frac{k}{a_0^2H_0^2}=\frac{k}{a_{\rm bi}^2H_{bi}^2}e^{-2N_e}\left(\frac{a(t_I)H_I}{a_0 H_0}\right)^2\cr
&=&(\Omega(t_{bi})-1)
e^{-2N_e}\left(\frac{a(t_I)H_I}{a_0 H_0}\right)^2.
\eea

Then to solve the flatness problem, and not have any fine-tuning, we assume that initially we had a large 
deviation from flatness, i.e., $\Omega(t_{\rm bi})-1>\Omega_0-1$, 
which leads to the same condition (\ref{condition}) on the number of e-folds as was obtained from the smoothness and the 
horizon problems. 

However, for the solution of the flatness problem in holographic cosmology, it is useful to instead put some numbers 
in the time evolution, and find what is the actual ratio of $\Omega-1$ at the end of the (would-be) inflationary time, 
now associated with the end of the holographic cosmology period, to the one today, $\Omega_0-1$.
Using the radiation domination evolution until $e^+e^-$ annihilation, we find that a value of order 1 of $\Omega-1$ 
today turns into  $\Omega-1\sim 10^{-16}$ at $e^+e^-$ annihilation. Further assuming the same radiation domination 
also down to the end of (would-be) inflation $t_I$, or the end of the holographic cosmology period, we find 
\be
(\Omega-1)_I=(\Omega-1)_{e^+e^-}\left(\frac{a_e H_e}{a_I H_I}\right)^2=10^{-16}
\left(\frac{T_e}{T_I}\right)^2\sim 10^{-54}\;,\label{fluct}
\ee
since $T_{e^+e^-}\sim 1 MeV$, and we assumed that the end of inflation, or of holographic 
cosmology, is at a temperature  $T_I=T_{\rm inflation}\sim 10^{16}GeV$.

It follows then that we would need a reduction factor of about $10^{-54}$ in $\Omega-1$ to solve the flatness problem, 
and this is the same factor that appears in the smoothness and horizon problems. 

3. {\bf Relic and monopole problem}

The question is now in two parts: why don't we see general relics in the Universe, and in particular, why don't we see
monopoles, which are generated during high scale phase transitions like GUT phase transitions, roughly one per horizon 
volume at the time. 

We know that there should be some phase transitions happening at high energies (high temperatures), either happening 
when compactifying a more fundamental supergravity or string theory, or at an intermediate stage, via a field theoretical 
grand unified theory (GUT) phase transition. But if the phase transition happens in an expanding Universe, with an expanding 
horizon size, the Kibble mechanism guarantees that one generates about one monopole per nucleon. Indeed, when the 
effective potential for the GUT symmetry breaking scalar goes through the phase transition (changes as the 
temperature drops), and the minimum is not at zero anymore, but at an arbitrary direction, and a
 value at the minimum of the potential, patches of arbitrary directions for the scalar, of horizon size, develop. When these
patches join, they generically form a monopole solution (with a topological charge given by the scalar orientation), in a 
volume of the order of the horizon size. A similar mechanism generates also a nucleon from constituent partons, leading 
to roughly one monopole per nucleon (assuming thermal equilibrium). More precisely, we know that the there are 
about $10^9$ photons per nucleon today, so one generates about one monopole per $10^9$ photons. 

However, direct experimental searches for monopoles in materials on Earth show that there are less than $10^{-30}$ 
monopoles per nucleon (\cite{Zeldovich:1978wj}, see also \cite{Weinberg:2008zzc}, chapter 4.1.C), 
so we need a reduction factor of at least $10^{-30}$ 
per volume, or $10^{-10}$ per linear size, for the density of monopoles in the Universe.

But the Kibble mechanism is also valid for non-magnetic relics generated in phase transitions, like for instance cosmic 
strings, domain walls, etc. In this case, the constraints on their existence don't come from direct searches, but rather 
from their gravitational effects on the Universe, in case they would exist in space. In order to not over-close the Universe, 
we need a reduction factor in the number density of relics of about $10^{-11}$ (the details are in \cite{Kolb:1990vq}, 
chapter 7.5). This is much less stringent than for monopoles.

Inflation dilutes the monopoles and relics, specifically their number density, through the period of exponential inflation. 
For that to be true, we need the GUT phase transition to happen either before, or during inflation.
Of course, any original density gets similarly diluted by the expansion, but now photons are mostly created at the end 
of inflation, during reheating, after which the Universe is in thermal equilibrium (before, it wasn't), so the net effect of 
inflation is to dilute the monopole (or relic) to photon ratio by the amount of expansion. 

Since we need the effect of inflation to be an increase of $10^{10}$ in linear size for the monopoles, the phase transition 
must occur at least a number of e-folds of 
\be
N_e>\ln 10^{10}\simeq 23
\ee
before the end of inflation. It is now redundant to impose the condition for generic relics, for which a reduction of only 
$10^{11}$ in volume, or about $10^4$ in linear size, is needed until the end of inflation. 

However, in the holographic cosmology case we will see that the dilution of generic relics and of monopoles has 
different origins, so we need to remember both results.

4. {\bf Entropy problem}

We want to understand why is the entropy in the Universe so large?

The entropy per baryon today is about $10^9$, for a total of about $10^{88}$ for the entropy inside the horizon volume 
today. But the total entropy increases with time due to the increase of the horizon volume, so we must consider the 
entropy within the horizon volume at the last time we understand very well, of the Big Bang Nucleosynthesis. 
Using the radiation and matter dominated evolution formulas, given that $s\propto a(t)^{-3}$ and $S_H=sH^{-3}
\propto (t/a(t))^3$, we find that $S_H(t_{\rm BBN})\sim 10^{63}$. On the other hand, 
at the end of a phase transition, we expect to have numbers of the order one per horizon. 

The solution of inflation to the entropy problem is that there is a large generation of entropy, in the form of photons per 
particle, during reheating, which transforms the initial quantum fluctuations. The exponential expansion leads to a  large 
volume, further increasing the entropy inside the horizon. 

Since the energy density of radiation in equilibrium is $\rho_R\propto T^4$, and the entropy per (comoving) volume 
also in equilibrium is $s\propto T^3$, during the reheating phase $s\propto \rho_R^{3/4}$, and one finds the 
energy density of radiation during reheating behaves as $\rho_R\propto a^{-3/2}$ (adiabatic expansion would give 
$s\propto a^{-4}$, so photons are created), the total entropy in a comoving volume increases,
\be
S\propto a^3\propto a^3 \rho_R^{3/4}\propto a^{15/8}.
\ee

5. {\bf Perturbations problem}

We want to understand how to generate perturbations in the Universe.

The puzzling point is that the perturbations that we see in the CMBR are classical, not quantum, and 
were super-horizon in the past, therefore they were always classical. Moreover, even the perturbations on smaller scales, 
responsible for creating structure (galaxies, etc.), if we go enough in the past, were super-horizon, so they were always 
classical. Indeed, scales grow with $a(t)$, but the (particle or Hubble) horizon size goes like $t$, which in the matter 
dominated era goes like $a(t)^{3/2}$, and in the radiation dominated era as $a(t)^2$, therefore as time goes by, scales 
fall {\em inside} the horizon. How were these perturbations created then? 

In inflation, the answer is that scales grow exponentially, but the (Hubble) horizon size $H^{-1}$ is approximately 
constant, which means that all scale are quickly blown up outside the horizon. Then initial quantum fluctuations, 
generated because of quantum field theory in curved spacetime, go outside the horizon, where they are frozen in, 
becoming classical, and growing with the scale. Eventually, they come back inside the horizon during regular 
cosmology, but now as classical fluctuations. 

6. {\bf Baryon asymmetry problem}

The last question is, why is the baryon number nonzero, and yet so small?

The baryon asymmetry  $(N_B-N_{\bar B})/N_B\sim 10^{-9}$ must have been created at some early time, usually 
considered to be around the time of a GUT transition, through some baryogenesis mechanism. But Sakharov gave the 
necessary and sufficient conditions for such a mechanism: 1) To have a mechanism for baryon number 
violation, which is true in a GUT theory, where proton can decay and be created through ``leptoquark'' transitions; 
2) to have a CP violation in the theory, which is true in the Standard Model, and perhaps enhanced in a GUT theory; and 
3) to have interactions out of equilibrium. If conditions 1 and 2 are about particle physics, condition 3 is about 
cosmology, so it needs to be explained in a cosmological theory. Usual cosmology was assumed to be in equilibrium, so 
it could not produce baryon asymmetry. 

But the essential point of inflation is that evolution is very fast. In exponential inflation, the result is that the Hubble 
time (Hubble horizon) $H^{-1}$ is very small and constant, smaller than the equilibration time for reactions, so 
reactions happen out of equilibrium, allowing for baryon asymmetry to be created. On the other hand, the smallness of the 
baryon asymmetry is due to the large entropy per baryon ($\sim 10^9$), leading to a small baryon asymmetry ($\sim 
10^{-9}$), so smallness of the latter is related to the solution of the entropy problem.

\section{Toy model for the monopole problem in holographic cosmology}

In this section we perform the main computation of the paper, which will be needed to solve the monopole 
problem in holographic cosmology, by ``diluting'' an initial monopole perturbation in the bulk cosmology.

The energy-momentum tensor in field theory $T_{\mu\nu}$ couples to the graviton perturbation $h_{\mu\nu}$
in the bulk, so 
gravity tensor and scalar fluctuation correlators and their evolution 
will be calculated from the correlators of $T_{\mu\nu}$ in field theory. 

Similarly, global symmetry currents in the field theory $j_\mu^a$ couple to gauge field perturbations $A_\mu^a$ 
in the bulk, so gauge field fluctuations evolution will be calculated from the correlators of the currents $j_\mu^a$. 
We will see in the next section what is the precise relation, but here we will  just say that the relevant issue is 
whether the currents are marginally relevant operators. We will understand this in a way derived from the 
generalized conformal structure of the correlators, which will allow us to extract something akin to the conformal 
dimension of the operator in a conformal field theory. 

The concrete calculation we are interested in therefore is the calculation of the (nonabelian) global symmetry current 
correlators in the class of super-renormalizable field theories that are used for the phenomenological holographic model. 

Since however we cannot consider generally the global nonabelian symmetry currents, the symmetries depending on the 
model, we have to choose a toy model for it. We will be looking for a model with $SU(N)$ fields in the adjoint, like the 
phenomenological holographic model.

\subsection{The model and its Feynman diagrams}

The simplest model will have gauge fields and scalars. Indeed, it was found in \cite{Afshordi:2016dvb,Afshordi:2017ihr}
that fitting with the CMBR data requires that there are more scalars than fermions (we can have zero fermions). In the 
absence of fermions, the only interaction allowed by the generalized conformal structure (and super-renormalizability in 
3 dimensions) is of the $\phi^4$ type. We want to have an interaction that preserves a global $SO(3)$ symmetry, but also 
to allow for a vortex type ansatz that minimizes the scalar potential, as we will see at the end of this section. 

The simplest possibility that we found was to have six complex scalar fields, $\phi_i^a$, $i=1,2$ and $a=1,2,3$ 
that also transform, in the index $a$, in the $3$ representation of the group $SO(3)$, and for the potential to 
be $|\vec{\phi}_1\times \vec{\phi}_2|^2$. Spacetime indices in 3 dimensions will be denoted in this section by $\mu$.
As we said, both $A_\mu$ and $\phi_i^a$ have also an index in the adjoint 
of $SU(N)$ that is implicit. The action in Minkowski space is 
\be
S=\int d^3x\Tr\left[-\frac{1}{2}F_{\mu\nu}F^{\mu\nu}-2\sum_{i=1,2}
|D_\mu\vec{\phi}_i|^2-4\lambda|\vec{\phi}_1\times \vec{\phi}_2|^2\right]\;,
\ee
so the scalar potential, giving the scalar self-interaction, is 
\be
V=4\lambda\Tr|\vec{\phi}_1\times \vec{\phi}_2|^2.
\ee
The trace is over the $SU(N)$ indices, and is normalized with $\Tr [T_A T_B]=\frac{1}{2}\delta_{AB}$, in order to 
compare with the general form (\ref{phenoaction}).

We will find that the result for the current 2-point function is independent, up to 2-loops, of the coupling $\lambda$, 
which means that the only purpose of the potential is to define the global symmetry, and to define a certain vortex ansatz.

The Euclidean space action is then 
\be
S=\int d^3x\Tr\left[\frac{1}{2}F_{\mu\nu}F^{\mu\nu}+2\sum_{i=1,2}|D_\mu\vec{\phi}_i|^2
+4\lambda|\vec{\phi}_1\times \vec{\phi}_2|^2\right]\;,\label{Euclideanaction}
\ee
After taking the trace (in components), we will get the familiar $1/4, 1, 1$ coefficients for the 3 terms. 

There is a global $SO(3)$ symmetry with Noether current (note that by multiplying it with a different normalization would 
only change the normalization of the current 2-point function, which is of no interest for our purposes)
\be
j_\mu^a=\sum_{i=1,2}\vec{\phi}_i^*T^aD_\mu\vec{\phi}_i+h.c.\;,
\ee
where $T_a$ are the $SO(3)$ generators.  Since the vector indices on $\vec{\phi}$ are also denoted 
by $a$, as they are in the adjoint (the 3 representation), it means that $(T_a)_{bc}=if_{abc}=i\epsilon_{abc}$, so 
\be
j_\mu^a=\sum_{i=1,2}i\epsilon^{abc}\phi_i^{b,*}D_\mu\phi_j^c+h.c.\label{current}
\ee

In the above, the $SU(N)$ adjoint indices $A$ were implicit, and also implicit was the notation with 
$\Tr[T_A T_B]=\delta_{AB}$. The covariant derivative on the scalar would be, explicitly, $D_\mu^{AB}=\d_\mu\delta^{AB}
-ig\sqrt{2}(T_C)^{AB}A_\mu^C$, and otherwise performing the traces we would find a gauge kinetic term of $-\frac{1}{4}
F_{\mu\nu}^A F^{A\mu\nu}$. 

Since we are at large $N$, we consider only planar diagrams. That also means that we will obtain a result of the standard type, 
$N^2 f(g^2N)$, as we explain in Appendix \ref{secmomFeyn}, where we also write the Feynman rules. 

We will work in Euclidean space, since we want to relate the Euclidean space on the boundary with the three dimensional spatial 
Euclidean space in the bulk.

\subsubsection{Feynman diagrams}

\begin{figure}[h]
\begin{center}
\includegraphics[width=140mm]{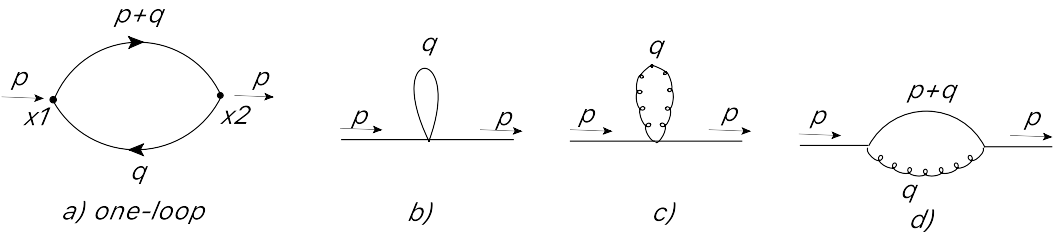}
\end{center}
\caption{One-loop diagrams: a) Unique one-loop diagram for the two-point function of currents. b),c),d) one-loop counterterm
diagrams. }
\label{fig:one-loop}
\end{figure}

\begin{figure}[h]
\begin{center}
\includegraphics[width=80mm]{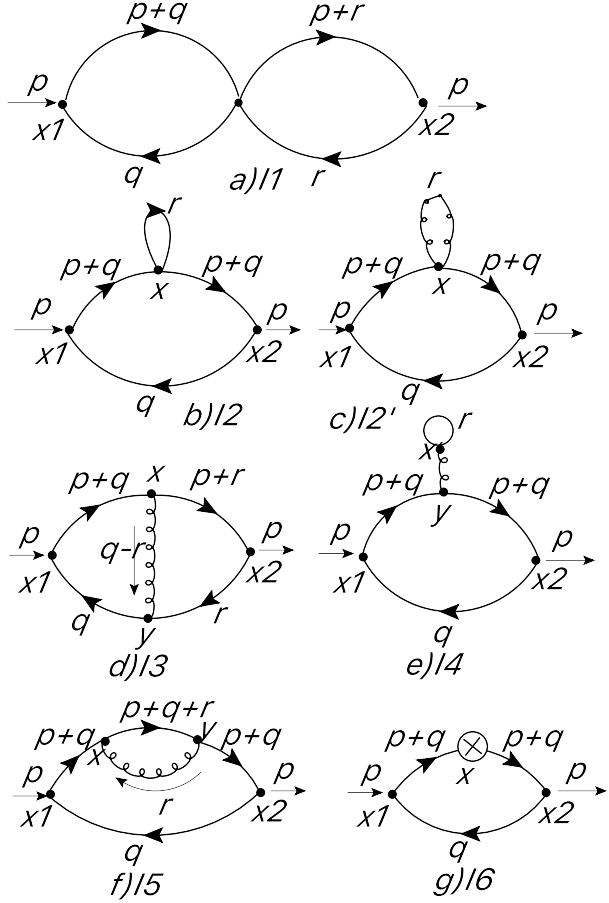}
\end{center}
\caption{Two-loop diagrams without external gauge field insertions: a) $I_1$. b) $I_2$. c) $I_2'$. d) $I_3$. e) $I_4$. 
f) $I_5$. g) $I_6$. }
\label{fig:two-loop}
\end{figure}

We write the 1-loop and 2-loop diagrams for the $\langle j_\mu^a (x_1)j_\nu ^b(x_2)\rangle $ correlator, i.e., 
up to order $\lambda$ or $g^2$. The relevant diagrams, with no external insertion of a gauge field,  
are drawn in Figs.\ref{fig:one-loop} and \ref{fig:two-loop}.

The nonzero diagrams among these (as we will see), are $I_3$ and $I_5$, which have two 3-point vertices coming from 
the action, so they will contribute with a factor of $1/2!$ to the 2-point function. 

\begin{figure}[h]
\begin{center}
\includegraphics[width=50mm]{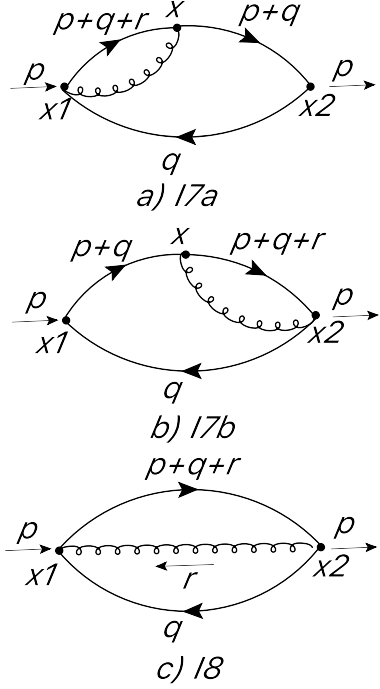}
\end{center}
\caption{Two-loop diagrams with external gauge field insertions: a) $I_{7a}$, with one external insertion. 
b) $I_{7b}$, mirror diagram with one external insertion. c) $I_8$, with two external insertions.}
\label{fig:Ttwo-loopGauge}
\end{figure}

To these, we must add diagrams with gauge field insertion in the external vertices, as in 
Fig.\ref{fig:Ttwo-loopGauge}: diagrams 7a and 7b, with  a gauge field 
line connecting the left/right external vertex with one of the lines of the one-loop diagram (momentum $r$ 
on the gauge line, $q$ on the line without gauge vertex, and momenta $p+q+r$, and $p+q$, for the line with 
gauge vertex), and diagram 8, with two external gauge field insertions, that is, 
with a gauge field line connecting the two external vertices (momenta $r$ on the gauge line,
and $q$, and $p+q+r$, on the scalar lines).

These two diagrams come from the order one or order zero action, with one or two insertions of the external gauge 
field, so they will contribute with a factor of $1!=0!=1$ to the 2-point function.

All diagrams will be considered in dimensional regularization. 

{\bf One-loop}

At one loop there is  a single Feynman diagram, as in Fig.\ref{fig:one-loop}a,
a loop with two external current insertions, with
a momentum $p$ coming in at one end, and the same $p$ coming out of the other. Considering a loop momentum $q$ on one 
of the legs, the other has $p+q$, which leads to the result (a factor of 2 comes from the sum over $i=1,2$ 
for the scalar running in the loop)
\be
I_{\mu\nu}^{ab}(p)=2\epsilon^{acd}\epsilon^{bdc}\int \frac{d^dq}{(2\pi)^d}\frac{(p+2q)_\mu(p+2q)_\nu}{q^2(p+q)^2}\equiv -4\delta^{ab}I_{\mu\nu}(p)\;,\label{oneloopdiag}
\ee

This is calculated in Appendix\ref{Oneloopres}, with the result

\be
I_{\mu\nu}^{ab}(p)=\frac{p}{4}\delta^{ab}\left(\eta_{\mu\nu}-\frac{p_\mu p_\nu}{p^2}\right).
\ee

This result is finite, which means it doesn't generate any extra momentum scale dependence (``anomalous dimension''), 
and the Noether current is marginal at one-loop.

{\bf Two-loop}

At 2-loops, there are 6 diagrams for no gauge field insertions at the external vertices, 
plus one counterterm diagram, as in Fig.\ref{fig:two-loop}. 
We can find them algorithmically by considering a 4-scalar vertex, of order $\lambda$, 
or two scalar-scalar-gauge vertices of order $g$ each, or one scalar-scalar-gauge-gauge vertex of order $g^2$ 
in between $x_1$ and $x_2$, and 
then connecting their legs in all possible ways with the external points (which have two scalar legs each). 

For the $\lambda$ vertex, we get the figure 8 diagram ($I_1$), with a sum over whether the scalars 
connected to $x_1$ are of type 1, and the ones to $x_2$ of
type 2, or vice versa; and a one-loop diagram, with a scalar blob on one of the legs ($I_2$), summed
 over which leg has the blob, over 
whether the one-loop diagram is with $i=1$ (and the blob with $i=2$) or with $i=2$ (and the blob with $i=1$). 

For the $g^2$ vertex, we get the same diagram as $I_2$, but with a gauge field blob for either 
$i=1$ or $i=2$ scalar loop ($I_2'$). 

For the two order $g$ vertices, we find three diagrams: one is the "theta" diagram, with the one-loop 
diagram (with either $i=1$ or $i=2$ in the loop) crossed 
vertically by a gauge field, connecting the two lines ($I_3$); the second is the one-loop diagram with a
 tadpole emanating from one of the legs: 
one gauge field propagator ending in a scalar blob, summed over $i=1$ or $i=2$ ($I_4$). The 
third diagram is the one-loop diagram, with a 
gauge propagator connecting two points, $x$ and $y$, on the upper scalar propagator ($I_5$). 

Finally, there is the counterterm diagram, with a counterterm vertex on one of the legs of the one-loop diagram ($I_6$).
Note that there would be in principle also a 2-loop diagram with a ghost loop, namely the diagram $I_4$, with the scalar loop 
on the tadpole replaced by a ghost loop, but this would vanish for the same reason that the original diagram was also zero.

Then there are two more diagrams with external gauge field insertion, as in Fig.\ref{fig:Ttwo-loopGauge},
one with a single insertion (diagram 7a for insertion 
on the left, and 7b, for insertion on the right), and one with two external gauge field insertions (diagram 8).

The respective diagrams are found as 
\bea
I_{1,\mu\nu}^{ab}&=&I_{2,\mu\nu}^{ab}=I_{4,\mu\nu}^{ab}=0\cr
I_{3,\mu\nu}^{ab}(p)&=&-4g^2\delta^{ab}
J_0\left\{\left(\eta_{\mu\nu}-d\frac{p^\mu p^\nu}{p^2}\right)\frac{2}{1-d}\left[\frac{3d-8}{d-4}-\frac{5}{3}\right]
-\frac{p^\mu p^\nu}{p^2}4\frac{d-2}{d-4}\right\}\cr
&&-4g^2\delta^{ab}\left\{
-4\frac{p_\mu p_\nu}{p^2}\left[-\frac{(3d-10)(3d-8)}{(d-4)^2}J_0+\frac{d-3}{d-4}p^2B_0^2\right]\right.\cr
&&-8\eta_{\mu\nu}\left[-\frac{d(d-3)}{(d-4)^2(d-1)}J_0+\frac{1}{2(d-4)(d-1)}p^2B_0^2\right]\cr
&&\left.-\frac{8p_\mu p_\nu}{p^2}\left[\frac{2(d-2)(2d^2-9d+8)}{(d-4)^2(d-1)}J_0 -\frac{(d-2)^2}{2(d-4)(d-1)}
p^2B_0^2\right]\right\}\cr
I^{ab}_{5,\mu\nu}(p)
&=&-16g^2\delta^{ab}\frac{J_0}{3(d-4)}\left\{-4\delta_{\mu\nu}
+d\frac{p^\mu p^\nu}{p^2}\right\}\cr
I^{ab}_{6,\mu\nu}(p)
&=&-8\lambda_{\rm ct}\delta^{ab}\frac{B_0}{d-1}\left[(3-d/4)\eta_{\mu\nu}-\left(\frac{d^2}{4}-2d+1\right)\frac{p_\mu 
p_\nu}{p^2}\right]\cr
I^{ab}_{7a,\mu\nu}(p)&=&I^{ab}_{7b,\mu\nu}
=+16g^2\delta^{ab}\frac{J_0}{d-4}\left\{-\delta_{\mu\nu}+(d-3)\frac{p^\mu p^\nu}{p^2}\right\}\cr
I_{8,\mu\nu}^{ab}(p)
&=&-16g^2\delta^{ab}\eta_{\mu\nu}B I_{1,1+\frac{2-d}{2}}=-16g^2
\delta^{ab}\eta_{\mu\nu}J_0\;,
\eea
where we have defined
\bea
G_1&=&\frac{\Gamma(2-d/2)\Gamma(d/2-1)^2}{\Gamma(d-2)}\cr
G_2&=&\frac{\Gamma(3-d)\Gamma(d/2-1)^3}{\Gamma(3d/2-3)}\cr
B_0&\equiv& \frac{p^{d-4}}{(4\pi)^{d/2}}G_1\cr
&\equiv& p^{d-4}B\cr
J_0&=&\frac{p^{2d-6}}{(4\pi)^d}G_2.
\eea
Besides this, we find that $\lambda_{\rm ct}$ is one-loop finite, so we actually don't need the counterterm. 

The calculations are described in Appendices \ref{secmomFeyn} and \ref{secdetails}. 
Moreover, we calculate the leading result in two independent ways, ensuring that the result is correct. 

Despite the highly nontrivial forms of the results above, they sum to a very simple and consistent result for any dimension $d$ 
(so to all orders in $\epsilon$), 
\be
I_{\rm 2-loop;\mu\nu}^{ab}(p)=\frac{16}{d-4}g^2\delta^{ab}\left(\delta_{\mu\nu}-\frac{p_\mu p_\nu}{p^2}\right)\left[2J_0-
\frac{d^3-5d^2-8d+8}{(d-1)(d-4)}J_0+\frac{B_0^2p^2}{d-1}\right].\label{totaltwoloop}
\ee

\subsection{From current correlator divergences to anomalous dimension}

Finally, the current 2-point function is (introducing also the factor of $N^2$,  and the $N$ multiplying $g^2$, which 
were neglected before)
\bea
\langle j_\mu^a (p) j_\nu^b(-p)\rangle&=&N^2\left[\frac{p}{4}\delta^{ab}
\left(\delta_{\mu\nu}-\frac{p_\mu p_\nu}{p^2}\right)
+(I_3+I_5+I_{7a}+I_{7b}+I_8)_{\mu\nu}^{ab}(g^2N)\right]\cr
&=&N^2\frac{p}{4}\delta^{ab}\left[\left(\delta_{\mu\nu}-\frac{p_\mu p_\nu}{p^2}\right)
-4\cdot 16 \frac{g^2N}{p}J_0\left(\delta_{\mu\nu}-\frac{p_\mu p_\nu}{p^2}\right)+{\rm finite}\right]\;,\cr
&&
\eea
where we have expanded in $\epsilon$, for $d=3+\epsilon$. We see that $J_0$ is divergent ($\propto 1/\epsilon$), but otherwise we have substituted $d=3$ 
in its coefficient in (\ref{totaltwoloop}).

But we are interested only in the $p$ dependence of the result. From the generalized conformal structure, we expect a 
result of the type (at $g^2_{\rm eff}=\frac{g^2N}{p}\ll 1$)
\be
\langle j_\mu^a (p) j_\nu^b(-p)\rangle=\frac{N^2p}{4}\pi_{\mu\nu}[1+cg^2_{\rm eff}\ln g^2_{\rm eff}+...]
=\frac{N^2p}{4}\pi_{\mu\nu}[1-cg^2_{\rm eff}\ln p+...] \;,
\ee
where the dots refer to subleading terms and $\pi_{\mu\nu}\equiv \delta_{\mu\nu}-\frac{p_\mu p_\nu}{p^2}$ is the
transverse projector. 

Even though we are not in a conformal theory, only in a theory with generalized conformal structure, we want to 
define a notion of anomalous dimension, defined by the proportionality
\be
\langle j_\mu^a (p) j_\nu^b(-p)\rangle \propto N^2 \pi_{\mu\nu}p^{1+2\delta}\label{deltacurrent}
\simeq N^2p\pi_{\mu\nu}[1+2\delta \ln p+...]\;,
\ee
which leads us to identify
\be
2\delta =-cg^2_{\rm eff}.
\ee

In the renormalized 2-loop result, which means for us just dropping the $1/\epsilon$ divergent term, the $p$ dependence 
comes exclusively from the finite part of the $J_0$ term, by identifying
\be
\frac{g^2N J_0}{p}= -\frac{2\pi g^2N}{(4\pi)^3p\epsilon}p^{2d-6}+...=-\frac{g^2N}{32 \pi^2p\epsilon}[1+2\epsilon \ln p]+...
={\rm divergent}-\frac{1}{16\pi^2}\frac{g^2N}{p} \ln p+....
\ee
with the similar term coming from the general dependence of the form
\be
g^2_{\rm eff}\ln g^2_{\rm eff}=\frac{g^2N}{p}\ln \frac{g^2N}{p}=-\frac{g^2N}{p}\ln p+...
\ee

We then deduce
\bea
\langle j_\mu^a (p) j_\nu^b(-p)\rangle&=&N^2\frac{p}{4}\delta^{ab}
\left[\left(\delta_{\mu\nu}-\frac{p_\mu p_\nu}{p^2}\right)
+\frac{4}{\pi^2} \frac{g^2N}{p}\ln p \left(\delta_{\mu\nu}-\frac{2}{3}\frac{p_\mu p_\nu}{p^2}\right)
+...\right]\cr
&=&N^2\frac{p}{4}\delta^{ab}
\left[\left(\delta_{\mu\nu}-\frac{p_\mu p_\nu}{p^2}\right)
-\frac{4}{\pi^2} g^2_{\rm eff}\ln g^2_{\rm eff} \left(\delta_{\mu\nu}-\frac{2}{3}\frac{p_\mu p_\nu}{p^2}\right)
+...\right]\;\cr
&&
\eea
which implies that
\be
\delta_j=\frac{2}{\pi^2}g^2_{\rm eff}>0\;,
\ee
which makes $j$ an irrelevant operator, {\em growing in the UV}. 

However, what we were really interested in was the {\em magnetic} current, dual to the electric one.

\subsection{Correlators of magnetic currents}

The vortex current is related to the electric current, at least in the Abelian-Higgs model, by the duality relation
\cite{Murugan:2014sfa}
\be
j^\mu_{\rm vortex}=\frac{1}{2\pi e \Phi_0^2}\epsilon^{\mu\nu\rho}\d_\nu j_\rho
\equiv\frac{1}{K}\epsilon^{\mu\nu\rho}\d_\nu j_\rho\;,\label{currentduality}
\ee
for a scalar VEV $\Phi_0$. If the scalar modulus is dynamical, we expect the coefficient to be different. Also, in general, 
maybe a function of the coupling could appear in front. The way this appears is reviewed in Appendix \ref{secPV}.
In \cite{Murugan:2014sfa} it was also shown that the same relation corresponds, via AdS/CFT, to the usual 
4 dimensional Maxwell duality in the bulk of the gravity dual.

The same relation would be expected to relate, for a scalar-gauge field model in 2+1 dimensions, the electric current 
and its particle-vortex dual, the ``magnetic'' current. Then, an electric current 2-point function of the type 
\be
\langle j_\mu (p) j_\nu(-p)\rangle =f\left(\delta_{\mu\nu}-\frac{p_\mu p_\nu}{p^2}\right)\;,
\ee
would be transformed to a vortex (magnetic) current 2-point function
\be 
\langle j^\mu_{\rm vortex}(p)j^\nu_{\rm vortex}(-p)\rangle=
\epsilon^{\mu\rho\sigma}\epsilon^{\nu\lambda \tau}\left(\delta_{\sigma\tau}-\frac{p_\sigma p_\tau}{p^2}\right)
\frac{f}{K^2}=\left(\delta_{\mu\nu}-\frac{p_\mu p_\nu}{p^2}\right)\frac{p^2}{K^2}f.
\ee

But a more precise relation was found by Witten in \cite{Witten:2003ya}, and then Herzog, Kovtun, Sachdev, Son 
in \cite{Herzog:2007ij} in the context of the duality of the ABJM model with $AdS_4\times \mathbb{CP}^3$. 

According to Witten, conformal structure in 2+1 dimensions fixes the current-current correlator to be of the form
\be
\langle j_i (p) j_j(-p)=\left(p^2\delta_{ij}-p_i p_j\right)\frac{t}{2\pi \sqrt{k^2}}+\epsilon_{ijk}p_k \frac{w}{2\pi}\;,
\ee
where $t$ and $w$ are functions only of the coupling (``constants''), and the contact term involving $w$ has no simple 
way to be fixed, but is defined by the theory. 

Herzog et al. say that, more generally for the (non-Abelian) current, 
\be
\langle j^a_\mu(p) j^b_\nu(-p)\rangle=\sqrt{p^2}\left(\delta_{\mu\nu}-\frac{p_\mu p_\nu}{p^2}\right) K_{ab}\;,
\ee
and that the contact term is absent in theories with no CS terms (in the ABJM model, there is, so it is considered). 
Either way, in our case there isn't such a term. 

But then the statement of both is that the action of $Sl(2;\mathbb{Z})$ on the theory can be found as follows. 
$T$ just shifts $w$ by 1. $S$, corresponding to S-duality, acts on $\tau=w+it$ as $\tau\rightarrow -1/\tau$, and changes the 
electric current with the magnetic current. 

In  \cite{Witten:2003ya}, the magnetic current is the topological current 
\be
\tilde j_i=\frac{1}{2\pi}\epsilon_{ijk}\d_j A_k\;,
\ee
where  $A_i$ is the  source for the electric current $J_i$, that becomes the gauge field on the boundary for the 
bulk theory. Then  the magnetic current correlator is 
\be
\langle \tilde j_i(p) \tilde j_j(-p)\rangle=\frac{p^2\delta_{ij}-p_ip_j}{2\pi \sqrt{p^2}}\frac{t}{t^2+w^2}
-\frac{\epsilon_{ijk}p_k}{2\pi}\frac{w}{t^2+w^2}.
\ee

If $w=0$, this amounts to just inverting $t\rightarrow 1/t$ in the correlator. 
Herzog et al, in the case with a nonabelian current $j_\mu^a$, find that the matrix $K_{ab}$ is inverted. 

Both find that the action of S-duality operator $S$ corresponds in the bulk to usual Maxwell duality for the gauge field
 $A_\mu$ that sources the currents.

As we see, the argument is based on: 

-the form of the correlator, which is also fixed in our case, of generalized 
conformal structure, and being close to conformality, by replacing the 
constant function $t$ of the couplings with a function of the effective coupling $g_{\rm eff}$ that now runs with the 
scale, and otherwise 

-on S-duality, which acts in the same way. 

Then, the inversion acts on $1+2\delta \ln p$ in the 2-point function of the electric current in 
(\ref{deltacurrent}), turning it into $\simeq 1-2\delta \ln p$, which 
means that the anomalous dimension of the dual vortex (magnetic) current is 
$\delta(\tilde j)=-\delta(j)$. Since as we saw, $j_\mu^a$ is understood as an irrelevant operator ($\delta(j)>0$), 
then the magnetic current $\tilde j_\mu^a$ is a relevant operator ($\delta(\tilde j)<0$), as we wanted.

One more needed element in the analysis is the fact that there are ``monopoles'' in the theory, so it can be S-dualized. 
Of course, ideally we would need to consider `t-Hooft-Polyakov-type monopoles in the bulk and the corresponding objects
on the boundary, that is, in a non-Abelian theory and with a long range magnetic $U(1)$ charge. Note that monopole number 
in the 3 spatial dimensions in the bulk implies, as is usual for topological solitons, vortex number on the boundary. 
Indeed, we know that magnetic charge in the bulk translates into magnetic charge on the boundary, as is usual in AdS/CFT. 
But moreover, a topological object in 2 spatial dimensions with magnetic charge {\em is} a vortex. Ideally, it should be a 
vortex solution like the Nielsen-Olesen vortex in the Abelian case, or more precisely a nonabelian generalization for it (which 
are much more difficult to find).

As a reminder, what one understands when talking about monopoles in cosmology are 't Hooft-Polyakov 
non-Abelian monopoles with gauge fields
for the gauge group $SO(3)$ (perhaps embedded in a larger gauge group), so that we can identify gauge and spatial coordinate indices, 
coupled to 3 real scalar fields in the same adjoint representation of $SO(3)$, so 
\bea
A^a_i&=&-\epsilon_{ija}\frac{x^j}{gr^2}[1-K(\phi_0 gr)]\cr
\phi^a&=& \frac{x^a}{r}\phi_0 h(\phi_0 gr)\;,
\eea
where $\phi_0$ is the vacuum $|\phi|$ solution for a Higgs potential, 
and the boundary conditions at infinity for the functions $K(x)$ and $h(x)$ are $h\rightarrow 1, 
K\rightarrow 0$, giving a monopole configuration. Moreover, even the electromagnetic component of the gauge field, 
\be
A_\mu =\frac{\phi^a_{\rm vac}A_\mu^a}{\phi_0}\;,
\ee
where $\phi^a_{\rm vac}$ is the solution for $\phi^a$ at infinity (in the vacuum), has a monopole 
structure, looking like a Dirac monopole at large $r$, 
resulting in a magnetic field (which is similarly projected)
\be
B_i\simeq \frac{g}{4\pi }\frac{x_i}{r^3}.
\ee
The difference from the Dirac monopole is that the non-Abelian monopole is actually smooth near $r=0$, where 
$1-K(\phi_0 g r)\simeq \phi_0 gr$, so $A_i^a\simeq -\epsilon_{ija} x^j/r$ 
remains finite, and so does the magnetic field. 

A non-Abelian version of the Nielsen-Olesen Abelian vortex (the object with magnetic "charge", or rather flux $\Phi=\int B dS$,  
in 2+1 dimensions, i.e., on the boundary of the 3+1 dimensional bulk)
was not found yet. Non-Abelian vortices already found are complicated solutions in complicated theories, not relevant for the generic 
action we want to consider (with all fields in the adjoint of an $SU(N)$ group). 

The Nielsen-Olesen vortex has vortex number and charge concentrated in a singular point (as for any 
type of vortex, see Appendix \ref{secPV}), 
but the magnetic field (or flux) is spread out, so it would seem like a good substitute for our problem. However, again we cannot use it for 
our problem, because we need an action for non-Abelian adjoint $SU(N)$ fields, and we can find no simple way to embed the 
Abelian Nielsen-Olesen
set-up into our action. Furthermore, it is not clear what would it correspond to in 3+1 dimensions, since in 3+1 dimensions both the magnetic 
field and the scalar field are non-singular at zero for the 't Hooft Polyakov monopole. 

The next possibility is for both the magnetic field and the vortex number to be concentrated in a singular point, something which we 
will call a "Dirac vortex", by analogy with "Dirac monopole", yet still embedded into a non-Abelian theory. This should indeed correspond 
in the bulk to a Dirac monopole embedded in the non-Abelian theory, meaning both magnetic charge and vortex number 
(singularity) in scalar field are 
concentrated in a point. Even more specifically, the magnetic field will be a delta function at $r=0$, of given flux, that will act as a source in 
the equations of motion of the scalar field (this is, in fact, the usual procedure called "adding a magnetic flux to a particle", which leads to 
anyonic statistics for the particle charged with respect to the Abelian gauge field).

Then, since we cannot consider the field theory dual of true ('t Hooft-Polyakov) monopoles,  
at the very least, we can consider the bulk monopoles to be Dirac monopoles instead 
(since, as we said, any monopole looks like a Dirac monopole at long distances), and corresponding on the boundary to this
``Dirac vortex'', which would likely be enough for the issue of principle. 
What that means specifically is that both the $U(1)$  magnetic field and the vortex number are located at $r=0$, and 
are neither spread out, nor non-Abelian, and the magnetic flux is a delta function source. That is fine, since in AdS/CFT a 
dynamical magnetic field in the bulk
corresponds to a magnetic field source on the boundary.

Keeping this in mind, we see that we can consider the model treated until now, because:

-it admits a $U(1)$ global symmetry (coupled to an external $A_\mu$, as needed for this AdS/CFT 
dual of Dirac monopoles), embedded in the 
$SO(3)$ we have been considering, therefore with current $j_\mu$ embedded in $j_\mu^a$, 
\be
j_\mu=i\sum_{i=1,2}\vec{\phi}_iD_\mu \vec{\phi}_i+h.c.
\ee

-this $U(1)$ rotates the two complex fields $\vec{\phi}_1,\vec{\phi}_2$ in the same way, as 
\be
\phi^a_1\rightarrow e^{i\b}\phi^a_1;\;\;\;
\phi^a_2\rightarrow e^{i\b}\phi^a_2\;,\;\;\forall \b
\ee
and, more importantly, allows us to write a vortex  ansatz that keeps the field in the vacuum, $V=0$, 
\be
\phi^a_1=\phi_1(r) f^a e^{i\a}\;,\;\; 
\phi^a_2=\phi_2(r) f^a e^{i\a}.
\ee
(another possibility would be to put $\phi_1^a=C\phi_2^a=C\phi_2^a(r)e^{i\a}$, with $C$ constant), where $\a$ is the polar angle in 
coordinate space and $f^a$ is a constant vector in $SO(3)$ group space.
Then, by choosing $\phi_1(r)=\phi_2(r)=\phi_0$ a constant but cutting out the point $r=0$, we obtain a solution with vortex 
number (nontrivial holonomy)
(see also the detailed analysis in \cite{Murugan:2014sfa}). To see this, note that 
the $A_\mu$ equation of motion is $\d_\mu\theta=A_\mu$, or $\d_\a \theta=A_\a$, so $\oint_C \theta(\a)
=2\pi$, and we have a holonomy=magnetic flux  $\int_S B\cdot dS=\oint_C A_\a d\a=2\pi$, even though the gauge field 
$A_\mu$ is pure gauge (outside of $r=0$).

Thus, to obtain the needed "Dirac vortex", we put a delta function magnetic field source for the gauge field $A_\mu$, that 
will create also the vortex number (this is the usual procedure of ``adding a magnetic flux'' to a particle, that leads to anyonic statistics). 

Then, any such solution of the equations of motion will have  vortex number (which just needs to be matched to the corresponding 
delta function magnetic flux source). Vortex number means there will be a vortex current, as we 
saw was needed in order to have a well defined quantum S-duality. 

Moreover, we concentrated on constant $|\phi_1|$ and $|\phi_2|$, but we can make them vary, leading to $\d_\mu\phi_{1,2}\neq 0$, 
which must be matched by a nonzero and varying potential $V$, which can be obtained by
putting $f^{a_1}$ and $f^{a_2}$ different directions in group space for the fields.  One could then obtain a nontrivial 
vortex solution, with $r$-dependence
and $\phi_{1,2}(r=0)=0$, but still with 
vortex number and magnetic field concentrated on $r=0$ (generalizing the discussion in section 3 of \cite{Murugan:2014sfa}), but we will 
not explore that.

To summarize, the model we considered admits a well defined quantum S-duality operation which, by a simple extension of 
Witten's argument, leads to an inverse behaviour for the magnetic current correlator, therefore to a relevant magnetic current
operator.

\section{Solutions to problems in holographic cosmology}

We saw that in inflationary cosmology, the solutions to the problems listed were basically due to the rapid 
(exponential or power law with $n>1$) expansion of space. In holographic cosmology however, we must give an 
answer from the point of view of the field theory on the boundary. The field theory has the property of generalized conformal 
structure, or the fact that the momentum dependence is contained in the dependence of the dimensionless effective 
coupling $g^2_{\rm eff}=\frac{g^2N}{q}$. The solutions of the problems will be then mostly 
based on the scaling of the 2-point
functions for the energy-momentum tensor $T_{ij}$ and of the nonabelian global currents $j_i^a$. 

1. {\bf Smoothness and horizon problems}

At first sight, in a strongly coupled gravity theory, the first instinct is to say that the notion of causality is ill-defined in the 
case of a non-geometric, or highly quantum, phase. However, while that might be true at the early stages, eventually 
the perturbations will transition into a geometric phase, and finally become classical outside the horizon, 
entering the horizon just now as CMBR perturbations. That means that certainly we need to explain them. 

A somewhat better answer is that the holographic map is {\em nonlocal}, but the fundamental theory is not gravity, but the field theory 
at the boundary, which is causal and local. The effect of the nonlocal map is to create, at the end of the non-geometric 
phase, an {\em apparent} nonlocality evident in the smoothness and horizon problems. Since the $\langle  h_{ij} h_{kl}\rangle$ 
correlators observed on the sky (in the CMBR) are derived from the $\langle T_{ij} T_{kl} \rangle $ correlators in the 
boundary field theory, which are causal and local, and the apparent non-locality only appears when we try to extrapolate 
to the past into a non-geometric phase. 

However, the correct answer about having correlations in the sky over regions that were initially out of causal contact 
(which in inflation is resolved by the fact that the inflationary period makes past lightcones meet before they hit the initial singularity)
is found by translating the initial correlation problem into field theory. It becomes the problem of having nonzero two-point 
functions of the dual field theory operators ${\cal O}$ when going in the deep IR. But in fact, $\langle {\cal O}(x_1){\cal O}(x_2)\rangle
\neq 0$ as $|x_1-x_2|$ becomes very large, or more precisely, is a decaying power law, and not an exponentially decaying function, in our 
theory with generalized conformal invariance, since there are no mass scales (other than 
$g_{YM}$, which appears only through the dimensionless $g^2_{eff}=g^2_{YM}N/q$).
Moreover, since the theory should be IR finite (there are proofs of IR finiteness for subclasses 
of theories, with no counterexamples in general, 
and in some cases there are lattice proofs), 
{\em there is no singularity from the point of view of field theory, which then would 
effectively resolve the dual cosmological singularity. }

There are 2 potential caveats to the above argument. The first is that the field theory is defined phenomenologically, yet 
by the implicit holography, we expect that for $g^2_{eff}\gg 1$, the correct description is in terms 
of a dual gravity (and perhaps string) theory, 
where we are back to a cosmology, and could have the same potential problem. As an example, 
for the case of the field theory on $N$ D2-branes, 
for $q<g^2_{YM}N$ (at low times in cosmology), the gravity description is correct, as shown 
in \cite{Itzhaki:1998dd}, and there is a cosmology 
with $a(t)\propto t^7$, see for instance \cite{Nastase:2018cbf}. The second caveat is that 
we also have the {\em smoothness problem}, namely 
not only that there are correlations over large regions, but also that there are no large fluctuations. But this is actually the same as the first 
caveat, since the absence of large fluctuations is correlated with the IR finiteness of the field theory, thus with the possibility for the 
field theory description to also make sense at $g^2_{\rm eff}\sim 1$ and beyond. If the field 
theory makes sense (it is IR finite) at all $g^2_{\rm eff}$, 
no matter how large, we have no constraint, and the smoothness and horizon problems are {\em always} solved, unlike in the usual inflation
case. Otherwise, we consider the starting point for the theory to make sense to be $g^2_{\rm eff}\sim 1$, corresponding to the beginning 
of the non-geometric phase in cosmology, and the constraint will be on the evolution during this non-geometric phase.

Then, that would leave the quantitative explanation of why we have smoothness over $N=72$ horizons, and how to 
translate that into a constraint in the field theory, like it was translated for inflation into at least 56 e-folds or so of inflation. 
The constraint must be on an a type and amount of RG flow (dual to -inverse- 
time evolution, therefore related to number of e-folds)
during which the field theory description is valid, after which we transition into usual radiation-dominated cosmology. 
We have seen that in inflation, the condition on $d_H$ translates into a condition on the number of e-folds (\ref{condition}),
which physically is written as (\ref{condition2}), but moreover, in generality, so away from inflation, becomes (\ref{fluct}).

The last condition, of an amplification of about $10^{-54}$ of perturbations, was found in the context of the flatness 
problem, which gave the same constraint as the smoothness and horizon problems on inflation, therefore we will 
delay here the exact form of the constraint until the flatness problem is resolved. Then the result must be an 
amplification of at least $10^{-54}$ of perturbations under the RG flow. 

To understand better why we have the same constraint as for the flatness problem, 
and the constraint on the number of e-folds is obtained 
from a constraint on the amount of RG flow in field theory, consider the following set-up. 
RG flow refers to the evolution of the field theory with respect to a momentum scale, which is the inverse of a 
spatial scale. Consider therefore a spatial scale, or separation, $L$ (connecting points $x$ and $y$), and consider its effect in the bulk of the 
gravitational space. This is defined by a geodesic that goes between $x$ and $y$ by moving 
into the bulk and then coming back in the boundary. 
We want to relate the "depth" of the dip of the geodesic into the bulk with the spatial scale $L$. 

We will consider the standard case of a Wick-rotated AdS space in the bulk, corresponding to inflationary cosmological evolution, 
$a(t)\propto e^{Ht}$. Then we can use the geodesic in the AdS space, which is Wick-rotated with respect
 to our cosmological space
(via the "domain wall/cosmology correspondence" \cite{Skenderis:2006jq}). 
The (renormalised) length of this geodesic provides the 2-point function of a dual operator inserted 
at each of the two points \cite{Graham:1999pm}.
This geodesic is exactly the one 
used in the calculation of the Wilson loop \cite{Maldacena:1998im}
(see also \cite{Rey:1998ik}), and is known to be independent of the dimensionality $d$ of the $AdS_{d+1}$ space, 
so the $AdS_5\times S^5$ calculation 
also applies to our $AdS_4$ case. 
Indeed, this is a spacelike geodesic in the gravitational space, therefore calculating the minimum path through the space. 
Since the string worldsheet calculating the 
Wilson loop is time-translation invariant, the minimal area calculation reduces to the minimum length, times the (large) total time $T$.  
To review, one has the metric
\be
ds^2=\a'\left[\frac{U^2}{\tilde R^2}(-d\tau^2+d\vec{x}_{d-1}^2)+\tilde R^2\frac{dU^2}{U^2}+...\right]\;,
\ee
(here $\tilde R^2$ is the dimensionless radius of $AdS_{d+1}$, in the $AdS_5$ case $\tilde R^2=R^2/\a`=\sqrt{4\pi g_s N}$, 
and $U=r/\a'$ has dimensions of energy)
which for the time-translation invariant ($\tau$ invariant) string results in the action (independent of $d$, since among the 
spatial directions, only the $x$, in which we have the spatial 
separation $L$, contributes)
\be
S_{\rm string}=\frac{T}{2\pi}\int_{-L/2}^{+L/2} dx\sqrt{(\d_x U)^2+\frac{U^4}{\tilde R^4}}.
\ee
Minimizing it, one obtains the implicit form of the geodesic, $x=x(U)$, via
\be
x=\frac{\tilde R^2}{U_0}\frac{dy}{y^2\sqrt{y^4-1}}.
\ee
Setting $x=L/2$, we obtain the desired relation, 
\be
\frac{L}{2}=\frac{\tilde R^2}{U_0}\frac{\sqrt{2}\pi^{3/2}}{\Gamma(1/4)^2}=\frac{R^2}{r_0}\frac{\sqrt{2}\pi^{3/2}}{\Gamma(1/4)^2}
\equiv \frac{c}{2}\frac{R^2}{r_0}.
\ee
Here $U_0=r_0/\a'$ is the minimum value of $U$ (the "dip") reached by the geodesic.

Wick rotating ($R^2\rightarrow -R^2$, $r^2=-\tilde r^2$, $\tau^2=-\tilde \tau^2$) the metric and writing it in usual cosmological 
time coordinates, we obtain the metric 
\be
ds^2=\frac{\tilde r^2}{R^2}(+d\tilde \tau^2+d\vec{x}_{d-1}^2)-R^2\frac{d\tilde r^2}{\tilde r^2}+...
=e^{t/R}(d\tilde \tau^2+d\vec{x}^2_{d-1})-dt^2+...\;,
\ee
where we have the relation $\tilde r=Re^{t/R}$, where therefore $R=1/H$ ($H$ is the Hubble constant during an inflationary-like phase). 
Then for the geodesic we obtain 
\be
L=cR e^{-t/R}\;,
\ee
or for momentum scales in field theory, corresponding to inflation,
\be
k=\frac{H}{c}e^{Ht}.
\ee
Thus at least in the geometric phase, $N_e$ e-folds of inflation corresponds to an $e^{N_e}$ factor multiplying the RG 
momentum scale in field theory.
In the non-geometric phase described by perturbative field theory on the boundary, we can therefore consider the factor 
multiplying the RG momentum scale
as being what takes the place of the $e^{N_e}$ factor for inflation.

2. {\bf Flatness problem}

We must understand why in the non-geometric phase (dual to field theory), a small fluctuation of the cosmology 
from the flat $\Omega=1$ case is made even smaller, down to a $10^{-54}$ precision, so that afterwards, in
the usual radiation dominated cosmology,  such a deviation can grow again to order one. 

In the holographic picture, time evolution corresponds to inverse RG flow, from the IR to the UV. We need to understand
why a small gravitational perturbation, deforming the space from the $\Omega=1$ flat one, is very small in the 
field theory UV, down to $10^{-54}$ order for the case of inflation at $10^{16}GeV$, whereas it is natural, of order 1, 
in the field theory IR. In the field theory picture, that amounts to having fluctuations for the energy-momentum tensor 
$T_{ij}$ growing from order $10^{-54}$ in the (almost free) field theory UV to of order one in the nontrivial field theory 
IR. That means that the energy-momentum tensor is a (marginally) relevant operator, taking us away from the 
naive field theory IR. 

As we saw, the field theory is super-renormalizable, and the generalized conformal structure means that all the 
momentum dependence arises through the effective coupling $g^2_{\rm eff}=\frac{g^2 N}{q}$. Super-renormalizability 
means that all couplings correspond to relevant or marginal operators. The energy-momentum tensor $T_{ij}$ is 
marginal (dimension 3 in 3 dimensions) at the classical level, but at the quantum level its 2-point function decomposes 
into a scalar and a tensor piece as we saw, but both of the form $q^3N^2f(g^2_{\rm eff})$, with the $q^3$ factor
matching the same factor in the CMBR power spectra. 

In the field theory UV, for $g^2_{\rm eff}\ll 1$, i.e. late times in the non-geometric phase, and small scales (large $l$)
in the CMBR, the function $f$ starts off at 1, corresponding to a scale invariant spectrum, and then deviates in a 
specific way, calculated at 2-loops in \cite{Afshordi:2017ihr}, namely both for the scalar and the tensor perturbations
we have
\be
f(g^2_{\rm eff})=f_0\left(1-f_1g^2_{\rm eff}\ln g^2_{\rm eff}+f_2 g^2_{\rm eff}+{\cal O}(g^2_{\rm eff})\right).
\ee

The essential feature, relevant for the flatness problem, is that $f_1<0$ both for the best fit to the CMBR data, and 
for most of the theoretical parameter space \cite{Afshordi:2016dvb}. Since at $g^2_{\rm eff}\ll 1$, the 
$f_1$ term dominates over the $f_2$ term, it amounts to a negative power law dependence on the momentum. 
Indeed, assuming this comes from the expansion of a power, 
\be
f(g^2_{\rm eff})\propto q^{2\delta}\sim 1+2\delta \ln q\sim 
1-2\delta \ln g^2_{\rm eff}+...
\ee
implies that the power is negative,
\be
2\delta \simeq f_1 g^2_{\rm eff}<0\;,
\ee
or that $T_{ij}$ is marginally relevant. Note that this dependence is also the one that corresponds, in the standard 
inflationary description of the CMBR data, to a red tilt, $n_s-1<0$, as was already observed (see for instance 
\cite{Afshordi:2016dvb,Afshordi:2017ihr}).

Of course, we use the terminology of conformal field theory for its deformations (relevant or irrelevant), though we 
only have generalized conformal structure, but the same results apply. For $\delta<0$ in the 2-point correlator of the 
energy-momentum tensor, this still leads to a dilution of the dual gravitational perturbations in the cosmology, along the 
inverse RG flow, as the sign of the power $\delta$ determines the sign of the relative variation $\delta h/h$ along the RG line.  
Indeed, consider the definition of the power spectra of scalar and tensor fluctuations (\ref{DeltaTS}), basically $\delta h_{ij}(q)/h_{ij}$, which 
is very close to 1 (in inflation, it gives $q^{n_s-1}$, here it gives $1/[1+(...)\log q]$). In it, $q$ 
refers to momentum scale in the sky, so large $q$ 
corresponds to small cosmological time $t$. From (\ref{DeltaTSAB}) and (\ref{TTAB}), we see that the power spectrum
is related to the the inverse of the correlator of $T_{ij}$ fluctuations. Then the power spectrum is $\propto 1/q^{2\delta}$ and $q$ grows with 
an inverse power of the cosmological time $t$, meaning that $\delta<0$ corresponds to dilution of gravitational perturbations
during the cosmological time $t$. 

From the point of view of the general field theory only, we can also see why $\delta<0$ corresponds to a relevant deformation, and 
that in turn means dilution of perturbations during the inverse RG flow. According to general Wilsonian renormalization theory, 
for an operator ${\cal O}$ of dimension $\Delta$, in our case the operator $T_{ij}$, added to the theory with a momentum
cut-off $1/\Lambda$, we have
\be
S=S_{\rm QFT}+\int d^dx \Lambda^{\Delta-d}\varphi {\cal O}\;,
\ee
where $\varphi$ is the dimensionless coupling, $h_{ij}$ in our case. Since the phenomenological 
QFT (\ref{phenoaction}) is super-renormalizable,
we are close to the free field fixed point, so when $\Delta <d$ (the operator ${\cal O}$ is relevant), in the UV (for $\Lambda\rightarrow 0$), 
the dimensionless coupling $\varphi$ goes to zero (so $\delta h_{ij}\rightarrow 0$), 
whereas if $\Delta>d$ (the operator ${\cal O}$ is irrelevant),
the coupling $\varphi$ dominates in the UV. The dimension of the operator is obtained from the 2-point function close to the 
fixed point, $\langle{\cal O}(x){\cal O}(0)
\rangle\sim 1/x^{2\Delta}$ for $x\rightarrow 0$ or $\langle {\cal O}(q){\cal O}(-q)\rangle\sim q^{2\Delta-d}$ for $q\rightarrow \infty$. 
In our case $\langle T_{ij}(q) T_{kl}(-q)\rangle \sim q^3 f(g^2_{\rm eff})\sim q^{3+2\delta}$, 
so $\Delta=3+\delta$, meaning the operator is relevant.

Finally then, for both the smoothness and horizon problems, and for the flatness problem, the quantitative issue, of 
diluting fluctuations to order $10^{-54}$, becomes a selection tool for the field theories: first, we must have a field 
theory that dilutes fluctuations, i.e. $\delta<0$, which turns into $f_1<0$, which is valid for most of the theoretical 
parameter space. 
Second, we must {\em restrict the amount of RG flow happening in the field theory during the non-geometric phase}
(corresponding in inflation to a bound on the number of e-folds), such as to go from a coefficient of order $10^{-54}$  in the UV to
a coefficient of order one in the IR. 

3. {\bf Relic and monopole problem}

The calculation of section 4 was related to this issue. 

At first sight, again we would say that in the absence of geometry, we cannot say what is a GUT phase transition, so the 
problem of diluting any created monopoles doesn't arise. But a phase transition in cosmology would be some sort of 
phase transition in field theory as well. The details of the Kibble mechanism in the context of the field theory is beyond the 
scope of this paper, though we will come back to it in further work. 

But the relevant issue is whether we can define a monopole abstractly, from the topology instead of a solution, in cosmology. 
That is certainly possible even in the absence of a geometrical description. Moreover, for the relic problem, the relevant 
issue is whether we can dilute the gravitational effects of the relic (since we are agnostic about its composition in this case). 
In the relic case then, we can consider it as just another type of gravitatinal perturbation $h_{\mu\nu}$. 

Using the AdS/CFT correspondence, the monopole in the bulk, defined as a topological configuration with a magnetic charge, 
corresponds to a monopole, or magnetic field configuration with some topological charge, on the boundary. 
More precisely, it means magnetic charge and vortex number in the 2+1 dimensional boundary, as analyzed in the previous
section. It should be really a ``'t Hooft monopole'' in the bulk, corresponding to a regular vortex on the boundary, 
but we considered the approximation of a  ``Dirac monopole'' in the bulk, corresponding to a ``Dirac vortex'' on the 
boundary. 

The regime we are interested in is of many monopoles in the bulk, corresponding to a weakly coupled situation in the 
boundary field theory, so the monopole gauge field perturbation $A_\mu^a$ is sourced by a field theory global current
(for a nonabelian global symmetry group corresponding to the bulk gauge symmetry group) $j_i^a$. 
Since however we have a monopole field, electric/magnetic dual to an electric field, in the field theory we need to have
a magnetic or topological current, electric/magnetic dual to the Noether (electric) current.
The monopoles, 
like other general relics, also induce a gravitational field perturbation, dual in the field theory to a perturbation in the 
energy-momentum tensor $T_{ij}$. 

The constraint for the general relics is then the dilution of the perturbation in $T_{ij}$ along the inverse RG flow (IR 
to UV), by an amount of $10^{-11}$ in volume, or $10^{-4}$ in linear size 
(so that the dual cosmology doesn't over-close). This is the same issue as for the 
smoothness, horizon and flatness problems, resolved by the same fact that $T_{ij}$ is marginally relevant. The constraint 
on the amount of RG flow needed is however much less stringent, so it doesn't introduce anything new. 

Similarly now, the constraint for monopoles is the dilution of the perturbation in $j_i^a$ along the inverse RG flow, 
by an amount of at least $10^{10}$ in linear scale, corresponding to an RG flow in energy of the same value, from the 
IR to the UV. The constraint on the possible field theories amounts to selecting the ones that have marginally relevant 
magnetic currents $\tilde j_i^a$, with $\delta(\tilde j)<0$. 

The calculation in section 4 showed that for a rather generic field theory, as in our toy model (indeed, as we saw, the 
form of the 4-point potential was irrelevant for the calculation of $\delta(j)$), and generic Noether current $j$, we obtained 
$\delta(j)>0$, an irrelevant operator. Its electric/magnetic dual, or rather S-dual, the magnetic current, was then found to 
have $\delta(\tilde j)<0$, a relevant operator, as we wanted. Of course, we don't expect the results to be fully general, 
just like they were not in the case of $T_{ij}$, the calculation was meant to just show that a large portion of the 
theoretical parameter space satisfies this condition. The condition $\delta(\tilde j)<0$ will restrict the possible models. 
And as for gravitational perturbations, the dilution of $\delta \tilde j$ 
by at least $10^{10}$ along the inverse RG flow is a condition on the 
amount of RG flow needed for the field theory before the end of the non-geometric phase.

4. {\bf Entropy problem and the arrow of time}

In inflation, the entropy problem was solved by reheating, which produced the desired large entropy. Now we still 
must have a period corresponding to reheating (at the very least in order to match the description of inflation, which is 
still part of the larger paradigm of holographic inflation), so we could explain it in the same way. 

However, it is easier to explain the solution in the field theory. To understand the analogy with inflation and reheating, 
we must understand how the energy is transferred from the gravitational degrees of freedom to the Standard Model 
ones. In the dual field theory, the Standard Model of particle physics is also part of the field theory on the boundary, 
just a part that doesn't become the gravitational sector. So the transfer of energy is just from a part of the field theory 
model to another. 

The question of the entropy problem can then be reformulated as, why is the entropy of the field theory so large, and 
why is it larger in the UV (dual to the late times in cosmology) than in the IR? The entropy in the field theory is large because
there is a large number of degrees of freedom (large group rank $N$ and large number of scalars and fermion
flavours), in order to have a classical gravitational space by holographic duality. And the entropy is larger in the UV 
than in the IR because that is what happens on a general RG flow. It is a natural expectation from the point of view 
of field theory, based on the theory of RG flows, and not a choice of model. 

Of course, we must still have only $S_1\sim 10^9$ in the UV of the field theory (corresponding to the end of 
reheating in the gravitational theory), and $S_1\sim 1$ in the IR, so this exact amount of entropy reduction
 would be a constraint on the field theory models. 

In this context, we note that, since time evolution is related to inverse RG flow, which increases the number of degrees of freedom
thus the entropy, holographic cosmology gives a very simple solution to an old problem: the arrow of time is now clearly defined, 
in a completely model-independent way, that is, independent of the particular realization of the holographic cosmology paradigm. 
The fact that the number of degrees of freedom is larger in the UV than in the IR is a general property of RG flows, and not a 
property of a particular model.

Moreover, universality of IR dynamics (corresponding to small time) means that having low entropy at initial times is natural, 
explaining why $S_1\sim1 $ in the IR, which are otherwise sometimes thought as a very special choice of initial conditions.
As we said, one is still left to explain the particular value of the entropy now (perhaps thought of as a constraint on field theory 
models). Indeed, as first emphasized by Penrose \cite{Penrose:1900mp} (see also \cite{Carroll:2004pn,Wald:2005cb}),
the entropy could have been a lot larger: collecting all the mass in the observable 
Universe in a single black hole would lead to an entropy of about $10^{121}$, much larger than the one observed today ($\sim 10^{88}$), 
meaning that we need a strong selection of models. However, that is related to the concrete
model for reheating, which is outside of the scope of this paper. 

5. {\bf Perturbations problem}

The perturbations are as simple to explain as in inflation, in a sense easier, since we don't need to make any assumptions 
about the physics (in the inflation case, we needed to assume that we can use quantum field theory in curved spacetime, 
and that we can use the Bunch-Davies vacuum as an initial condition). The gravitational correlators $\langle h_{ij} h_{kl}
\rangle$ that we observe in the CMBR are dual to the energy-momentum correlators $\langle T_{ij} T_{kl}\rangle$ 
on the boundary, which means that the classical gravitational perturbations that we observe are dual to regular quantum 
field theory perturbations for the energy-momentum tensor. Since classical is mapped to quantum in AdS/CFT, the 
solution of the perturbations problem is more natural now.

6. {\bf Baryon asymmetry problem}

The solution of the baryon asymmetry problem, like in the inflation case, has the same resolution as the entropy problem:
the baryon number asymmetry is related to the $S_1\sim 10^9$ that is created during reheating. And the fact that the 
Sakharov conditions need to be satisfied is the same: the particles physics condition of baryon number violation and CP 
violation are the same, since as we said, the Standard Model of particle physics is also part of the dual field theory; and the 
fact that we need to have reactions out of thermal equilibrium, as  function of time, is mapped to the fact that 
thermal equilibrium is not reached along the RG flow until the point corresponding to reheating, 
because of the rapid change in number of degrees of freedom. 

\section{Conclusions and discussion}

In this paper we have considered the solutions to the pre-inflationary problems in holographic cosmology. 
The smoothness and horizon problems, and the flatness problem, which in inflationary cosmology are solved by an 
exponential expansion, and give a bound on the number of e-folds, in holographic cosmology are solved by the fact that, 
on the majority of the theoretical parameter space of the phenomenological field theory model, the energy-momentum 
tensor is marginally relevant, so dual gravitational perturbations get diluted. For general relics, in the relic problem, 
we have a similar solution in holographic cosmology. Most of the paper was devoted to proving that in a simple toy 
model for the phenomenological model (and general enough, since the calculation was independent of the 
specific form of the potential for the scalars of the model),
 magnetic currents, S-dual to global electric currents, are marginally relevant 
operators as well, so the dual monopole perturbations get diluted. 
The entropy problem is solved by the fact that the field theory dual to a gravitational space has a large number of degrees 
of freedom, so has a large entropy per baryon, and the UV entropy is larger than the IR entropy, by general properties of 
RG flow, leading to a general arrow of time. The exact value of this ratio is a constraint on the field theory model. 
The perturbation problem was solved since classical gravitational perturbations are dual to quantum field theory perturbations. 
Finally, the baryon asymmetry problem is solved in the same way as the entropy problem, and by using the fact that the RG
flow is not in thermal equilibrium until reaching the point corresponding to reheating. 

During the calculation of the 2-point function of global symmetry currents in the toy model, relevant to the monopole problem, 
we have also calculated several integrals in dimensional regularization, and found an algorithmic way to calculate the 
general divergences of the integrals.

In this paper we have dealt with the regime of perturbative field theory, but the holographic cosmology paradigm 
encompasses also the case of usual inflation, when the dual field theory is not perturbative. Together with the 
fact that the CMBR fluctuations give as good a fit for holographic cosmology in the perturbative field theory regime as for the 
$\Lambda$ CDM model with inflation, the results in this paper mean that holographic cosmology in this regime
is as good a model as  usual inflationary cosmology. 

It would be interesting to understand further a number of details. The extension of the monopole analysis in this paper 
from the Dirac monopole/vortex type to `t Hooft monpole/vortex, as well as the Kibble mechanism in the field theory, 
must be understood. Another issue is how to turn the numerical constraints proposed in this paper into constraints on the 
parameters of the phenomenological model, but that would require extensive field theory calculations.

%%%%%%%%%%%%%%%%%%%%%%%%%%%%%%%%%%%%%%%%%%%%%%%%%%%%%%%%%%%%%%%%%%%%%%%%%%%%%%%%%%%%%%%%
\section*{Acknowledgements}
%%%%%%%%%%%%%%%%%%%%%%%%%%%%%%%%%%%%%%%%%%%%%%%%%%%%%%%%%%%%%%%%%%%%%%%%%%%%%%%%%%%%%%%%
This article is an extended version of the letter with Kostas Skenderis \cite{Nastase:2019rsn}, so all of the ideas and calculational set-ups 
were done in collaboration with him. I thank him also for direct collaboration at the early stages of this paper, and also for 
providing the basis of integrals from the \cite{KostasSoon} used in Appendix C.2 and B.4 to check parts  of the Feynman integrals.
We would like to thank Juan Maldacena and Paul McFadden for useful discussions. 
%KS would like to the ICTP-SAIFR and USP for hospitality during the initial stage of this work,
%and acknowledges support from FAPESP grants nr. 2016/01343-7, and nr. 2014/18634-
%9. This project has received funding/support from the European Union's Horizon 2020
%research and innovation programme under the Marie Sklodowska-Curie grant agreement
%No 690575. KS is also supported in part by the Science and Technology Facilities Council
%(Consolidated Grant ``Exploring the Limits of the Standard Model and Beyond").
%The work of HN 
My work is supported in part by  CNPq grant 301491/2019-4 and FAPESP grants 2019/21281-4 
and 2019/13231-7. I would also like to thank the ICTP-SAIFR for their support through FAPESP grant 2016/01343-7.

\appendix

\section{Coordinate ($x$) space integrals}

In this Appendix, for
completeness, we consider the  Feynman rules, and the Feynman diagrams for the current 2-point function, also in 
$x$ space, though the calculation of the integrals will be done only in the momentum space form.

First, the Feynman rules are as follows. 

{\bf External current insertions}

 -vertex insertion of $j_\mu^a$ with $\phi_i^{b}(x)\phi_j^{*c}(x)$ of value
\be
-i\epsilon^{abc}\d_\mu^\phi+i\epsilon^{abc}\d_\mu^{\phi^*}=-i\epsilon^{abc}(\d_\mu^\phi-\d_\mu^{\phi^*}).
\ee

--vertex insertion of $j_\mu^a$ with $A_\mu$, $\phi^b$ and $\phi^{*c}$, has the same value as in momentum space, 
\be
-2g\epsilon^{abc}\delta_{ij}\eta_{\mu\nu}.
\ee

{\bf Propagators}

-scalar propagator
\be
\Delta_{ij}^{ab}(x-y)=\frac{1}{4\pi}\frac{\delta_{ij}\delta^{ab}}{|x-y|}\;,
\ee

-gauge field propagator in Feynman gauge
\be
\Delta_{\mu\nu}(x-y)=\frac{1}{4\pi}\frac{\delta_{\mu\nu}}{|x-y|}\;,
\ee

{\bf Vertices}

-the vertex for $\phi^a_i\phi_j^{b,*}A_\mu$
\be
-ig\delta_{ij}\delta^{ab}(\d_\mu^\phi-\d_\mu^{\phi^*}).
\ee

-the vertex for $\phi_i^a\phi_j^{b,*}A_\mu A^\nu$
\be
-g^2\delta_\nu^\mu\delta_{ij}\delta^{ab}.
\ee

-the vertex for $\phi_1^a\phi_1^{c,*}\phi_2^b\phi_2^{d,*}$
\be
-\lambda \epsilon^{abe}\epsilon^{cde}=-2\lambda\delta_{ab}^{cd}.
\ee

{\bf One-loop}

With these Feynman rules, we can write the one-loop Feynman diagram result, which is simple,
\bea
I^{ab}_{\mu\nu}(x_1,x_2)
&=&\sum_{i=1,2}i^2\epsilon^{acd}\epsilon^{bdc}\frac{1}{(4\pi)^2}\frac{1}{|x_1-x_2|}(\d_{1,\mu}^{\leftarrow}-\d_{1,\mu}^{\rightarrow})(\d_{2,\nu}
^{\leftarrow}-\d_{2,\nu}^{\rightarrow})\frac{1}{|x_1-x_2|}\cr
&=&2\delta^{ab}\frac{2}{(4\pi)^2}\frac{4x_{1,\mu}x_{2,\nu}}{|x_1-x_2|^6}.
\eea

{\bf Two-loops}

At two loops, the formulas become more complicated. 

We can still find the vanishing of the same Feynman diagrams as in momentum space:

-The {\bf $I_1$ diagram} gives (a factor of 2 for the two types of scalars in the loops, 1,2)
\bea
I_{1,\mu\nu}^{ab}&=&2i^2\epsilon^{acd}\epsilon^{bc'd'}(-\lambda)(\delta_d^c\delta_{d'}^{c'}-\delta^c_{d'}\delta^{c'}_d) 
\int \frac{d^3x}{(4\pi)^4}\frac{1}{|x-x_1|}(\d_{1,\mu}^\leftarrow-\d_{1,\mu}^\rightarrow)\frac{1}{|x-x_1|}\cr
&&\times \frac{1}{|x-x_2|}(\d_{2,\nu}^\leftarrow -\d_{2,\nu}
^\rightarrow)\frac{1}{|x-x_2|}
=2\lambda \delta^{ab}\times 0=0.
\eea

-The {\bf $I_2$ and $I_2'$ diagrams} give, formally (vertex factors 
$-2 \epsilon^{acd}\epsilon^{bdc}(\lambda+g^2)=2(\lambda+g^2)\delta^{ab}$),
\be
I_{2,\mu\nu}^{ab}=2\delta^{ab}(\lambda+g^2)\frac{1}{(4\pi)^4}\int \frac{d^3x}{|x-x|}
\frac{1}{|x-x_1|}(\d_{1,\mu}^\leftarrow-\d_{1,\mu}^\rightarrow)
\frac{1}{|x_1-x_2|}(\d_{2,\mu}^\leftarrow-\d_{2,\mu}^\rightarrow)\frac{1}{|x-x_2|}\;,
\ee

This diagram could be removed by renormalization, but we saw that in momentum space, 
in dimensional regularization, it vanishes. The factor multiplying 
the divergent propagator is nonzero, namely
\bea
&&\frac{x_1\cdot x_2}{|x-x_1||x-x_2||x_1-x_2|}\left[\frac{1}{|x_1-x_2|^2|x-x_1|^2}
+\frac{1}{|x_1-x_2|^2|x-x_2|^2}\right.\cr
&&\left.-\frac{1}{|x_1-x_2|^4}
-\frac{1}{|x-x_1|^2|x-x_2|^2}\right].
\eea

-The {\bf $I_{4,\mu\nu}^{ab}(p)$ diagram} is proportional to 
\bea
&&\int d^dx \int d^dy\left[ \frac{1}{|x_1-x_2|} (\d_{1\mu}^\leftarrow-\d_{1\mu}^\rightarrow)
(\d_{2\nu}^\leftarrow-\d_{2\nu}^\rightarrow)\frac{1}{|x-x_1|}
(\d_{x,\rho}^\leftarrow-\d_{x,\rho}^\rightarrow)\frac{1}{|x-x_2|}\right]\times\cr
&&\times \frac{1}{|x-y|^2}(\d_{x,\rho}-\d_{x,\rho})\frac{1}{|x-x|}=0\;,
\eea
which vanishes because the last factor gives zero.

The nonzero diagrams are, however, more complicated than their momentum space counterparts. 

-The {\bf $I_3$ diagram} is nonzero, and gives 
\bea
I_{3,\mu\nu}^{ab}&=&2(-ig)^2\epsilon^{acd}\epsilon^{bdc}
\int \frac{d^3x d^3y}{(4\pi)^5|x-y|} \left\{\left[\frac{1}{|x_1-x|}(\d_{1,\mu}^\leftarrow
-\d_{1,\mu}^\rightarrow)\frac{1}{|y-x_1|}\right]\times\right.\cr
&&\left.\times(\d_{x,\rho}^\leftarrow-\d_{x,\rho}^\rightarrow)(\d_{y,\rho}^\leftarrow -\d_{y,\rho}^\rightarrow)
\left[\frac{1}{|x-x_2|}
(\d_{2,\mu}^\leftarrow-\d_{2,\mu}^\rightarrow)\frac{1}{|y-x_2|}\right]\right\}\cr
&=&-4g^2\delta^{ab}\int \frac{d^3x d^3y}{(4\pi)^5|x-y|}
\left\{\frac{3x\cdot y}{|x-x_1|^3|y-x_1|^3|x-x_2||y-x_2|}\times\right.\cr
&&\left.\times \left(\frac{1}{|x-x_1|^2}-\frac{1}{|y-x_1|^2}\right)
\left(\frac{1}{|x-x_2|^2}-\frac{1}{|y-x_2|^2}\right)\right.\cr
&&+\frac{3x\cdot y}{|x-x_1||y-x_1||x-x_2|^3|y-x_2|^3}\times\cr
&&\times\left(\frac{1}{|x-x_1|^2}-\frac{1}{|y-x_1|^2}\right)
\left(\frac{1}{|x-x_2|^2}-\frac{1}{|y-x_2|^2}\right)\cr
&&+\frac{9x\cdot y}{|x-x_1|^3|y-x_2|^3|y-x_1||x-x_2|}\times\cr
&&\times\left(\frac{1}{|x-x_1|^2}-\frac{1}{|y-x_1|^2}\right)
\left(\frac{1}{|x-x_2|^2}-\frac{1}{|y-x_2|^2}\right)\cr
&&+\frac{9x\cdot y}{|y-x_1|^3|x-x_2|^3|x-x_1||y-x_2|}\times\cr
&&\left.\times \left(\frac{1}{|x-x_1|^2}-\frac{1}{|y-x_1|^2}\right)
\left(\frac{1}{|x-x_2|^2}-\frac{1}{|y-x_2|^2}\right)\right\}.
\eea

It has UV divergences, just like the momentum space one, now corresponding to 
$x,y\rightarrow x_1,x_2$.  

-The {\bf $I_5$ diagram} is also nonzero, and is (there is a factor of two from the two scalars $i=1,2$ running in the loop)
\bea
I_{5,\mu\nu}^{ab}&=&4(-ig)^2(-i)\epsilon^{acd}(i)\epsilon^{adc}\int\frac{d^dxd^dy}{(4\pi)^5}\frac{1}{|x-y|}\times\cr
&&\times\left[\frac{1}{|x_1-x|}(\d_{1,\mu}^\leftarrow
-d_{1,\mu}^\rightarrow)(\d_{x,\rho}^\leftarrow-\d_{x,\rho}^\rightarrow)\frac{1}{|x_1-x_2||x-y|}\right.\cr
&&\left.(\d_{y,\rho}^\leftarrow-\d_{y,\rho}^\rightarrow)(\d_{2,\mu}^\leftarrow
-\d_{2,\mu}^\rightarrow)\frac{1}{|x_2-y|}\right]\cr
&=&8g^2\delta^{ab}\int\frac{d^dxd^dy}{(4\pi)^5}\frac{1}{|x-y|}\frac{(x_1\cdot x_2)(x\cdot y)}
{|x_1-x||x_1-x_2||x_2-y||x-y|}\times\cr
&&\times\left[\frac{3}{|x_1-x_2|^4|x-y|^2}\left(\frac{3}{|x-y|^2}-\frac{1}{|x_1-x|^2}\right)\right.\cr
&&+\frac{3}{|x_2-y|^4}\left(\frac{3}{|x_1-x|^4}+\frac{1}{|x_1-x_2|^2|x-y|^2}\right.\cr
&&\left.-\frac{1}{|x_1-x|^2|x-y|^2}-\frac{1}{|x_1-x_2|^2|x_1-x|^2}\right)\cr
&&-\frac{1}{|x-y|^2|x_2-y|^2}\left(\frac{3}{|x_1-x|^4}+\frac{3}{|x_1-x_2|^2|x-y|^2}\right.\cr
&&\left.-\frac{3}{|x_1-x|^2|x-y|^2}-\frac{1}{|x_1-x|^2|x_1-x_2|^2}\right)\cr
&&-\frac{1}{|x_2-y|^2|x_1-x_2|^2}\left(\frac{3}{|x_1-x|^4}+\frac{3}{|x_1-x_2|^2|x-y|^2}\right.\cr
&&\left.\left.-\frac{1}{|x_1-x|^2|x-y|^2}-\frac{3}{|x_1-x_2|^2|x_1-x|^2}\right)\right]\cr
&&
\eea

-The {\bf $I_6$ diagram} with the one-loop counterterm is the one-loop diagram, with the counterterm on one of the legs,
giving
\bea
I_{6,\mu\nu}^{ab}&=& 2\lambda_{ct}\delta^{ab}\int \frac{d^dx}{(4\pi)^3}\frac{1}{|x_1-x|}
(\d_{1,\mu}^\leftarrow -\d_{1,\mu}^\rightarrow)\frac{1}{|x_1-x_2|}
(\d_{2,\mu}^\leftarrow -d\_{2,\mu}^\rightarrow)\frac{1}{|x_2-x|}\cr
&=&2\lambda_{ct}\delta^{ab}\int \frac{d^dx}{(4\pi)^3}\frac{x_1\cdot x_2}
{|x_1-x||x_1-x_2||x_2-x|}\left[\frac{1}{|x_1-x|^2|x_1-x_2|^2}-\frac{3}{|x_1-x_2|^4}\right.\cr
&&\left.-\frac{1}{|x_2-x|^2}\left(\frac{1}{|x_1-x|^2}-\frac{1}{|x_1-x_2|^2}\right)\right].
\eea

-There are also the integrals for $I_7$ and $I_8$, but  we have 
already found that momentum space is easier, so we will not write expressions for them.

We realize now that a) the integrals for the Feynman diagrams are more complicated than in momentum space and b) it is 
more complicated to deal with potential divergences. That means that the standard treatment, of the calculation of Feynman 
diagrams in  momentum space, and in dimensional regularization, is an easier one, therefore will be adopted.

\section{Calculation of the Feynman diagrams and relevant integrals}\label{secmomFeyn}

The model was defined in section 4, with action (\ref{Euclideanaction}) and global symmetry current (\ref{current}). 

\subsubsection{Feynman rules}\label{Feynrules}

The Feynman rules for the model will be considered dropping the adjoint gauge indices. Indeed, we will only consider the 
planar diagrams, which are leading at large $N$, the case we are interested in, in which case as usual, using `t Hooft's double 
line notation for the adjoint fields, we find that the factors of $N$ pair up with $g^2$ to form the `t Hooft coupling $g^2N$.
Also, as we said, we will find that the result in this planar limit is independent of $\lambda$, so the only relevant coupling 
will be $g^2N$. Of course, there will be an overall factor of $N^2$ coming from the 2 index loops at 1-loop, where there is 
no coupling constant for the current 2-point function. 
So we will find a result of the type $N^2 f(g^2N)$. 

Explicitly, since the 1-loop diagram has no gauge field vertices, its factor is $\delta_A^A\simeq N^2$ (in the large $N$ 
limit), whereas the nonzero 2-loop diagrams come with two gauge field vertices (coupling to the scalars), order $g^2$, 
giving a factor of $(T_A)^{BC}(T_A)_{BC}\simeq N^3$. In both cases, we see the form $N^2f(g^2N)$ appearing.

We will work in Euclidean space, since we want to relate the 3 dimensional Euclidean space with the spatial part in the bulk.

{\bf External current insertions}

We will calculate current 2-point functions, so we start with the external current insertions.

The external current $j_\mu^a$ has both a $\phi^b\phi^{*c}$ part, and a $A_\nu \phi^b\phi^{*c}$ part, so we have 
the two possible insertions:

-vertex insertion of $j_\mu^a$ for momentum $k_1$ on $\phi^b$ and $k_2$ on $\phi^{*c}$ of value
(note that a possible overall minus sign in the vertex is irrelevant, as these vertices come in pairs in the relevant diagrams, 
but the {\em relative} sign with respect to the insertion with $A_\mu$ below is important)
\be
\epsilon^{abc}(k_1^\mu-k_2^\mu).
\ee

-vertex insertion of $j_\mu^a$ with $A_\mu$, $\phi^b$ and $\phi^{*c}$ of value (the 
factor of 2 comes from having 2 hermitian conjugate terms)
\be
-2g\epsilon^{abc}\delta_{ij}\eta_{\mu\nu}.
\ee

{\bf Propagators}

-scalar propagator
\be
\Delta_{ij}^{ab}(p)=\frac{\delta_{ij}\delta^{ab}}{p^2}.
\ee

-gauge field propagator in Feynman gauge
\be
\Delta_{\mu\nu}(p)=\frac{\delta_{\mu\nu}}{p^2}.
\ee

{\bf Vertices}

The relevant terms in the Euclidean action are 
\be
\Tr\left\{ig A_\mu (\vec{\phi}_i^*\d^\mu\vec{\phi}_i-\vec{\phi}_i\d^\mu \vec{\phi}_i^*)
+g^2A_\mu A^\mu\phi\phi^*+\lambda|\vec{\phi}_1\times \vec{\phi}_2|^2\right\}\;,
\ee
so we obtain the rules:

-the vertex for $\phi_i^a(k_1)\phi_j^{b,*}(k_2)A_\mu(k_3)$ (convention with all momenta in)
\be
g\delta_{ij}\delta^{ab}(2\pi)^3\delta^3(k_1+k_2+k_3)(k_1-k_2)_\mu.
\ee

-the vertex for $\phi_i^a(k_1)\phi_j^{b,*}(k_2)A_\mu(k_3)A^\nu(k_4)$  (all momenta in)
\be
-\delta_\nu^\mu \delta_{ij}\delta^{ab}g^2(2\pi)^3\delta^3(k_1+k_2+k_3+k_4)
\ee

-the vertex for $\phi_1^a(k_1)\phi_1^{c,*}(k_3)\phi_2^b(k_2)\phi_2^{d,*}(k_4)$ (all momenta in)
\be
-\lambda (2\pi)^3\delta^3(k_1+k_2+k_3+k_4)\epsilon^{abe}\epsilon_{cde}=-2\lambda\delta_{ab}^{cd}(2\pi)^3\delta^3(k_1+k_2+k_3+k_4).
\ee

\subsection{Calculation of the one-loop result}\label{Oneloopres}

Here we calculate the one-loop integral (\ref{oneloopdiag}, which we can do as follows.

First, expand
\bea
I_{\mu\nu}(p)&=&\int \frac{d^dq}{(2\pi)^d}\frac{(p+2q)_\mu(p+2q)_\nu}{q^2(p+q)^2}
=p_\mu p_\nu \int\frac{d^dq}{(2\pi)^d}\frac{1}{q^2(q+p)^2}\cr
&&+2p^\mu \int\frac{d^dq}{(2\pi)^d}\frac{q_\nu}{q^2(p+q)^2}+(\mu\leftrightarrow \nu)
+4\int \frac{d^dq}{(2\pi)^d}\frac{q_\mu q_\nu}{q^2(q+p)^2}\cr
&\equiv & p_\mu p_\nu I_0+2p_\mu I^a_\nu(p)+2p_\nu I^a_\mu(p)+4I^b_{\mu\nu}(p).
\eea

Then, from Lorentz invariance,
and using the fact that 
\be
I=\int \frac{d^d q}{(2\pi)^d 	q^n}\;,\;\; \forall n\;,
\ee
as well as 
\be
\int \frac{d^dq}{(2\pi)^d}\frac{(q-p)^2}{q^n}=\int \frac{d^dq}{(2\pi)^d}\frac{q^2}{(q+p)^n}\;,
\ee
both vanish in dimensional regularization (for any $n$, not necessarily integer),  by the formula
\be
\int \frac{d^dq}{(2\pi)^d}\frac{1}{(q^2+m^2)^n}=\frac{\Gamma(n-d/2)}{(4\pi)^n\Gamma(n)}\left(\frac{m^2}{4\pi}
\right)^{\frac{d}{2}-n}
\ee
at $m\rightarrow 0$, we find
\bea
I^a_\mu(p)&=&p^\mu I_a\;,\;\; 2p^2I_a=\int\frac{d^dq}{(2\pi)^d}\frac{(q+p)^2-q^2-p^2}{q^2(q+p)^2}=
-p^2I_0(p)\Rightarrow\cr
 I^a_\mu(p)&=&-\frac{p^\mu I_0(p)}{2}\cr
I^b_{\mu\nu}(p)&=&I_{b,1}\frac{p_\mu p_\nu}{p^2} +I_{b,2}\eta_{\mu\nu}\cr
I^{b,\mu}_\mu(p)&=&0=I_{b,1}+dI_{b,2}\cr
I^b_{\mu\nu}\frac{p^\mu p^\nu}{p^2}&=& I_{b,1}+I_{b,2}\cr
&=&\frac{1}{4p^2} \int \frac{d^dq}{(2\pi)^d}\frac{[(p+q)^2-p^2-q^2]^2}{q^2(p+q)^2}=\frac{p^2}{4}I_0(p)
\Rightarrow\cr
I^b_{\mu\nu}&=&\left(\eta_{\mu\nu}-d\frac{p_\mu p_\nu}{p^2}\right)\frac{p^2I_0(p)}{4(1-d)} .
\eea

From the general Feynman parametrization, we obtain
\be
I_0(p)\equiv I_{1,1}(p)=p^{d-4}\frac{\Gamma(2-d/2)\b(d/2-1,d/2-1)}{(4\pi)^{d/2}}=p^{d-4}\frac{\Gamma(2-d/2)
[\Gamma(d/2-1)]^2}{(4\pi)^{d/2}\Gamma(d-2)}.
\ee

In terms of it, the one-loop Feynman diagram is
\be
I^{ab}_{\mu\nu}(p)=-4\delta^{ab}\frac{(-p^2)}{2}I_0(p)\left(\eta_{\mu\nu}-\frac{p_\mu p_\nu}{p^2}\right).
\ee

For $d=3$, we obtain 
\be
I_0(p)=\frac{1}{8p}
\ee
and
\be
I^{ab}_{\mu\nu}(p)=\frac{p}{4}\delta^{ab}\left(\eta_{\mu\nu}-\frac{p_\mu p_\nu}{p^2}\right).
\ee

A priori, one should consider also the counterterm diagrams in Fig.\ref{fig:one-loop}b,c,d, but since the one-loop result
is finite, it is not necessary.

\subsection{Calculation of the two-loop result}

In this section we calculate the Feynman diagrams at two-loops, and the final result for their sum, 
relegating some more technical details to the next section.

-the $I_1$ {\bf diagram } (momenta $p$ coming in and out at the external points, 
momenta $q$ and $r$ on the two lower lines of the two loops, 
and $p+q$ and $p+r$ on the upper lines; we use dimensional regularization) gives
\bea
I_{1,\mu\nu}^{ab}(p)&=&
-i^22\lambda\delta^{ab} \int \frac{d^d q}{(2\pi)^d}\frac{(p+2q)_\mu}{q^2(p+q)^2}\int \frac{d^dr}{(2\pi)^d}\frac{(p+2r)_\nu}
{r^2(p+r)^2}\cr
&=&2\lambda\delta^{ab}[p_\mu I_0(p)+2p_\mu I_a(p)][p_\nu I_0(p)+2p_\nu I_a(p)]\;,
\eea
where, as at one-loop, we have used Lorentz invariance to say that the integral with 
$q_\mu$ in the numerator is proportional to $p_\mu$, and moreover we already calculated that $I_a(p)=-I_0(p)/2$, 
so that $I_{1,\mu\nu}^{ab}=0$ in dimensional regularization. 

-the {\bf $I_2$ diagram} gives 
\be
I_{2,\mu\nu}^{ab}(p)=2(\lambda+g^2)\delta^{ab}
\left(\int \frac{d^dr}{(2\pi)^d}\frac{1}{r^2}\right)\int \frac{d^dq}{(2\pi)^d}
\frac{(p+2q)_\mu(p+2q)_\nu}{q^2(p+q)^4}\;,
\ee
but as we already noted, in dimensional regularization the factorized integral vanishes, since
\be
\int \frac{d^dr}{(2\pi)^d[r^2+m^2]}=\frac{\Gamma(1-d/2}{4\pi }\left(\frac{m^2}{4\pi}\right)^{d/2-1}
\rightarrow 0\;\;{\rm as}\;\; m\rightarrow 0.\label{pformula}
\ee

-the {\bf $I_4(p)$ diagram} is proportional to
\be
I_{4,\mu\nu}^{ab}(p)\propto \int\frac{d^d q}{(2\pi)^d}\frac{2q^\mu}{q^2}=0\;,
\ee
so vanishes in dimensional regularization.

We now turn to the integrals that are nonzero, and need to be calculated. 

-the {\bf $I_3$ diagram}. Considering that momenta in and out of the diagram are equal to $p$, and on the two lower 
loop lines they are $q$ and $r$ (in the same direction, $q$ to the left) and $p+q$, $p+r$ on the upper lines, 
while the vertical gauge line has $q-r$, downwards, the result of the Feynman diagram is 
(the coefficient has a factor of 2 from the sum over $i=1,2$ running 
in the loops, a 2 from the exchange of the two gauge field vertices, up and down,
and a $-2\delta^{ab}$ coming from $\epsilon^{acd}\epsilon^{bdc}$, and a $1/2!$ from the fact that it comes from 
the action square, $S^2$, term in $e^{-S}$, as we explained)
\be
I_{3,\mu\nu}^{ab}(p)=-4\delta^{ab}g^2\int \frac{d^dq}{(2\pi)^d}\int \frac{d^dr}{(2\pi)^d}\frac{(p+2q)_\mu (p+2r)_\nu(2p+q+r)\cdot (q+r)}{(p+q)^2
q^2(q-r)^2r^2 (p+r)^2}.\label{I3munu}
\ee

-the {\bf $I_5$ diagram}. Considering that 
the momentum $p$ comes in and out of the diagram, and $q$ is the loop momentum 
on the lower scalar propagator, and $r$ the loop momentum on the gauge propagator, 
and $p+q$, and $p+q+r$ the momenta on the upper scalar
propagators, we obtain 
(again, the sum over $i=1,2$ running in the loops gives a factor of 2, a 2 from the exchange of the two 
vertices, left and right,
but now also a factor of 2 from choosing 
on which scalar propagator to have the self-energy, and a $-2\delta^{ab}$ from $\epsilon^{acd}\epsilon^{bdc}$, and 
the same $1/2!$ factor since it comes from a $S^2$ term in $e^{-S}$)
\be
I_{5,\mu\nu}^{ab}(p)=-8g^2\delta^{ab}\int\frac{d^dq}{(2\pi)^d}\frac{d^dr}{(2\pi)^d}
\frac{(p+2q)_\mu (p+2q)_\nu (2p+2q+r)^2}{q^2(p+q)^4r^2(p+q+r)^2}.\label{I5}
\ee

-the {\bf $I_6$ diagram} is the diagram with the counterterm vertex. Considering that the
 momentum $p$ is coming in and out of the diagram, and that we have 
 $q$ on the lower scalar propagator, and $p+q$ on the upper ones, we obtain 
 (we have a factor of 2 from the sum over $i=1,2$ in the scalar loop, another 2 from the sum over which scalar
propagator receives the vertex, and a $-2\delta^{ab}$ from $\epsilon^{acd}\epsilon^{bdc}$)
\be
I_{6,\mu\nu}^{ab}(p)=-8\lambda_{\rm 
ct}\delta^{ab}\int \frac{d^dq}{(2\pi)^d}\frac{(p+2q)_\mu(p+2q)_\nu}{q^2(p+q)^4}.
\ee

However, at one-loop we don't have any UV divergences. We have seen that the one-loop diagram for the current 2-point 
function is finite, but the same is true for the whole theory. 

Indeed, the one-loop diagrams giving the counterterm are: 
a momentum ($p$) line on which we have a scalar blob (a 4-scalar interaction), or a gauge glob 
(a 2-scalar-2-gauge interaction), and a one-loop diagram 
formed by a gauge propagator starting and ending on the scalar line (two 2-scalar-one-gauge interactions), 
for a total of 
\be
\lambda_{\rm ct}=
(\lambda+g^2)\int \frac{d^dq}{(2\pi)^dq^2}+g^2\int\frac{d^dq}{(2\pi)^d}\frac{(2p+q)^2}{q^2(p+q)^2}.
\ee

As we said, the first term vanishes in dimensional regularization, and the second then becomes 
\be
g^2\int \frac{d^dq}{(2\pi)^d}\frac{2(p+q)^2-q^2+2p^2}{q^2(p+q)^2}\;,
\ee
in which the first two terms vanish in dimensional regularization, and the last gives 
\bea
\lambda_{\rm ct}&=&+2g^2[p^2]^{d/2-1}\frac{\Gamma(2-d/2)}{(4\pi)^{d/2}}\int_0^1[\a(1-\a)]^{d/2-2}\cr
&=&+2g^2p^{d-2}\frac{\Gamma(2-d/2)[\Gamma(d/2-1)]^2}{(4\pi)^{d/2}\Gamma(d-2)}.
\eea

We note that this value is finite, so we don't actually need the counterterm. Moreover, as we saw, the 
one-loop diagram was finite as well.

Replacing $d=3$, we get
\be
\lambda_{\rm ct}=+\frac{g^2p}{4}\;,
\ee
but since this is finite, we can absorb it in the renormalization conditions, and just replace $\lambda_{\rm ct}$ with zero.

-the {\bf $I_{7a}$ diagram} gives (there is a factor of
2 from the sum over $i=1,2$ in the loop, 2 from the sum over which propagator gets the vertex,
and $-2\delta^{ab}$ from $\epsilon^{acd}\epsilon^{bdc}$, and a $(-2g)$ factor from the external gauge field insertion)
\be
I_{7,\mu\nu}^{ab}(p)=+16g^2\delta^{ab}\int\frac{d^dq}{(2\pi)^d}\frac{d^dr}{(2\pi)^d}\frac{(2p+2q+r)_\mu 
(p+2q)_\nu}{q^2r^2(p+q)^2(p+q+r)^2}\;,\label{I7}
\ee
whereas diagram $I_{7b}$ gives the same result, just with $\mu$ and $\nu$ interchanged.

-the {\bf $I_8$ diagram} gives (there is 
a factor of 2 from summing over $i=1,2$ and a $-2\delta^{ab}$ from $\epsilon^{acd}\epsilon^{bdc}$, 
and two $(-2g)$ factors from the external gauge field insertion)
\be
I_{8,\mu\nu}^{ab}(p)=-16(-g)^2 \delta^{ab}\eta_{\mu\nu}\int \frac{d^dq}{(2\pi)^d}\frac{d^dr}{(2\pi)^d}
\frac{1}{q^2r^2(p+q+r)^2}.\label{I8}
\ee

{\bf Standard Denominators}

Thus the integrals we need to calculate are $I_3,I_5,I_6,I_{7a}$ and $I_8$. There is a basis of integrals with denominators
$q^2,r^2,(q+p)^2,(r+p)^2,(q-r)^2$, so we rewrite the nonzero 
integrals in terms of them. 

We make the changes: in $I_5$ and $I_7$, $p+r=-\tilde r$ (and drop the tilde), obtaining the nonzero terms:
\bea
I_{3,\mu\nu}^{ab}(p)&=&-4\delta^{ab}g^2\int \frac{d^dq}{(2\pi)^d}\int \frac{d^dr}{(2\pi)^d}\frac{(p+2q)_\mu (p+2r)_\nu(2p+q+r)\cdot (q+r)}{(p+q)^2
q^2(q-r)^2r^2 (p+r)^2}\cr
I_{5,\mu\nu}^{ab}(p)&=&-8g^2\delta^{ab}\int\frac{d^dq}{(2\pi)^d}\frac{d^dr}{(2\pi)^d}
\frac{(p+2q)_\mu(p+2q)_\nu(p+2q-r)^2}{q^2(q+p)^4(r+p)^2(q-r)^2}\cr
I_{6,\mu\nu}^{ab}(p)&=&-8\lambda_{\rm ct}\delta^{ab}\int \frac{d^dq}{(2\pi)^d}
\frac{(p+2q)_\mu(p+2q)_\nu}{q^2(p+q)^4}\cr
I_{7a,\mu\nu}^{ab}(p)&=&+16g^2\delta^{ab}\int\frac{d^dq}{(2\pi)^d}\frac{d^dr}{(2\pi)^d}
\frac{(p+2q-r)_\mu(p+2q)_\nu}{q^2(r+p)^2(q+p)^2(q-r)^2}\cr
I_{7b,\mu\nu}^{ab}(p)&=&+16g^2\delta^{ab}\int\frac{d^dq}{(2\pi)^d}\frac{d^dr}{(2\pi)^d}
\frac{(p+2q-r)_\nu(p+2q)_\mu}{q^2(r+p)^2(q+p)^2(q-r)^2}=I_{7a,\nu\mu}^{ab}(p)\cr
\lambda_{\rm ct}&=&+2g^2p^{d-2}\frac{\Gamma(2-d/2)[\Gamma(d/2-1)]^2}{(4\pi)^{d/2}\Gamma(d-2)}\cr
I_{8,\mu\nu}^{ab}&=&-16g^2\delta^{ab}\eta_{\mu\nu}2\int \frac{d^dq}{(2\pi)^d}\int \frac{d^dr}{(2\pi)^d}\frac{1}
{q^2r^2(p+q+r)^2}.\label{integrals}
\eea

\subsection{The full integrals in dimensional regularization}

In Appendix \ref{secdetails} we give the full details of the calculation of the integrals in dimensional
regularization, using two methods to calculate the integrals: a direct one, and an indirect
one, using a basis of integrals from \cite{KostasSoon}.
Here we only summarize the results. First, one needs to define some quantities that
will be useful in describing the results,
\bea
G_1&=&\frac{\Gamma(2-d/2)\Gamma(d/2-1)^2}{\Gamma(d-2)}\cr
G_2&=&\frac{\Gamma(3-d)\Gamma(d/2-1)^3}{\Gamma(3d/2-3)}\cr
B_0&\equiv& \frac{p^{d-4}}{(4\pi)^{d/2}}G_1\cr
&\equiv& p^{d-4}B\cr
J_0&=&\frac{p^{2d-6}}{(4\pi)^d}G_2.\label{divquant}
\eea

We see that $B_0$ is finite and $J_0$ is divergent, and we will find that the combinations appearing in the integrals 
will be $B_0^2$ and $J_0$, and then only the latter is of interest for the calculation of the anomalous dimension. 
As usual in dimensional regularization, the UV and IR divergences mix in the same integral, though if an integral has only 
one type of divergence, we could find out which it is by whether $d=3\pm \epsilon$ finds a positive coefficient.

However, since the one-loop result for the 2-point current correlator is finite and will therefore not interfere at 
two-loops, we can unambigously simply drop the coefficient of $1/\epsilon$ regardless of whether it comes from a UV or an IR 
divergence, and this will be our renormalization scheme at two-loops. The finite pieces will contain then a $\log p^2$ 
term that is IR divergent in the usual sense: as $p^2$ becomes on-shell and its mass becomes zero, this result will diverge, 
and one should add diagrams with soft emission from the external lines, but here we just want to extract the correct $\log 
p^2$ terms, for which nothing else is required. 

We will then define $d = 3+\epsilon$  as above in the following.
For $I_8$, we find
\bea
I_{8,\mu\nu}^{ab}(p)
&=&-16g^2\delta^{ab}\eta_{\mu\nu}B I_{1,1+\frac{2-d}{2}}=-16g^2
\delta^{ab}\eta_{\mu\nu}J_0\;,
\eea
using the notation in Appendix B, and is divergent.

For $I_6$, we find
\bea
I^{ab}_{6,\mu\nu}(p)
&=&-8\lambda_{\rm ct}\delta^{ab}\frac{B_0}{d-1}\left[(3-d/4)\eta_{\mu\nu}-\left(\frac{d^2}{4}-2d+1\right)\frac{p_\mu 
p_\nu}{p^2}\right].
\eea
but it is finite, so it doesn't contribute to the $\log p^2$ term, and can be ignored.
For $I_5$, we find
\bea
I^{ab}_{5,\mu\nu}(p)
&=&-16g^2\delta^{ab}\frac{J_0}{3(d-4)}\left\{-4\delta_{\mu\nu}
+d\frac{p^\mu p^\nu}{p^2}\right\}\;,
\eea
so it is divergent, and contributes to the $\log p^2$ term. Expanding in $\epsilon$, we get
\be
I^{ab}_{5,\mu\nu}(p)=-16g^2\delta^{ab}J_0\left\{\frac{4}{3}\delta_{\mu\nu}\left(1+\epsilon\right)-
\frac{p^\mu p^\nu}{p^2}\left[1+\frac{4\epsilon}{3}\right]\right\}.
\ee

For $I_7$, we find
\bea
I^{ab}_{7a,\mu\nu}(p)
&=&+16g^2\delta^{ab}\frac{J_0}{d-4}\left\{-\delta_{\mu\nu}+(d-3)\frac{p^\mu p^\nu}{p^2}\right\}\;,
\eea
and then also $I^{ab}_{7a,\mu\nu}=I^{ab}_{7b,\mu\nu}$.
Expanding in $\epsilon$, we find
\be
I^{ab}_{7a,\mu\nu}(p)=+16g^2\delta^{ab}J_0\left\{\eta_{\mu\nu}
\left(1+\epsilon\right)-\epsilon\frac{p^\mu p^\nu}{p^2}\right\}.
\ee

For $I_3$, we find
\bea
I_{3,\mu\nu}^{ab}(p)&=&-4g^2\delta^{ab}
J_0\left\{\left(\eta_{\mu\nu}-d\frac{p^\mu p^\nu}{p^2}\right)\frac{2}{1-d}\left[\frac{3d-8}{d-4}-\frac{5}{3}\right]
-\frac{p^\mu p^\nu}{p^2}4\frac{d-2}{d-4}\right\}\cr
&&-4g^2\delta^{ab}\left\{
-4\frac{p_\mu p_\nu}{p^2}\left[-\frac{(3d-10)(3d-8)}{(d-4)^2}J_0+\frac{d-3}{d-4}p^2B_0^2\right]\right.\cr
&&-8\eta_{\mu\nu}\left[-\frac{d(d-3)}{(d-4)^2(d-1)}J_0+\frac{1}{2(d-4)(d-1)}p^2B_0^2\right]\cr
&&\left.-\frac{8p_\mu p_\nu}{p^2}\left[\frac{2(d-2)(2d^2-9d+8)}{(d-4)^2(d-1)}J_0 -\frac{(d-2)^2}{2(d-4)(d-1)}
p^2B_0^2\right]\right\}.
\eea
Finally, the sum of all the diagrams gives the full 2-loop result
\be
I_{\rm 2-loop}(p)=\frac{16}{d-4}g^2\delta^{ab}\left(\delta_{\mu\nu}-\frac{p_\mu p_\nu}{p^2}\right)\left[2J_0-
\frac{d^3-5d^2-8d+8}{(d-1)(d-4)}J_0+\frac{B_0^2p^2}{d-1}\right].
\ee

\subsection{Divergences of the model}

As a consistency check for our calculation, and as a way to extract simply just the needed
divergence, we check that the two ways of calculating the integrals used in Appendix \ref{secdetails} give
the same divergent piece for the most complicated integral, $I_3$.
Focusing on the coefficient of $J_0$ for $d=3$, we obtain in the second way of calculating, 
using the basis of integrals (the
notation is explained in Appendix \ref{secdetails}.2)
\bea
K_1&=&\frac{J_0}{p^4}\;,\;\;\; I_\mu ^\a=p_\mu K_1.\cr
K_{00}^{11}&=&K_{00}^{22}=0=K_{00}^{12}=K_{00}^{21}\cr
K_{11}^{12}&=&K_{11}^{21}=-\frac{J_0}{p^4}\cr
K_{11}^{11}&=&K_{11}^{22}=-\frac{J_0}{p^4}.\Rightarrow\;\;\;
I_{\mu\nu}^{11}=I_{\mu\nu}^{22}=-\frac{p_\mu p_\nu}{p^2}\frac{J_0}{p^2}\cr
I_{\mu\nu}^{12}&=&I_{\mu\nu}^{21}=-J_0\frac{p_\mu p_\nu}{p^2}\Rightarrow\;\;\;
I_{\mu\nu}^{ss}=-4\frac{p_\mu p_\nu}{p^2} \frac{J_0}{p^2}.\cr
K_{001}^{111}&=&0\cr
K_{111}^{111}&=&\frac{J_0}{p^4}\Rightarrow\;\;\;
I_{\mu\nu\rho}^{111}=\frac{p_\mu p_\nu p_\rho}{p^2}\frac{J_0}{p^2}.\cr
K_{001;1}^{112}&=&K_{001;2}^{112}=0\cr
K_{111}^{112}&=&\frac{J_0}{p^2}\Rightarrow\;\;\;
I_{\mu\nu\rho}^{112}=\frac{p_\mu p_\nu p_\rho}{p^2}J_0\cr
K_{0011;1}^{1112}&=&K_{0011;2}^{1122}=0\cr
K_{0000}^{1112}&=&\frac{J_0}{30}\;,\;\;\;
K_{1111}^{1112}=-\frac{J_0}{p^4}\Rightarrow \cr
I_{\mu\nu\rho\sigma}^{1112}&=&(\delta_{\mu\nu}\delta_{\rho\sigma}+\delta_{\mu\rho}\delta_{\nu \sigma}+
\delta_{\mu\sigma}\delta_{\nu\rho})\frac{J_0}{30}-\frac{p_\mu p_\nu}{p^2}\frac{p_\rho p_\sigma}{p^2}J_0\Rightarrow\cr
I_{\mu\rho\rho\sigma}^{1112}&=&J_0\left(\frac{\delta_{\mu\sigma}}{6}-\frac{p_\mu p_\sigma}{p^2}\right).\cr
K_{0011;1}^{1122}&=&K_{0011;2}^{1122}=K_{0011;3}^{1122}=0\cr
K_{0000;1}^{1122}&=&\frac{J_0}{10}\;,\;\;\;
K_{0000;2}^{1122}=\frac{J_0}{60}\cr
K_{1111}^{1122}&=&-\frac{J_0}{p^4}\Rightarrow\cr
I_{\mu\nu\rho\sigma}^{1122}&=&\frac{J_0}{10}\delta_{\mu\nu}\delta{\rho\sigma}+(\delta_{\mu\rho}\delta_{\nu\sigma}
+\delta_{\mu\sigma}\delta_{\nu\rho})\frac{J_0}{60}-\frac{p_\mu p_\nu}{p^2}\frac{p_\rho p_\sigma}{p^2}J_0
\Rightarrow\cr
I_{\mu\rho\rho\sigma}^{1122}&=&
=J_0\left(\frac{\delta_{\mu\sigma}}{6}-\frac{p_\mu p_\sigma}{p^2}\right).
\eea
Substituting in the expansion of $I_3$ in the basis of integrals,
\bea
I^{ab}_{3\mu\nu}(p)&=&-4\delta^{ab}g^2\left[4p_\mu p_\nu p^2K_1+p_\mu p_\nu I^{ss}_{\rho\rho}(p)
+4p_\rho p_\mu (I^{21}_{\nu\rho}+I^{22}_{\nu \rho})+4p_\rho p_\nu (I^{11}_{\mu\rho}+I^{12}_{\mu\rho})\right.\cr
&&\left. +2p_\mu (I^{112}_{\rho\rho\nu}+I^{111}_{\rho\rho \nu}+2I^{112}_{\nu \rho \rho})+
2p_\nu (I^{112}_{\rho\rho\mu}+I^{111}_{\rho\rho \mu}+2I^{112}_{\mu \rho \rho})\right.\cr
&&\left.+8p_\rho (I^{112}_{\mu\rho\nu}+I^{112}_{\nu \rho \mu})
+4(I^{1112}_{\mu\rho\rho\nu}+I^{1112}_{\nu \rho\rho\mu}+2I^{1122}_{\mu\rho\nu\rho})\right],\cr
&&
\eea
we get
\be
I_3(p)=-4\delta^{ab}g^2J_0\left[\frac{8}{3}\delta_{\mu\nu}\right].
\ee

This is the same result as is obtained from the first method of calculating, from (\ref{firstfive}) plus (\ref{lastone}).

For $I_8$, the result (\ref{I8result}) is already just a $J_0$ piece in $d=3$, 
\be
I_{8,\mu\nu}^{ab}(p)=-16g^2
\delta^{ab}\delta_{\mu\nu}J_0.
\ee

For $I_5$, we have calculated the result in (\ref{I5result}), with divergent part given by 
\be
I_{5,\mu\nu}^{ab}(p)=-16g^2\delta^{ab}J_0\left[\frac{4}{3}\delta_{\mu\nu}-\frac{p_\mu p_\nu}{p^2}\right].
\ee

For $I_7$, we calculated it as well in (\ref{I7result}), given by 
\be
I_{7,\mu\nu}^{ab}(p)=+16g^2\delta^{ab}J_0\delta_{\mu\nu}.
\ee

This completes the calculation of the divergences of the current 2-point function.

\section{Details for the calculation of the momentum space integrals in dimensional regularization}\label{secdetails}

\subsection{Direct calculation of integrals}

We begin with a set of notations for the objects that will appear in the calculations, as defined in (\ref{divquant}).

Expanding in $d=3+\epsilon$, with $\psi(x)\equiv \Gamma'(x)/\Gamma(x)$, we find
\bea
G_1&\simeq & \pi^{3/2}\left[1+\epsilon\left(-\psi(1)+\frac{1}{2}\psi(1/2)\right)\right]+...\cr
G_2&\simeq & -\frac{2}{\epsilon}\left[1+\epsilon\left(-\psi(1)+\frac{3}{2}\psi(1/2)-\frac{3}{2}\psi(3/2)\right)\right]+...
\Rightarrow \cr
B_0&\simeq & \frac{p^{\epsilon-1}}{8}\left[1+\epsilon\left(-\frac{1}{2}\ln(4\pi)-\psi(1)+\frac{1}{2}\psi(1/2)\right)\right]+...\cr
J_0&\simeq & -\frac{2\pi p^{2\epsilon}}{(4\pi)^3\epsilon}\left[1+\epsilon\left(-\ln (4\pi)-\psi(1)+\frac{3}{2}\psi(1/2)
-\frac{3}{2}\psi(3/2)\right)\right]+...
\eea

We will then define $d=3+\epsilon$ as above in the following. We write some formulas for integrals in 
dimensional regularization that will be used in the following. 

Using Feynman parametrization, we have
\bea
I_{1,1}(p)&\equiv& \int \frac{d^dq}{(2\pi)^d}\frac{1}{(q^2+m_1^2)((q+p)^2+m_2^2)}\cr
&=&\int_0^1 d\a \int \frac{d^d\tilde q}{(2\pi)^d}\frac{1}{[\tilde q^2+\a(1-\a)p^2+\a m_1^2+(1-\a)m_2^2]^2}\cr
&=&\frac{\Gamma(2-d/2)}{(4\pi)^{d/2}}\int_0^1[\a(1-\a)p^2+\a m_1^2+(1-\a)m_2^2]^{\frac{d}{2}-2}.
\eea

Then, puting $m_1=m_2=0$, we first obtain what we called $I_0(p)$ previously,
\be
I_{1,1}(p)\equiv I_0(p)=p^{d-4}\frac{\Gamma(2-d/2)\b(d/2-1,d/2-1)}{(4\pi)^{d/2}}=
p^{d-4}\frac{\Gamma(2-d/2)[\Gamma(d/2-1)]^2}{(4\pi)^{d/2}\Gamma(d-2)}.\label{I11}
\ee

More generally, taking $(\d^2/\d m_2^2)^n$ and then putting $m_1=m_2=0$, we obtain 
\bea
I_{1,1+n}(p)&\equiv & \int \frac{d^dq}{(2\pi)^d}\frac{1}{q^2(p+q)^{2(n+1)}}\cr
&=&\frac{\Gamma(n+2-d/2) p^{d-4-2n}}{(4\pi)^{d/2}\Gamma(n)}\int^1_0d\a \a^{\frac{d}{2}-2-n}(1-\a)^{\frac{d}
{2}-2}\cr
&=&\frac{p^{d-4-2n}}{(4\pi)^{d/2}}\frac{\Gamma(n+2-d/2)\Gamma(d/2-1-n)\Gamma(d/2-1)}
{\Gamma(n+1)\Gamma(d-2-n)}.\label{I11n}
\eea
We then note that the formula is not only valid for integer $n$, but by analytical continuation (just like usual dimensional 
regularization) can be defined for any real $n$. 

In particular, for $n=1$ we get
\be
I_{1,2}(p)=p^{d-6}\frac{\Gamma(2-d/2)(d/2-2)[\Gamma(d/2)]^2}{(4\pi)^{d/2}(d/2-1)\Gamma(d-1)}.
\ee

Other relevant values of $n$ for the following are given by:
\bea
p^2I_{1,1+\frac{4-d}{2}}(p)&=&\frac{p^{2d-6}}{(4\pi)^{d/2}}\frac{\Gamma(3-d)\Gamma(d-2)\Gamma(d/2-1)}{\Gamma(2-
-d/2)\Gamma(3d/2-3)}\times \left[\frac{\frac{3d}{2}-4}{\frac{d}{2}-2}\right]\cr
I_{1,1+\frac{2-d}{2}}(p)&=&\frac{p^{2d-6}}{(4\pi)^{d/2}}\frac{\Gamma(3-d)\Gamma(d-2)\Gamma(d/2-1)}{\Gamma(2-
d/2)\Gamma(3d/2-3)}\times \left[1\right]\cr
\frac{1}{p^2}I_{1,1+\frac{-d}{2}}(p)&=&\frac{p^{2d-6}}{(4\pi)^{d/2}}\frac{\Gamma(3-d)\Gamma(d-2)\Gamma
(d/2-1)}{\Gamma(2-d/2)\Gamma(3d/2-3)}\times \left[\frac{1}{3}\right].
\eea

{\bf Calculation of Feynman integrals}

-For $I_8$, doing the integral over $r$ in (\ref{I8}), using (\ref{I11}) and then (\ref{I11n}),
we obtain
\bea
I_{8,\mu\nu}^{ab}(p)&=&
-16g^2 \delta^{ab}\eta_{\mu\nu}\frac{\Gamma(2-d/2)}{(4\pi)^{d/2}}\frac{[\Gamma(d/2-1)]^2}{\Gamma(d-2)}
\int \frac{d^dq}{(2\pi)^d}\frac{1}{q^2(p+q)^{4-d}}\cr
&=&-16g^2\delta^{ab}\eta_{\mu\nu}B I_{1,1+\frac{2-d}{2}}=-16g^2
\delta^{ab}\eta_{\mu\nu}J_0\;,\label{I8result}
\eea
so is divergent (has a $J_0$ piece).

-For $I_6$, we expand the integral as
\bea
I_{6,\mu\nu}&=&-8\lambda_{\rm ct}\delta^{ab}\int \frac{d^dq}{(2\pi)^d}\frac{(p+2q)_\mu (p+2q)_\nu}{q^2
(p+q)^4}\cr
&=&-8\lambda_{\rm ct}\delta^{ab}\left[p_\mu p_\nu I_{1,2}(p)+2p_\mu I^\nu_{1,2}(p)+2p_\nu I^\mu_{1,2}(p)
+4I^{\mu\nu}_{1,2}(p)\right].
\eea

Then, by Lorentz invariance and vanishing of the various integrals in dimensional regulararization, we obtain
\bea
I^\mu_{1,2}(p)&\equiv & \int \frac{d^dq}{(2\pi)^d}\frac{q^\mu}{q^2(q+p)^4}=\frac{p^\mu}{2p^2}
\int \frac{d^dq}{(2\pi)^d}\frac{(q+p)^2-q^2p^2}{q^2(q+p)^2}\cr
&=&\frac{p^\mu}{2p^2}[I_{1,1}(p)-p^2I_{1,2}(p)]\cr
&=&-\frac{p^\mu p^{d-6}}{4(4\pi)^{d/2}}\frac{\Gamma(2-d/2)[\Gamma(d/2-1)]^2}{\Gamma(d-2)}\cr
I^{\mu\nu}_{1,2}(p)&\equiv& \frac{d^dq}{(2\pi)^d}\frac{q_\mu q_\nu}{q^2(q+p)^4}=\eta_{\mu\nu}J_1
+\frac{p^\mu p^\nu}{p^2}J_2\Rightarrow\cr
I^{\mu\mu}_{1,2}(p)&=&dJ_1+J_2=\int \frac{d^dq}{(2\pi)^d}\frac{1}{(q+p)^4}=0\Rightarrow J_2=-dJ_1\cr
\frac{p_\mu p_\nu}{p^2}I^{\mu\nu}_{1,2}(p)&=&J_1+J_2=(1-d)J_1=\frac{1}{4p^2}\int \frac{d^dq}{(2\pi)^d}
\frac{[(q+p)^2-q^2-p^2]^2}{q^2(q+p)^4}\cr
&=&\frac{1}{4}[p^2I_{1,2}(p)-2I_{1,1}(p)]\Rightarrow\cr
I^{\mu\nu}_{1,2}(p)&=&J_1\left(\eta_{\mu\nu}-d\frac{p_\mu p_\nu}{p^2}\right)\cr
&=&-\frac{1}{4(d-1)}[p^2I_{1,2}-2I_{1,1}]\left(\eta_{\mu\nu}-d\frac{p_\mu p_\nu}{p^2}\right).
\eea

Finally, we obtain for the integral
\bea
I^{ab}_{6,\mu\nu}&=&-8\lambda_{\rm ct}\delta^{ab}p^{d-4}\frac{\Gamma(2-d/2)[\Gamma(d/2-1)]^2}{(4\pi)^{d/2}
(d-1)\Gamma(d-2)}\left[(3-d/4)\eta_{\mu\nu}+\left(\frac{d^2}{4}-2d-1\right)\frac{p_\mu p_\nu}{p^2}\right]\cr
&=&-8\lambda_{\rm ct}\delta^{ab}\frac{B_0}{d-1}\left[(3-d/4)\eta_{\mu\nu}-\left(\frac{d^2}{4}-2d+1\right)\frac{p_\mu 
p_\nu}{p^2}\right].
\eea

This is finite (it has no $J_0$ piece), 
so even if we would consider a nonzero $\lambda_{\rm ct}$, it would not contribute to the 
$\log p^2$ term. We can therefore safely ignore it.

For completeness, note that in $d=3+\epsilon$, we obtain 
\bea
I^{ab}_{6,\mu\nu}&=&-\frac{\lambda_{\rm ct}\delta^{ab}}{2p}\left[\frac{9-\epsilon}{4}\eta_{\mu\nu}
-\frac{11+2\epsilon}{4}\frac{p_\mu p_\nu}{p^2}\right]\times\cr
&&\times\left\{1+\epsilon\left[\ln p -\frac{1}{2}\ln (4\pi)
-\frac{1}{2}+\frac{1}{2}\psi(1/2)-\psi(1)\right]\right\}.
\eea

-For $I_5$, we first expand (\ref{I5}) as
\bea
I_{5,\mu\nu}^{ab}(p)&=&
-8g^2\delta^{ab}\int\frac{d^dq}{(2\pi)^d}\frac{d^dr}{(2\pi)^d}(p+2q)_\mu(p+2q)_\nu \left[\frac{1}{q^2(p+q)^4r^2}\right.\cr
&&\left.+\frac{3}{q^2(p+q)^2r^2(p+q+r)^2}
+\frac{2(p+q)^\rho r_\rho}{q^2(p+q)^4r^2(p+q+r)^2}\right]\;,
\eea
where we have used $(2p+q+r)^2=(p+q+r)^2+3(p+q)^2+2(p+q)\cdot r$.

The first term in the square bracket above contains the integral
$\int \frac{d^dr}{(2\pi)^d}\frac{1}{r^2}$,
which is zero in dimensional regularization, as we have shown above. The third (last) term in the square bracket contains an  integral already calculated,
\be
\int \frac{d^dr}{(2\pi)^d}\frac{r_\rho}{r^2(p+q+r)^2}=-\frac{(p+q)_\rho}{2} I_0(p+q)\;,
\ee
leading to the term being 
\be
-\int \frac{d^dq}{(2\pi)^d}\frac{(p+2q)_\mu(p+2q)_\nu}{r^2(p+q)^2}\int\frac{d^dr}{(2\pi)^d}\frac{1}{q^2(p+q+r)^2}.
\ee

The second term in the square bracket then gives $+3$ times the same integral (instead of the $-1$), for a total factor
of 2. Using the form of $I_0(p)=I_{1,1}(p)$ in (\ref{I11}), 
we finally obtain 
\be
I_{5,\mu\nu}^{ab}(p)=-16g^2\delta^{ab}\frac{\Gamma(2-d/2)[\Gamma(d/2-1)]^2}{(4\pi)^{d/2}\Gamma(d-2)}
\int \frac{d^dq}{(2\pi)^d}\frac{(p+2q)_\mu(p+2q)_\nu}{q^2(p+q)^{6-d}}\;,
\ee
and then we expand the integral as 
\be
\int \frac{d^dq}{(2\pi)^d}\frac{(p+2q)_\mu(p+2q)_\nu}{q^2(p+q)^{6-d}}
=p_\mu p_\nu I_{1,1+\frac{4-d}{2}}+2p_\mu I^\mu_{1,1+\frac{4-d}{2}}+2p_\nu I^\nu_{1,1+\frac{4-d}{2}}
+4I^{\mu\nu}_{1,1+\frac{4-d}{2}}.
\ee

Similarly to what we did before in the case of $n=2$ in $I_{1,1+n}$, we 
calculate first 
\bea
I^\mu_{1,1+\frac{4-d}{2}}&\equiv&\int \frac{d^dq}{(2\pi)^d}\frac{q^\mu}{p^2(q+p)^{6-d}}=\frac{p^\mu}{2p^2}
\int\frac{d^dq}{(2\pi)^d}\frac{(q+p)^2-q^2-p^2}{q^2(q+p)^{6-d}}\cr
&=&\frac{p^\mu}{2p^2}\left[I_{1,1+\frac{2-d}{2}}-p^2I_{1,1+\frac{4-d}{2}}\right]\;,\label{Imu}
\eea
then (using, as before, Lorentz invariance and vanishing of some dimensional regularization integrals)
\bea
I^{\mu\nu}_{1,1+\frac{4-d}{2}}&\equiv& \int \frac{d^dq}{(2\pi)^d}\frac{q^\mu q^\nu}{q^2(q+p)^{6-d}}
=\frac{p^\mu p^\nu}{p^2}J_{b1}+\delta_{\mu\nu}J_{b2}\Rightarrow\cr
I^{\mu\mu}_{1,1+\frac{4-d}{2}}&=&J_{b1}+dJ_{b2}=0\Rightarrow J_{b1}=-dJ_{b2}\cr
\frac{p^\mu p^\nu}{p^2}I^{\mu\nu}_{1,1+\frac{4-d}{2}}&=&J_{b1}+J_{b2}=\frac{1}{4p^2}\int \frac{d^dq}{(2\pi)^d}
\frac{[(q+p)^2-q^2-p^2]^2}{q^2(q+p)^{6-d}}\cr
&=&\frac{1}{4p^2}\left[I_{1,1+\frac{-d}{2}}+p^4I_{1,1+\frac{4-d}{2}}-2p^2I_{1,1+\frac{2-d}{2}}\right]\Rightarrow\cr
I^{\mu\nu}_{1,1+\frac{4-d}{2}}&=&\left(\delta_{\mu\nu}-d\frac{p^\mu p^\nu}{p^2}\right)\frac{1}{4p^2(1-d)}
\left[I_{1,1+\frac{-d}{2}}+p^4I_{1,1+\frac{4-d}{2}}-2p^2I_{1,1+\frac{2-d}{2}}\right].\cr
&&\label{Imunu}
\eea

We then find for the $I_5$ Feynman diagram
\bea
I^{ab}_{5,\mu\nu}&=&-16g^2\delta^{ab}\frac{\Gamma(2-d/2)[\Gamma(d/2-1)]^2}{(4\pi)^{d/2}\Gamma(d-2)}
\left\{\frac{p^\mu p^\nu}{p^2}p^2 I_{1,1+\frac{4-d}{2}}\right.\cr
&&\left.+2\frac{p^\mu p^\nu}{p^2}\left[I_{1,1+\frac{2-d}{2}}
-p^2 I_{1,1+\frac{4-d}{2}}\right]\right.\cr
&&\left.+\left(\delta_{\mu\nu}-d\frac{p^\mu p^\nu}{p^2}\right)\frac{1}{p^2(1-d)}
\left[I_{1,1+\frac{-d}{2}}+p^4I_{1,1+\frac{4-d}{2}}-2p^2I_{1,1+\frac{2-d}{2}}\right]\right\}\cr
&=&-16g^2\delta^{ab}\frac{p^{2d-6}[\Gamma(d/2-1)]^3\Gamma(3-d)}{(4\pi)^d\Gamma(3d/2-3)}\left\{\frac{\delta_
{\mu\nu}}{1-d}
\left[\frac{1}{3}+\frac{3d/2-4}{d/2-2}-2\right]\right.\cr
&&\left.+\frac{p^\mu p^\nu}{p^2}\left[+\frac{1}{d-1}\frac{3d/2-4}{d/2-2}-\frac{2}{d-1}+\frac{d}{3(d-1)}\right]\right\}\cr
&=&-16g^2\delta^{ab}\frac{J_0}{3(d-4)}\left\{-4\delta_{\mu\nu}
+d\frac{p^\mu p^\nu}{p^2}\right\}.\label{I5resultd}
\eea

Substituting $d=3+\epsilon$, we get
\be
I^{ab}_{5,\mu\nu}=-16g^2\delta^{ab}J_0\left\{\frac{4}{3}\delta_{\mu\nu}\left(1+\epsilon\right)-
\frac{p^\mu p^\nu}{p^2}\left[1+\frac{4\epsilon}{3}\right]\right\}.\label{I5result}
\ee

We see that the result is divergent, containing a $J_0$ piece, so it will contribute to the $\log p^2$ term.

-For the $I_7$ diagram in (\ref{I7}), using the fact, calculated previously, that $I_\mu^a(p+q)=-\frac{(p+q)_\mu}{2}
I_0(p+q)$ and the form of $I_0(p+q)=I_{1,1}(p+q)$ in (\ref{I11}), we find
\bea
I^{ab}_{7a,\mu\nu}&=&+16g^2\delta^{ab}\left\{\int \frac{d^dq}{(2\pi)^d}\frac{(2p+2q)_\mu (2p+q)_\nu}{q^2(p+q)^2}
\int \frac{d^dr}{(2\pi)^d}\frac{1}{r^2(p+q+r)^2}\right.\cr
&&\left.+\int \frac{d^dq}{(2\pi)^d}\frac{ (2p+q)_\nu}{q^2(p+q)^2}
\int \frac{d^dr}{(2\pi)^d}\frac{r_\mu}{r^2(p+q+r)^2}\right\}\cr
&=&-+24g^2\delta^{ab}\frac{\Gamma(2-d/2)[\Gamma(d/2-1)]^2}{(4\pi)^{d/2}\Gamma(d-2)}
\int \frac{d^dq}{(2\pi)^d}\frac{(p+q)_\mu (p+2q)_\nu}{q^2(p+q)^{6-d}}\cr
&=&+24g^2\delta^{ab}\frac{\Gamma(2-d/2)[\Gamma(d/2-1)]^2}{(4\pi)^{d/2}\Gamma(d-2)}
\left[p_\mu p_\nu I_{1,1+\frac{4-d}{2}}(p)+2p_\mu I^\nu_{1,1+\frac{4-d}{2}}(p)\right.\cr
&&\left.+p_\nu I^\mu_{1,1+\frac{4-d}{2}}(p)+2I^{\mu\nu}_{1,1+\frac{4-d}{2}}(p)\right].
\eea

Using the integrals in (\ref{Imu}) and (\ref{Imunu}), we find 
\bea
I^{ab}_{7a,\mu\nu}&=&+24g^2\delta^{ab}\frac{\Gamma(2-d/2)[\Gamma(d/2-1)]^2}{(4\pi)^{d/2}\Gamma(d-2)}
\left[\frac{p_\mu p_\nu}{p^2} p^2 I_{1,1+\frac{4-d}{2}}(p)\right.\cr
&&\left.+\frac{3}{2}
\frac{p^\mu p^\nu}{p^2}\left[I_{1,1+\frac{2-d}{2}}
-p^2 I_{1,1+\frac{4-d}{2}}\right]\right.\cr
&&\left.+\left(\delta_{\mu\nu}-d\frac{p^\mu p^\nu}{p^2}\right)\frac{1}{2p^2(1-d)}
\left[I_{1,1+\frac{-d}{2}}+p^4I_{1,1+\frac{4-d}{2}}-2p^2I_{1,1+\frac{2-d}{2}}\right]\right]\cr
&=&+24g^2\delta^{ab}\frac{p^{2d-6}[\Gamma(d/2-1)]^3\Gamma(3-d)}{(4\pi)^d\Gamma(3d/2-3)}\left\{
\frac{\delta_{\mu\nu}}{2(1-d)}
\left[+\frac{1}{3}+\frac{3d/2-4}{d/2-2}-2\right]\right.\cr
&&\left.+\frac{p^\mu p^\nu}{p^2}\left[+\frac{1}{2(d-1)}\frac{3d/2-4}{d/2-2}+\frac{(d-3)}{2(d-1)}
+\frac{d}{6(d-1)}\right]\right\}\cr
&=&+16g^2\delta^{ab}\frac{J_0}{d-4}\left\{-\delta_{\mu\nu}+(d-3)\frac{p^\mu p^\nu}{p^2}\right\}.\label{I7resultd}
\eea

Since the result is symmetric in $(\mu\nu)$, $I^{ab}_{7a,\mu\nu}=I^{ab}_{7b,\mu\nu}$.

Substituting $d=3+\epsilon$, we find
\be
I^{ab}_{7a,\mu\nu}=+16g^2\delta^{ab}J_0\left\{\eta_{\mu\nu}
\left(1+\epsilon\right)-\epsilon\frac{p^\mu p^\nu}{p^2}\right\}.\label{I7result}
\ee

-For the $I_3$ diagram, the most complicated, we substitute 
\be
(2p+q+r)\cdot (q+r)=(p+q)^2+(p+r)^2-2p^2+q^2+r^2-(q-r)^2\;
\ee
in (\ref{I3munu}), to write
\bea
I^{ab}_{3,\mu\nu}(p)&=&-4g^2\delta^{ab}\int \frac{d^dq}{(2\pi)^d}(p+2q)_\mu\int \frac{d^dr}{(2\pi)^d}(p+2r)_\nu\cr
&&\times\left[\frac{1}{q^2(q-r)^2r^2(p+r)^2}+\frac{1}{(p+q)^2q^2(q-r)^2r^2}\right.\cr
&&\left.+\frac{1}{(p+q)^2(q-r)^2r^2(p+r)^2}+\frac{1}{(p+q)^2q^2(q-r)^2(p+r)^2}\right.\cr
&&\left.-\frac{1}{(p+q)^2q^2r^2
(p+r)^2}-\frac{2p^2}{(p+q)^2q^2(q-r)^2r^2(p+r)^2}\right].\label{I3}
\eea

In this integral, in the square bracket 
we can do the first 5 terms in the same way as we did for $I_5,I_6,I_7$, doing first one integral, then the 
other. But for the last term, the scalar integral with nontrivial denominator $\Delta$, we need something else. 
This integral, 
\be
I_{11111}\equiv \int\frac{d^dk_1}{(2\pi)^d}\int \frac{d^dk_2}{(2\pi)^d}\frac{1}{\Delta}=
\int\frac{d^dk_1}{(2\pi)^d}\int \frac{d^dk_2}{(2\pi)^d}\frac{1}{(k_1+p)^2k_1^2(k_1-k_2)^2k_2^2(k_2+p)^2}\;,
\ee
was calculated in \cite{Grozin:2005yg} (in section 5.1), as 
\be
I_{11111}=\frac{(p^2)^{d-5}}{(4\pi)^d}G(1,1,1,1,1)\;,
\ee
where, since
\be
\frac{p^{2d-10}}{(4\pi)^d}G_1^2=\frac{p^2B_0^2}{p^4}\;,\;\;\;
\frac{p^{2d-10}}{(4\pi)^d}G_2=\frac{J_0}{p^4}\;,
\ee
we have
\be
I_{11111}=+\frac{2}{(d-4)p^4}\left[-(d-3)p^2B_0^2+\frac{(3d-8)(3d-10)}{d-4}J_0\right]\equiv \frac{K}{p^4}.
\ee

We calculate then the terms in (\ref{I3}) one by one. The first term has the integral 
$\int \frac{d^dq}{(2\pi)^d}\frac{q_\mu}{q^2(q-r)^2}=
\frac{r_\mu}{2}I_0(-r)=\frac{r_\mu}{2}r^{d-4}B$.
Then the first term is
\be
B\int \frac{d^dr}{(2\pi)^d}\frac{(p+2r)_\nu(p+r)_\mu}{(p+r)^2r^{6-d}}=B\int \frac{d^dq}{(2\pi)^d}\frac{r_\mu
(2r-p)_\nu}{r^2(r-p)^{6-d}}\;,
\ee
under a change of $r\rightarrow r-p$ in the second equality.

The second term is 
\bea
&&\int \frac{d^dq}{(2\pi)^d}\frac{(p+2q)_\mu}{q^2(p+q)^2}\int\frac{d^dq}{(2\pi)^d}\frac{(p+2r)_\nu}{r^2(r-q)^2}
=\int \frac{d^dq}{(2\pi)^d}\frac{(p+2q)_\mu}{q^2(p+q)^2}(p+q)_\nu I_0(-q)\cr
&=&B \int \frac{d^dq}{(2\pi)^d}\frac{(p+2q)_\mu (p+q)_\nu}{(q+p)^2q^{6-d}}\;,
\eea
which means it is obtained from exchanging $\mu\leftrightarrow \nu$ in the first term.

The third term has the integral
\be
\int \frac{d^dq}{(2\pi)^d}\frac{(p+2q)_\mu}{(q-r)^2(q+p)^2}=\int \frac{d^dq}{(2\pi)^d}\frac{(p+2q+2r)_\mu}
{q^2(q+p+r)^2}=(p+2r-(p+r))I_0(p+r)=r_\mu I_0(p+r)\;,
\ee
giving 
\be
B \int \frac{d^d r}{(2\pi)^d}\frac{r_\mu(p+2r)_\nu}{r^2(r+p)^{6-d}}.
\ee 

The fourth term has 
\bea
&&\int\frac{d^dr}{(2\pi)^d}\frac{(p+2r)_\nu}{(r-q)^2(r+p)^2}=
\int\frac{d^dr}{(2\pi)^d}\frac{(p+2q+2r)_\nu}{r^2(r+p+q)^2}\cr
&=&(p+2q-(p+q))_\nu I_0(p+q)=q_\nu I_0(p+q)\;,
\eea
so is obtained from exchanging $\mu\leftrightarrow\nu$ in the third term. 

The fifth term is 
\be
\int \frac{d^dq}{(2\pi)^d}\frac{(p+2q)_\mu}{q^2(q+p)^2}\int \frac{d^dr}{(2\pi)^d}\frac{(p+2r)_\nu}{r^2(r+p)^2}\;,
\ee
and we can easily see that both integrals give $p_\mu I_0(p)-p_\mu I_0(p)=0$, so this term vanishes.

Then the sum of the first five factors gives
\bea
&&B\int \frac{d^dq}{(2\pi)^d}\left[\frac{q_\mu(2q-p)_\nu +(2q-p)_\mu q_\nu}{q^2(q-p)^{6-d}}
+\frac{(2q+p)_\mu q_\nu +q_\mu (2q+p)_\nu}{q^2(q+p)^{6-d}}\right]\cr
&=&B\int \frac{d^dq}{(2\pi)^d}
\frac{8q_\mu q_\nu +2q_\mu p_\nu +2p_\mu q_\nu}{q^2(q+p)^{6-d}}.
\eea

Using (\ref{Imu}) and (\ref{Imunu}), we obtain
for the sum of the first 5 terms
\bea
&&B\left\{\left(\eta_{\mu\nu}-d\frac{p^\mu p^\nu}{p^2}\right)\frac{8}{4p^2(1-d)}
\left[I_{1,1+\frac{-d}{2}}+p^4I_{1,1+\frac{4-d}{2}}-2p^2I_{1,1+\frac{2-d}{2}}\right]\right.\cr
&&\left.
+\frac{2p^\mu p^\nu}{p^2}\left[I_{1,1+\frac{2-d}{2}}-p^2I_{1,1+\frac{4-d}{2}}\right]\right\}.
\eea

As before, we construct $J_0$, and finally obtain for the sum of the first five terms in (\ref{I3}) 
(still to be multiplied by $-4g^2\delta^{ab}$)
\be
J_0\left\{\left(\eta_{\mu\nu}-d\frac{p^\mu p^\nu}{p^2}\right)\frac{2}{1-d}\left[\frac{3d-8}{d-4}-\frac{5}{3}\right]
-\frac{p^\mu p^\nu}{p^2}4\frac{d-2}{d-4}\right\}.\label{firstfive}
\ee

Substituting $d=3+\epsilon$, we get:
\be
J_0\left\{\left(\eta_{\mu\nu}-(3+\epsilon)\frac{p^\mu p^\nu}{p^2}\right)\left(\frac{8}{3}+\frac{20\epsilon}{3}\right)
+4(1+2\epsilon)\frac{p_\mu p_\nu}{p^2}\right\}
\ee

The last term in $I_3$ in (\ref{I3}) is (still to be multiplied by $-4g^2\delta^{ab}$)
\bea
&&-2p^2\int \frac{d^dq}{(2\pi)^d}\int \frac{d^dr}{(2\pi)^d}\frac{(p+2q)_\mu (p+2r)_\nu}{\Delta}\cr
&=&-2p^2p^\mu p^\nu \int \int \frac{1}{\Delta}-4p^2p^\nu
\int \int \frac{r_\nu}{\Delta}-4p^2p^\nu \int\int \frac{q_\mu}{\Delta}
-8p^2\int\int\frac{q_\mu r_\nu}{\Delta}\;,
\eea
where we find that 
\bea
&&\int \int\frac{r_\mu}{\Delta}=p_\mu J\Rightarrow J=
2p_\mu \times \int \int \frac{r_\mu}{\Delta}=\int\int \frac{(p+q)^2-p^2-r^2}{\Delta}\cr
&=&\int\int \left[\frac{1}{(p+q)^2q^2
(q-r)^2r^2}-\frac{1}{(p+q)^2q^2(q-r)^2(p+r)^2}\right]-p^2I_{11111}.
\eea

But both the last integrals give $BI_{1,1+\frac{4-d}{2}}(p)$ (the first gives the integral of $-p$, but that is the same as 
the one of $p$), so they cancel agains each other, finally leading to 
\be
\int \int \frac{r_\mu}{\Delta}=-\frac{p_\mu}{2}I_{11111}=\int \int \frac{q_\mu}{\Delta}\;,\label{qmurmu}
\ee
since the denominator is symmetric under the exchange of $q_\mu$ with $r_\mu$. 

For the tensor integral, by Lorentz invariance we write
\be
\int \int \frac{q_\mu r_\nu}{\Delta}=\delta_{\mu\nu}K_{00}^{(12)}+\frac{p_\mu p_\nu}{p^2}K_{11}^{(12)}
\ee
and we multiply it by  $2\delta^{\mu\nu}$ to obtain 
\bea
&&2d K_{00}^{12}+2K_{11}^{12}=\int\int \frac{r^2+q^2-(r-q)^2}{\Delta}\cr
&=&\int\int \left[\frac{1}{(p+q)^2q^2(q-r)^2
(p+r)^2}+\frac{1}{(p+q)^2(q-r)^2r^2(p+r)^2}-\frac{1}{(p+q)^2q^2r^2(p+r)^2}\right]\cr
&=&B\int\left[\frac{1}{q^2(q+p)^{6-d}}+\frac{1}{r^2(r+p)^{6-d}}-\frac{p^{d-4}}{q^2(q+p)^2}\right]\cr
&=&2B I_{1,1+\frac{4-d}{2}}(p)-B^2p^{2d-8}.
\eea

Similarly, we multiply it by  $p^\mu p^\nu/p^2$ to obtain 
\bea
&&K_{00}^{12}+K_{11}^{12}=\frac{1}{4p^2}\int \frac{((p+q)^2-p^2-q^2)((p+r)^2-p^2-r^2)}{\Delta}\cr
&=&\frac{1}{4p^2}\int\int\left[\frac{1}{q^2r^2(q-r)^2}-\frac{1}{q^2(q-r)^2(p+r)^2}-\frac{1}{(p+q)^2(q-r)^2r^2}
+\frac{1}{(p+q)^2(q-r)^2(p+r)^2}\right.\cr
&&\left. -\frac{p^2}{(p+q)^2q^2(r-q)^2r^2}+\frac{p^2}{(p+q)^2q^2(q-r)^2(p+r)^2}\right.\cr
&&\left.-\frac{p^2}{q^2(q-r)^2r^2
(p-r)^2}+\frac{p^2}{(p+q)^2(q-r)^2r^2(p+r)^2}\right]+\frac{p^2}{4}I_{11111}\cr
&=&\frac{1}{4p^2}\int \left[-\frac{2B}{(r+p)^2r^{4-d}}+\frac{B}{(r+p)^{6-d}}\right]+\frac{p^4}{4}I_{11111}\cr
&=&-\frac{B}{2p^2}I_{1,1+\frac{2-d}{2}}(p)+\frac{p^2}{4}I_{11111}\;,
\eea
where we have used that some integrals vanish (as being independent of $p$ after a shift), and have then cancelled several 
terms against each other.

The solution of these two equations is 
\bea
&&K_{00}^{(12)}=-\frac{(d-3)d}{p^2(d-4)^2(d-1)}J_0+\frac{1}{2(d-4)(d-1)}B_0^2\cr
&&K_{11}^{(12)}=\frac{2(d-2)(2d^2-9d+8)}{p^4 (d-4)^2(d-1)}J_0-\frac{(d-2)^2}{2p^2(d-4)(d-1)}B_0^2.\label{K11K00}
\eea

Putting everything together, we find that the 
 last term in $I_3$ in (\ref{I3}) gives (still to be multiplied by $-4g^2\delta^{ab}$)
\bea
&&-4\frac{p_\mu p_\nu}{p^2}\left[-\frac{(3d-10)(3d-8)}{(d-4)^2}J_0+\frac{d-3}{d-4}p^2B_0^2\right]\cr
&&-8\eta_{\mu\nu}\left[-\frac{d(d-3)}{(d-4)^2(d-1)}J_0+\frac{1}{2(d-4)(d-1)}p^2B_0^2\right]\cr
&&-\frac{8p_\mu p_\nu}{p^2}\left[\frac{2(d-2)(2d^2-9d+8)}{(d-4)^2(d-1)}J_0 -\frac{(d-2)^2}{2(d-4)(d-1)}p^2B_0^2
\right].\label{lastone}
\eea

Adding together $I_3$ in (\ref{firstfive}) and (\ref{lastone}), $I_5$ in (\ref{I5resultd}), $I_{7a}$ in (\ref{I7resultd})
twice (for diagrams $7a$ and $7b$) and $I_8$ in (\ref{I8result}), we obtain in total the 2-loop result
\be
I_{\rm 2-loop}=\frac{16}{d-4}g^2\delta^{ab}\left(\delta_{\mu\nu}-\frac{p_\mu p_\nu}{p^2}\right)\left[2J_0-
\frac{d^3-5d^2-8d+8}{(d-1)(d-4)}J_0+\frac{B_0^2p^2}{d-1}\right].
\ee

\subsection{Calculation using basis of integrals}

Some of these integrals can also be obtained using the formulas in the \cite{KostasSoon} 
for the basis of integrals with given denominators. 

First, in the formula
 (for $\a=1,2$)
\be
I^\a_\mu(p)\equiv \int \frac{d^d k_1}{(2\pi)^d}\frac{d^d k_2}{(2\pi)^d}\frac{(k_\a)_\mu}{\Delta}=p_\mu K_1^{\a}\;,
\ee
we find 
\be
K_1\equiv K_1^{(1)}=K_1^{(2)}=-\frac{(3d-10)(3d-8)}{(d-4)^2p^4}J_0+\frac{d-3}{(d-4)p^2}B_0^2
\ee
which is consistent with (\ref{qmurmu}), which implies $K_1p^4=-K/2$.

Next, for the tensor integrals 
\be
I^{\a\b}_{\mu\nu}(p)\equiv \int \frac{d^d k_1}{(2\pi)^d}\frac{d^d k_2}{(2\pi)^d}\frac{(k_\a)_\mu (k_\b)_\nu}{\Delta}=
\delta_{\mu\nu}K_{00}^{(\a\b)}+p_\mu p_\nu K_{11}^{(\a\b)}\;,
\ee
we have 
\bea
&&K_{00}^{(11)}=K_{00}^{(22)}=-\frac{2(d-3)(d-2)}{p^2(d-4)^2(d-1)}J_0+\frac{d-3}{2(d-4)(d-1)}B_0^2\cr
&&K_{00}^{(12)}=-\frac{(d-3)d}{p^2(d-4)^2(d-1)}J_0+\frac{1}{2(d-4)(d-1)}B_0^2\cr
&&K_{11}^{(11)}=K_{11}^{(22)}=\frac{(5d^3-33d^2+64d -32}{p^4(d-4)^2(d-1)}J_0-\frac{(d-3)d}{2p^2(d-4)(d-1)}
B_0^2\cr
&&K_{11}^{(12)}=\frac{2(d-2)(2d^2-9d+8)}{p^4 (d-4)^2(d-1)}J_0-\frac{(d-2)^2}{2p^2(d-4)(d-1)}B_0^2\;,
\eea
and the formulas for $K_{11}^{(12)}$ and $K_{00}^{(12)}$ are consistent with (\ref{K11K00}).

Further, or three equal momenta in the numerator,
\be
I^{111}_{\mu\nu\rho}(p)\equiv
\int \frac{d^d k_1}{(2\pi)^d}\frac{d^d k_2}{(2\pi)^d}\frac{(k_1)_\mu (k_1)_\nu (k_1)_\rho}{\Delta}
=(\delta _{\mu\nu}p_\rho +\delta_{\nu\rho} p_\mu +\delta_{\rho \mu}p_\nu) K_{001}^{(111)}+p_\mu p_\nu 
p_\rho K_{111}^{(111)}\;,
\ee
where
\bea
K_{001}^{(111)}&=&\frac{(d-3)(d-2)}{p^2(d-4)^2(d-1)}J_0-\frac{(d-3)}{4(d-4)(d-1)}B_0^2\cr
K_{111}^{(111)}&=&-\frac{3d^3-18d^2+29d-8}{p^4(d-4)^2(d-1)}J_0+\frac{(d-3)(d+2)}{4p^2(d-4)(d-1)}B_0^2.
\eea

Also, for two momenta equal and a third unequal in the numerator we have
\be
I^{112}_{\mu\nu\rho}(p)\equiv
\int \frac{d^d k_1}{(2\pi)^d}\frac{d^d k_2}{(2\pi)^d}\frac{(k_1)_\mu (k_1)_\nu (k_2)_\rho}{\Delta}
=\delta _{\mu\nu}p_\rho K_{001;1}^{(112)} +(\delta_{\nu\rho} p_\mu +\delta_{\rho \mu}p_\nu) K_{001;2}^{(112)}
+p_\mu p_\nu p_\rho K_{111}^{(112)}\;,
\ee
where
\bea
K_{001;1}^{(112)}&=& \frac{(d-3)(d-2)}{p^2(d-4)^2(d-1)}J_0-\frac{(d-3)}{4(d-4)(d-1)}B_0^2\cr
K_{001;2}^{(112)}&=&\frac{(d-3)d}{2p^4(d-4)^2(d-1)}J_0-\frac{1}{4(d-4)(d-1)}B_0^2\cr
K_{111}^{(112)}&=&-\frac{2d^2-9d+8}{(d-4)^2p^2}J_0+\frac{(d-2)}{4(d-4)p^2}B_0^2.
\eea

Moving on to the most complicated integrals, with four momenta in the numerator, in the case of three equal and one 
not equal, we get
\bea
I^{1112}_{\mu\nu\rho\sigma}(p)&\equiv
&\int \frac{d^d k_1}{(2\pi)^d}\frac{d^d k_2}{(2\pi)^d}\frac{(k_1)_\mu (k_1)_\nu (k_1)_\rho (k_2)_\sigma}{\Delta}
=(\delta_{\mu\nu}\delta_{\rho\sigma}+\delta_{\mu \rho }\delta_{\nu\sigma}+\delta_{\mu \sigma}\delta_{\nu\rho})
K_{0000}^{1112}\cr
&&+\delta_{(\mu\nu}p_{\rho)}p_\sigma K_{0011;1}^{1112}+\delta_{\sigma(\mu}p_\nu p_{\rho)}
K_{0011;2}^{1112}+p_\mu p_\nu p_\rho p_\sigma K_{1111}^{1112}\;,
\eea
where
\bea
K_{0000}^{1112}&=& \frac{d^2-d-4}{3(d-4)^2(d+1)(3d-4)}J_0-\frac{p^2}{8(d-4)(d-1)(d+1)}B_0^2\cr
K_{0011;1}^{1112}&=&-\frac{(d-3)(4d^3-7d^2-2d+8)}{3p^2(d-4)^2(d-1)(d+1)(3d-4)}J_0+
\frac{d^2-2d-2}{8(d-4)(d-1)(d+1)}B_0^2\cr
K_{0011;2}^{1112}&=&-\frac{(d-3)d(5d+4)}{6p^2(d-4)^2(d+1)(3d-4)}J_0+\frac{(d+2)}{8(d-4)(d-1)(d+1)}B_0^2\cr
K_{1111}^{1112}&=&\frac{d(10d^4-47d^3+39d^2+44d-64)}{3p^4(d-4)^2(d-1)(d+1)(3d-4)}J_0
-\frac{(d-2)d(d+2)}{8p^2(d-4)(d-1)(d+1)}B_0^2.
\eea

Finally, for two pairs of momenta in the numerator,
\bea
I^{1122}_{\mu\nu\rho\sigma}(p)&\equiv
&\int \frac{d^d k_1}{(2\pi)^d}\frac{d^d k_2}{(2\pi)^d}\frac{(k_1)_\mu (k_1)_\nu (k_2)_\rho (k_2)_\sigma}{\Delta}
=\delta_{\mu\nu}\delta_{\rho\sigma}K_{0000;1}^{1122}+(
+\delta_{\mu \rho }\delta_{\nu\sigma}+\delta_{\mu \sigma}\delta_{\nu\rho})
K_{0000;2}^{1122}\cr
&&+\delta_{\mu\nu}p_{\rho}p_\sigma K_{0011;1}^{1122}+\delta_{\rho\sigma}p_\mu p_\nu K_{0011;2}^{1122}
+p_{(\mu}\delta_{\nu)(\rho}p_{\sigma)}K_{0011;3}^{1122}+p_\mu p_\nu p_\rho p_\sigma K_{1111}^{1122}\;,
\eea
where
\bea
K_{0000;1}^{1122}&=&\frac{2d^4-13d^3+23d^2+6d-24}{3(d-2)(d-4)^2(d-1)(d+1)(3d-4)}J_0
-\frac{(d^2-3d-2)p^2}{8(d-4)(d-1)(d-2)(d+1)}B_0^2\cr
K_{0000;2}^{1122}&=&\frac{d^4-d^3-6d^2-4d+16}{6(d-2)(d-4)^2(d-1)(d+1)(3d-4)}J_0
-\frac{p^2}{4(d-4)(d-1)(d-2)(d+1)}B_0^2\cr
K_{0011;1}^{1122}&=&-\frac{(d-3)(5d^3-14d^2-4d+16)}{3(d-2)p^2(d-4)^2(d+1)(3d-4)}J_0
+\frac{(d^2-2d-4)}{8(d-4)(d-2)(d+1)}B_0^2\cr
&=&K_{0011;2}^{1122}\cr
K_{0011;3}^{1122}&=&-\frac{(d-3)d(2d^2-d-4)}{3(d-2)p^2(d-4)^2(d+1)(3d-4)}J_0+\frac{d}{8(d-4)(d-2)(d+1)}B_0^2\cr
K_{1111}^{1122}&=&\frac{4d(2d^3-6d^2-3d+4)}{3p^4(d-4)^2(d+1)(3d-4)}J_0-\frac{d^2}{8(d-4)(d+1)d^2}B_0^2.
\eea

Putting together all the integrals above, the Feynman integral $I_3$ is 
\bea
I^{ab}_{3\mu\nu}(p)&=&-8\delta^{ab}g^2\left[4p_\mu p_\nu p^2K_1+p_\mu p_\nu I^{ss}_{\rho\rho}(p)
+4p_\rho p_\mu (I^{21}_{\nu\rho}+I^{22}_{\nu \rho})+4p_\rho p_\nu (I^{11}_{\mu\rho}+I^{12}_{\mu\rho})\right.\cr
&&\left. +2p_\mu (I^{112}_{\rho\rho\nu}+I^{111}_{\rho\rho \nu}+2I^{112}_{\nu \rho \rho})+
2p_\nu (I^{112}_{\rho\rho\mu}+I^{111}_{\rho\rho \mu}+2I^{112}_{\mu \rho \rho})\right.\cr
&&\left.+8p_\rho (I^{112}_{\mu\rho\nu}+I^{112}_{\nu \rho \mu})
+4(I^{1112}_{\mu\rho\rho\nu}+I^{1112}_{\nu \rho\rho\mu}+2I^{1122}_{\mu\rho\nu\rho})\right].\cr
&&
\eea
The final result looks too complicated, so we will only verify, in Appendix B.4, that the correct result is obtained 
for the divergences of the integral (the coefficient of $J_0$ in $d=3$).

We can also verify the integral $I_7$, but for it we need formulas for another denominator, since we have 
\bea
I^{ab}_{7a,\mu\nu}&=&+16g^2\delta^{ab}\int\frac{d^dq}{(2\pi)^d}\frac{d^dr}{(2\pi)^d}
\frac{(2q-r)_\mu(2q+\tilde p)_\nu}{\Delta_2}\cr
&=&+16g^2\delta^{ab}\int\frac{d^dq}{(2\pi)^d}\frac{d^dr}{(2\pi)^d}\frac{1}{\Delta_2}[4q_\mu q_\nu -2r_\mu q_\nu
+2q_\mu \tilde p_\nu -r_\mu \tilde p_\nu]\cr
&=&
+16g^2\delta^{ab}[4\bar I^{22}_{\mu\nu}(\tilde p)-2\bar I^{21}_{\nu \mu}(\tilde p)+2\tilde p_\nu \bar I^2_\mu (\tilde p)
-\tilde p_\nu \bar I^1_\mu(\tilde p)]\;,\label{intermed}
\eea
where now
\be
\Delta_2=(q+\tilde p)^2q^2r^2(q-r)^2
\ee

Here  $\bar I^{22}, \bar I^{21}, \bar I^1 $ and $\bar I^2$, with  denominator $\Delta_2$, are found as follows.

We have for the tensor integrals
\be
\bar I^{\a\b}_{\mu\nu}(p)\equiv \int \frac{d^d k_1}{(2\pi)^d}\frac{d^d k_2}{(2\pi)^d}
\frac{(k_\a)_\mu (k_\b)_\nu}{\Delta_2}=
\delta_{\mu\nu}Y_{00}^{(\a\b)}+p_\mu p_\nu Y_{11}^{(\a\b)}\;,
\ee
where
\bea
Y^{11}_{00}&=&-\frac{d-3}{3(d-4)(d-1)}J_0\cr
Y^{12}_{00}&=&-\frac{1}{6(d-4)}J_0\cr
Y^{22}_{00}&=&-\frac{1}{3(d-4)}J_0\cr
Y^{11}_{11}&=&\frac{(d-3)d}{3p^2(d-4)(d-1)}J_0\cr
Y^{12}_{11}&=&\frac{2(d-3)}{3(d-4)p^2}J_0\cr
Y^{22}_{11}&=&\frac{4(d-3)}{3(d-4)p^2}J_0.
\eea

Finally, for the vector integrals,
\be
\bar I^\a_\mu(p)\equiv \int \frac{d^d k_1}{(2\pi)^d}\frac{d^d k_2}{(2\pi)^d}\frac{(k_\a)_\mu}{\Delta_2}=p_\mu
Y_1^{\a}.
\ee
where
\be
Y^1_1=-\frac{(d-3)}{(d-4)p^2}J_0\;, \;\;\; Y^2_1=2Y^1_1.
\ee

Substituting all these results into (\ref{intermed}), we find
\be
I_{7,\mu\nu}^{ab}=16g^2\delta^{ab}\frac{J_0}{d-4}\left[-\delta_{\mu\nu}+
(d-3)\frac{p_\mu p_\nu}{p^2}\right]\;
\ee
which matches against the result of the direct calculation.

For $I_5$, we start from the formula in (\ref{integrals}), and change variables as $q\rightarrow q-p, r\rightarrow r-p$, after
which we rename $\tilde p=-p$, to obtain 
\bea
I_{5,\mu\nu}^{ab}&=&-8g^2\delta^{ab}\int \frac{d^dq}{(2\pi)^2}\int \frac{d^dr}{(2\pi)^d}\frac{(2q+\tilde p)_\mu 
(2q+\tilde p)_\nu (2q-r)^2}{(q+\tilde p)^2q^4r^2(q-r)^2}\cr
&=&-8g^2\delta^{ab}\int \frac{d^dq}{(2\pi)^2}\int \frac{d^dr}{(2\pi)^d}(2q+\tilde p)_\mu 
(2q+\tilde p)_\nu \left[\frac{2}{(q+\tilde p)^2q^2r^2(q-r)^2}\right.\cr
&&\left.-\frac{1}{(q+\tilde p)^2q^4(q-r)^2}+\frac{2}{(q+\tilde p)^2q^4r^2}\right]\cr
&=&-16 g^2\delta^{ab}\int \frac{d^dq}{(2\pi)^2}\int \frac{d^dr}{(2\pi)^d}\frac{(2q+\tilde p)_\mu 
(2q+\tilde p)_\nu}{\Delta_2(\tilde p)}\cr
&=&-16 g^2\delta^{ab}\left[4I_{\mu\nu}^{22}(\tilde p)+2\tilde p_\mu \bar I_\nu^2(\tilde p)+2\tilde p_\nu \bar I_\mu^1
(\tilde p)+\tilde p_\mu \tilde p_\nu I_1\right]\;,
\eea
where we have used that in the third line, the two terms vanish under integration over $r$, we have obtained 
the same integrals used for $I_7$, and additionally we have defined the integral
\bea
I_1&\equiv&\int \frac{d^dq}{(2\pi)^2}\int \frac{d^dr}{(2\pi)^d}\frac{1}{(q+\tilde p)^2q^2r^2(q-r)^2}\cr
&=&\frac{\Gamma(2-d/2)[\Gamma(d/2-1)]^2}{(4\pi)^{d/2}\Gamma(d-2)}I_{1,1+\frac{4-d}{2}}(p)\cr
&=&\frac{p^{2d-8}}{(4\pi)^d}\frac{\Gamma(3-d)[\Gamma(d/2-1)]^3}{\Gamma(3d/2-3)}\times\frac{3d-8}{d-4}\cr
&=&\frac{3d-8}{d-4}J_0.
\eea

Substituting $\bar I^{22}_{\mu\nu}, \bar I^2_\mu, \bar I^1_\mu$ and $I_1$, we obtain 
\be
I_{5,\mu\nu}^{ab}=-16g^2\delta^{ab}\frac{J_0}{3(d-4)}\left[-4\delta_{\mu\nu}+d\frac{p_\mu p_\nu}
{p^2}\right]\;,
\ee
which matches the result of direct calculation.

\section{Review of particle-vortex duality}\label{secPV}

Here we review the action of particle-vortex duality at the level of the path integral, as defined in \cite{Murugan:2014sfa}.

Particle-vortex duality acts on the action of the Abelian-Higgs model, with the gauge field considered 
as external (though it is needed, in order to have some vortex configuration), 
\be
S=\int d^3x \left[-\frac{1}{2}|(\d_\mu -iea_\mu)\Phi|^2 -V\right]\;,
\ee
rewritten by $\Phi=\Phi_0 e^{i\theta}$ as 
\bea
S&=&-\frac{1}{2}\int d^3x [(\d_\mu \Phi_0)^2+\Phi_0^2(\d_\mu \theta+e a_\mu)^2+2V]\cr
&=&-\frac{1}{2}\int d^3x [(\d_\mu \Phi_0)^2+\Phi_0^2(\d_\mu \theta_{\rm smooth}+\d_\mu \theta_{\rm vortex}
+e a_\mu)^2+2V]
\eea

Here we have decomposed the variable $\theta$ into a smooth component $\theta$, that doesn't contain vortices, and a {\em singular component}
$\theta_{\rm vortex}$ which contains vortices, due to the holonomy conditions (global, for total vortex number, and on patches, due to the presence of 
vortices and anti-vortices separated in space),
\be
\theta=\theta_{\rm smooth}+\theta_{\rm vortex}\;,
\ee
following \cite{Burgess:2001sy}.

Indeed, the one-vortex ansatz for a complex scalar with a vortex at $r=0$ is $\Phi=\Phi_0(r) e^{i\a}$, meaning $\theta=\a$ ($\a$ is 
the polar angle on the spatial 2 dimensional plane), a condition that gives a holonomy, and associates a singularity, since 
\be
\epsilon^{ij}\d_i \d_j \theta=2\pi \delta^2(x).\label{thetasing}
\ee

To see that this is true, integrate
 over a small disk the relation, and using the Green-Riemann theorem (Stokes theorem in 2d), 
\be
\int d^2z (\d_x f_y-\d_y f_x)=\oint_C d\vec{l}\cdot \vec{f}\;,
\ee
we get 
\be
2\pi=\int d^2z \epsilon^{ij}\d_i \d_j\theta=\oint_C d\a \d_\a \theta=\theta(2\pi)-\theta(0)=2\pi.
\ee

If the potential is the one 
of the Abelian-Higgs model, $\lambda(|\Phi|^2-v^2)^2$, there is a vortex (the Nielsen-Olesen vortex), but 
in general, the point is that 
the vortex ansatz must be (and is in the Abelian-Higgs case) compatible with $V=0$ at infinity. 

Now more generally, for $N$ vortices slowly moving, and located at positions $x_a(t)$, $a=1,..,N$, we define the vortex current
in the same way as for the electric current of Dirac delta function electrons (the way Dirac intended when he invented his 
delta function), 
\be
j^\mu_{\rm vortex}=\frac{1}{2\pi}\epsilon^{\mu\nu\rho}\d_\nu \d_\rho \theta=
\frac{1}{2\pi}\epsilon^{\mu\nu\rho}\d_\nu \d_\rho \theta_{\rm vortex}=\sum_{a=1}^N N_a \dot x^\mu_a\delta^2
[x-x_a(t)]\;,\label{vortexcurr}
\ee
where for completeness we wrote a vortex charge (number) $N_a$ at each vortex location. We note that indeed, the 
charge density $j^0$ is, for a single static vortex, equal to simply $\delta^2(x)$. Also note that the contribution to the vortex current 
is only from the vortex part of $\theta$, $\theta_{\rm vortex}$, the smooth (non-singular) part not contributing.

Then we can perform the particle-vortex duality in the path integral as usual (following the procedure for, say, T-duality
in 2 dimensions). 
We trade $\d_\mu \theta$ for a new field $\lambda_\mu$, constrained to satisfy $\epsilon^{\mu\nu\rho}\d_\nu
\lambda_\rho=0$ 
(constraint that is solved by the previous form), then  $\lambda_{\mu}$ also splits into a vortex (singular) part and a 
smooth part, except that actually, because of (\ref{thetasing}), only the smooth part satisfies the constraint, 
\be
\epsilon^{\mu\nu\rho}\d_\nu \lambda_{\rho, {\rm smooth}}=0\label{constra}
\ee
that we want to impose, whereas the vortex (singular) part will generate a vortex current, since for {\em one} vortex, 
\be
\epsilon^{0ij}\d_i \lambda_{j, {\rm vortex}}=\epsilon^{ij}\d_i \d_j \theta=2\pi \delta^2(x)\;,
\ee
and more generally, by replacing $\d_\rho\theta_{\rm vortex}$ with $\lambda_{\rm \rho,{\rm vortex}}$ in (\ref{vortexcurr}).

The constraint (\ref{constra})
is imposed with a Lagrange multiplier $b_\mu$, so the 
equivalent first order action is
\be
S=-\frac{1}{2}\int d^3x \left[(\d_\mu \Phi_0)^2+\Phi_0^2(\lambda_{\mu,{\rm smooth}}+\lambda_{\mu,{\rm vortex}}
+e a_\mu)^2+2V-2\epsilon^{\mu\nu\rho}b_\mu \d_\nu \lambda_{\rho,{\rm smooth}}\right].
\ee

Now solving for $\lambda_{\mu, {\rm smooth}}$ instead of $b_\mu$, we find
\be
(\lambda_{\mu,{\rm smooth}}+\lambda_{\mu,{\rm vortex}}+e a_\mu)\Phi_0^2=\epsilon_{\mu\nu\rho}\d^\nu b^\rho.
\ee

Replacing back in the action, we find the dual action,
\be
S=\int d^3x \left[-\frac{1}{4\Phi_0^2}(\d_\mu b_\nu-\d_\nu b_\mu)^2-e\epsilon^{\mu\nu\rho}b_\mu \d_\nu 
a_\rho-\frac{2\pi}{e}j^\mu_{\rm vortex}(t)b_\mu-\frac{1}{2}(\d_\mu\Phi_0)^2-V\right]\;,
\ee
so $\theta$ has been dualized to the gauge field $b_\mu$, and we have obtained in the action
 a coupling of it to the vortex current, 
which therefore acts as an electric current  after the duality. 

The duality performed is an S-duality, of the strong/weak type, since $\Phi_0$ acted as $1/g$ for the field $\theta$ 
($\Phi_0^2$ in front of the kinetic action), but for the field $b_\mu$ we have $\Phi_0$ acting as $g$ (we have $1/\Phi_0^2$
in front of the kinetic action).

Note that all the manipulations above are not classical, but rather quantum, since they are still valid at the level of the 
path integral: indeed, the actions are quadratic in the terms considered, so still valid as path integral transformations, 
as we can easily check, so we have a fully quantum equivalence. 

Thus we have a quantum S-duality, leading to an action on the currents (\ref{currentduality}). Note that it is strictly speaking
only true for constant $\Phi_0$, otherwise there is an ambiguity for where the $\Phi_0$ appears in the formula (outside, or 
inside the derivative).

%%%%%%%%%%%%%%%%%%%%%%%%%%%%%%%%%%%%%%%%%%%%%%%%%%%%%%%%%%%%%%%%%%%%%%%%%%%%%%%%%%%%%%%%
\bibliography{HoloCosmoPaper}

\providecommand{\href}[2]{#2}\begingroup\raggedright\begin{thebibliography}{10}

\bibitem{McFadden:2009fg}
P.~McFadden and K.~Skenderis, ``{Holography for Cosmology},''
  \href{http://dx.doi.org/10.1103/PhysRevD.81.021301}{{\em Phys. Rev.} {\bf
  D81} (2010)  021301},
\href{http://arxiv.org/abs/0907.5542}{{\tt arXiv:0907.5542 [hep-th]}}.
%%CITATION = ARXIV:0907.5542;%%.

\bibitem{Brout:1977ix}
R.~Brout, F.~Englert, and E.~Gunzig, ``{The Creation of the Universe as a
  Quantum Phenomenon},''
  \href{http://dx.doi.org/10.1016/0003-4916(78)90176-8}{{\em Annals Phys.} {\bf
  115} (1978)  78}.

\bibitem{Starobinsky:1979ty}
A.~A. Starobinsky, ``{Spectrum of relict gravitational radiation and the early
  state of the universe},'' {\em JETP Lett.} {\bf 30} (1979)  682--685.

\bibitem{Starobinsky:1980te}
A.~A. Starobinsky, ``{A New Type of Isotropic Cosmological Models Without
  Singularity},'' \href{http://dx.doi.org/10.1016/0370-2693(80)90670-X}{{\em
  Adv. Ser. Astrophys. Cosmol.} {\bf 3} (1987)  130--133}.

\bibitem{Sato:1980yn}
K.~Sato, ``{First Order Phase Transition of a Vacuum and Expansion of the
  Universe},'' {\em Mon. Not. Roy. Astron. Soc.} {\bf 195} (1981)  467--479.

\bibitem{Guth:1980zm}
A.~H. Guth, ``{The Inflationary Universe: A Possible Solution to the Horizon
  and Flatness Problems},''
  \href{http://dx.doi.org/10.1103/PhysRevD.23.347}{{\em Phys. Rev.} {\bf D23}
  (1981)  347--356}.
[Adv. Ser. Astrophys. Cosmol.3,139(1987)].
%%CITATION = PHRVA,D23,347;%%.

\bibitem{Linde:1981mu}
A.~D. Linde, ``{A New Inflationary Universe Scenario: A Possible Solution of
  the Horizon, Flatness, Homogeneity, Isotropy and Primordial Monopole
  Problems},'' \href{http://dx.doi.org/10.1016/0370-2693(82)91219-9}{{\em Phys.
  Lett.} {\bf 108B} (1982)  389--393}.
[Adv. Ser. Astrophys. Cosmol.3,149(1987)].
%%CITATION = PHLTA,108B,389;%%.

\bibitem{Albrecht:1982wi}
A.~Albrecht and P.~J. Steinhardt, ``{Cosmology for Grand Unified Theories with
  Radiatively Induced Symmetry Breaking},''
  \href{http://dx.doi.org/10.1103/PhysRevLett.48.1220}{{\em Phys. Rev. Lett.}
  {\bf 48} (1982)  1220--1223}.
[Adv. Ser. Astrophys. Cosmol.3,158(1987)].
%%CITATION = PRLTA,48,1220;%%.

\bibitem{Nastase:2019rsn}
H.~Nastase and K.~Skenderis, ``{Holography for the very early Universe and the
  classic puzzles of Hot Big Bang cosmology},''
  \href{http://dx.doi.org/10.1103/PhysRevD.101.021901}{{\em Phys. Rev. D} {\bf
  101} (2020) no.~2, 021901}, \href{http://arxiv.org/abs/1904.05821}{{\tt
  arXiv:1904.05821 [hep-th]}}.

\bibitem{Brandenberger:2012aj}
R.~H. Brandenberger and J.~Martin, ``{Trans-Planckian Issues for Inflationary
  Cosmology},'' \href{http://dx.doi.org/10.1088/0264-9381/30/11/113001}{{\em
  Class. Quant. Grav.} {\bf 30} (2013)  113001},
  \href{http://arxiv.org/abs/1211.6753}{{\tt arXiv:1211.6753 [astro-ph.CO]}}.

\bibitem{Starobinsky:2001kn}
A.~A. Starobinsky, ``{Robustness of the inflationary perturbation spectrum to
  transPlanckian physics},'' \href{http://dx.doi.org/10.1134/1.1381588}{{\em
  Pisma Zh. Eksp. Teor. Fiz.} {\bf 73} (2001)  415--418},
  \href{http://arxiv.org/abs/astro-ph/0104043}{{\tt arXiv:astro-ph/0104043}}.

\bibitem{Dvali:2020cgt}
G.~Dvali, A.~Kehagias, and A.~Riotto, ``{Inflation and Decoupling},''
  \href{http://arxiv.org/abs/2005.05146}{{\tt arXiv:2005.05146 [hep-th]}}.

\bibitem{Obied:2018sgi}
G.~Obied, H.~Ooguri, L.~Spodyneiko, and C.~Vafa, ``{De Sitter Space and the
  Swampland},'' \href{http://arxiv.org/abs/1806.08362}{{\tt arXiv:1806.08362
  [hep-th]}}.

\bibitem{Bedroya_2020}
A.~Bedroya, R.~Brandenberger, M.~Loverde, and C.~Vafa,
  \href{http://dx.doi.org/10.1103/physrevd.101.103502}{``Trans-planckian
  censorship and inflationary cosmology,''{\em Physical Review D} {\bf 101}
  (May, 2020)  }.

\bibitem{Bedroya:2019tba}
A.~Bedroya, R.~Brandenberger, M.~Loverde, and C.~Vafa, ``{Trans-Planckian
  Censorship and Inflationary Cosmology},''
  \href{http://dx.doi.org/10.1103/PhysRevD.101.103502}{{\em Phys. Rev. D} {\bf
  101} (2020) no.~10, 103502}, \href{http://arxiv.org/abs/1909.11106}{{\tt
  arXiv:1909.11106 [hep-th]}}.

\bibitem{Brahma:2019vpl}
S.~Brahma, ``{Trans-Planckian censorship conjecture from the swampland distance
  conjecture},'' \href{http://dx.doi.org/10.1103/PhysRevD.101.046013}{{\em
  Phys. Rev. D} {\bf 101} (2020) no.~4, 046013},
  \href{http://arxiv.org/abs/1910.12352}{{\tt arXiv:1910.12352 [hep-th]}}.

\bibitem{Bernardo:2019bbi}
H.~Bernardo, ``{Trans-Planckian censorship conjecture in holographic
  cosmology},'' \href{http://dx.doi.org/10.1103/PhysRevD.101.066002}{{\em Phys.
  Rev. D} {\bf 101} (2020) no.~6, 066002},
  \href{http://arxiv.org/abs/1912.00100}{{\tt arXiv:1912.00100 [hep-th]}}.

\bibitem{Afshordi:2016dvb}
N.~Afshordi, C.~Coriano, L.~Delle~Rose, E.~Gould, and K.~Skenderis, ``{From
  Planck data to Planck era: Observational tests of Holographic Cosmology},''
  \href{http://dx.doi.org/10.1103/PhysRevLett.118.041301}{{\em Phys. Rev.
  Lett.} {\bf 118} (2017) no.~4, 041301},
\href{http://arxiv.org/abs/1607.04878}{{\tt arXiv:1607.04878 [astro-ph.CO]}}.
%%CITATION = ARXIV:1607.04878;%%.

\bibitem{Afshordi:2017ihr}
N.~Afshordi, E.~Gould, and K.~Skenderis, ``{Constraining holographic cosmology
  using Planck data},''
  \href{http://dx.doi.org/10.1103/PhysRevD.95.123505}{{\em Phys. Rev.} {\bf
  D95} (2017) no.~12, 123505},
\href{http://arxiv.org/abs/1703.05385}{{\tt arXiv:1703.05385 [astro-ph.CO]}}.
%%CITATION = ARXIV:1703.05385;%%.

\bibitem{tHooft:1993dmi}
G.~'t~Hooft, ``{Dimensional reduction in quantum gravity},'' {\em Conf. Proc.}
  {\bf C930308} (1993)  284--296,
\href{http://arxiv.org/abs/gr-qc/9310026}{{\tt arXiv:gr-qc/9310026 [gr-qc]}}.
%%CITATION = GR-QC/9310026;%%.

\bibitem{Susskind:1994vu}
L.~Susskind, ``{The World as a hologram},''
  \href{http://dx.doi.org/10.1063/1.531249}{{\em J. Math. Phys.} {\bf 36}
  (1995)  6377--6396},
\href{http://arxiv.org/abs/hep-th/9409089}{{\tt arXiv:hep-th/9409089
  [hep-th]}}.
%%CITATION = HEP-TH/9409089;%%.

\bibitem{Maldacena:1997re}
J.~M. Maldacena, ``{The Large N limit of superconformal field theories and
  supergravity},'' \href{http://dx.doi.org/10.1023/A:1026654312961,
  10.4310/ATMP.1998.v2.n2.a1}{{\em Int. J. Theor. Phys.} {\bf 38} (1999)
  1113--1133}, \href{http://arxiv.org/abs/hep-th/9711200}{{\tt
  arXiv:hep-th/9711200 [hep-th]}}.
[Adv. Theor. Math. Phys.2,231(1998)].
%%CITATION = HEP-TH/9711200;%%.

\bibitem{Nastase:2015wjb}
H.~Nastase, {\em {Introduction to the ADS/CFT Correspondence}}.
\newblock Cambridge University Press, Cambridge,
2015.
\newblock
%%CITATION = INSPIRE-1415011;%%.

\bibitem{Ammon:2015wua}
M.~Ammon and J.~Erdmenger, {\em {Gauge/gravity duality}}.
\newblock Cambridge University Press, Cambridge,
2015.
\newblock
%%CITATION = INSPIRE-1376202;%%.

\bibitem{Witten:2001kn}
E.~Witten, ``{Quantum gravity in de Sitter space},'' in {\em {Strings 2001:
  International Conference Mumbai, India, January 5-10, 2001}}.
\newblock 2001.
\newblock
\href{http://arxiv.org/abs/hep-th/0106109}{{\tt arXiv:hep-th/0106109
  [hep-th]}}.
\newblock
%%CITATION = HEP-TH/0106109;%%.

\bibitem{Strominger:2001pn}
A.~Strominger, ``{The dS / CFT correspondence},''
  \href{http://dx.doi.org/10.1088/1126-6708/2001/10/034}{{\em JHEP} {\bf 10}
  (2001)  034},
\href{http://arxiv.org/abs/hep-th/0106113}{{\tt arXiv:hep-th/0106113
  [hep-th]}}.
%%CITATION = HEP-TH/0106113;%%.

\bibitem{Strominger:2001gp}
A.~Strominger, ``{Inflation and the dS / CFT correspondence},''
  \href{http://dx.doi.org/10.1088/1126-6708/2001/11/049}{{\em JHEP} {\bf 11}
  (2001)  049},
\href{http://arxiv.org/abs/hep-th/0110087}{{\tt arXiv:hep-th/0110087
  [hep-th]}}.
%%CITATION = HEP-TH/0110087;%%.

\bibitem{Maldacena:2002vr}
J.~M. Maldacena, ``{Non-Gaussian features of primordial fluctuations in single
  field inflationary models},''
  \href{http://dx.doi.org/10.1088/1126-6708/2003/05/013}{{\em JHEP} {\bf 05}
  (2003)  013},
\href{http://arxiv.org/abs/astro-ph/0210603}{{\tt arXiv:astro-ph/0210603
  [astro-ph]}}.
%%CITATION = ASTRO-PH/0210603;%%.

\bibitem{Maldacena:2011nz}
J.~M. Maldacena and G.~L. Pimentel, ``{On graviton non-Gaussianities during
  inflation},'' \href{http://dx.doi.org/10.1007/JHEP09(2011)045}{{\em JHEP}
  {\bf 09} (2011)  045},
\href{http://arxiv.org/abs/1104.2846}{{\tt arXiv:1104.2846 [hep-th]}}.
%%CITATION = ARXIV:1104.2846;%%.

\bibitem{Hartle:2012qb}
J.~B. Hartle, S.~W. Hawking, and T.~Hertog, ``{Accelerated Expansion from
  Negative $\Lambda$},''
\href{http://arxiv.org/abs/1205.3807}{{\tt arXiv:1205.3807 [hep-th]}}.
%%CITATION = ARXIV:1205.3807;%%.

\bibitem{Hartle:2012tv}
J.~B. Hartle, S.~W. Hawking, and T.~Hertog, ``{Quantum Probabilities for
  Inflation from Holography},''
  \href{http://dx.doi.org/10.1088/1475-7516/2014/01/015}{{\em JCAP} {\bf 1401}
  (2014) no.~01, 015},
\href{http://arxiv.org/abs/1207.6653}{{\tt arXiv:1207.6653 [hep-th]}}.
%%CITATION = ARXIV:1207.6653;%%.

\bibitem{Schalm:2012pi}
K.~Schalm, G.~Shiu, and T.~van~der Aalst, ``{Consistency condition for
  inflation from (broken) conformal symmetry},''
  \href{http://dx.doi.org/10.1088/1475-7516/2013/03/005}{{\em JCAP} {\bf 1303}
  (2013)  005},
\href{http://arxiv.org/abs/1211.2157}{{\tt arXiv:1211.2157 [hep-th]}}.
%%CITATION = ARXIV:1211.2157;%%.

\bibitem{Bzowski:2012ih}
A.~Bzowski, P.~McFadden, and K.~Skenderis, ``{Holography for inflation using
  conformal perturbation theory},''
  \href{http://dx.doi.org/10.1007/JHEP04(2013)047}{{\em JHEP} {\bf 04} (2013)
  047},
\href{http://arxiv.org/abs/1211.4550}{{\tt arXiv:1211.4550 [hep-th]}}.
%%CITATION = ARXIV:1211.4550;%%.

\bibitem{Mata:2012bx}
I.~Mata, S.~Raju, and S.~Trivedi, ``{CMB from CFT},''
  \href{http://dx.doi.org/10.1007/JHEP07(2013)015}{{\em JHEP} {\bf 07} (2013)
  015},
\href{http://arxiv.org/abs/1211.5482}{{\tt arXiv:1211.5482 [hep-th]}}.
%%CITATION = ARXIV:1211.5482;%%.

\bibitem{Garriga:2013rpa}
J.~Garriga and Y.~Urakawa, ``{Inflation and deformation of conformal field
  theory},'' \href{http://dx.doi.org/10.1088/1475-7516/2013/07/033}{{\em JCAP}
  {\bf 1307} (2013)  033},
\href{http://arxiv.org/abs/1303.5997}{{\tt arXiv:1303.5997 [hep-th]}}.
%%CITATION = ARXIV:1303.5997;%%.

\bibitem{McFadden:2013ria}
P.~McFadden, ``{On the power spectrum of inflationary cosmologies dual to a
  deformed CFT},'' \href{http://dx.doi.org/10.1007/JHEP10(2013)071}{{\em JHEP}
  {\bf 10} (2013)  071},
\href{http://arxiv.org/abs/1308.0331}{{\tt arXiv:1308.0331 [hep-th]}}.
%%CITATION = ARXIV:1308.0331;%%.

\bibitem{Ghosh:2014kba}
A.~Ghosh, N.~Kundu, S.~Raju, and S.~P. Trivedi, ``{Conformal Invariance and the
  Four Point Scalar Correlator in Slow-Roll Inflation},''
  \href{http://dx.doi.org/10.1007/JHEP07(2014)011}{{\em JHEP} {\bf 07} (2014)
  011},
\href{http://arxiv.org/abs/1401.1426}{{\tt arXiv:1401.1426 [hep-th]}}.
%%CITATION = ARXIV:1401.1426;%%.

\bibitem{Garriga:2014ema}
J.~Garriga and Y.~Urakawa, ``{Holographic inflation and the conservation of
  $\zeta$},'' \href{http://dx.doi.org/10.1007/JHEP06(2014)086}{{\em JHEP} {\bf
  06} (2014)  086},
\href{http://arxiv.org/abs/1403.5497}{{\tt arXiv:1403.5497 [hep-th]}}.
%%CITATION = ARXIV:1403.5497;%%.

\bibitem{Kundu:2014gxa}
N.~Kundu, A.~Shukla, and S.~P. Trivedi, ``{Constraints from Conformal Symmetry
  on the Three Point Scalar Correlator in Inflation},''
  \href{http://dx.doi.org/10.1007/JHEP04(2015)061}{{\em JHEP} {\bf 04} (2015)
  061},
\href{http://arxiv.org/abs/1410.2606}{{\tt arXiv:1410.2606 [hep-th]}}.
%%CITATION = ARXIV:1410.2606;%%.

\bibitem{Garriga:2014fda}
J.~Garriga, K.~Skenderis, and Y.~Urakawa, ``{Multi-field inflation from
  holography},'' \href{http://dx.doi.org/10.1088/1475-7516/2015/01/028}{{\em
  JCAP} {\bf 1501} (2015) no.~01, 028},
\href{http://arxiv.org/abs/1410.3290}{{\tt arXiv:1410.3290 [hep-th]}}.
%%CITATION = ARXIV:1410.3290;%%.

\bibitem{McFadden:2014nta}
P.~McFadden, ``{Soft limits in holographic cosmology},''
  \href{http://dx.doi.org/10.1007/JHEP02(2015)053}{{\em JHEP} {\bf 02} (2015)
  053},
\href{http://arxiv.org/abs/1412.1874}{{\tt arXiv:1412.1874 [hep-th]}}.
%%CITATION = ARXIV:1412.1874;%%.

\bibitem{Arkani-Hamed:2015bza}
N.~Arkani-Hamed and J.~Maldacena, ``{Cosmological Collider Physics},''
\href{http://arxiv.org/abs/1503.08043}{{\tt arXiv:1503.08043 [hep-th]}}.
%%CITATION = ARXIV:1503.08043;%%.

\bibitem{Kundu:2015xta}
N.~Kundu, A.~Shukla, and S.~P. Trivedi, ``{Ward Identities for Scale and
  Special Conformal Transformations in Inflation},''
  \href{http://dx.doi.org/10.1007/JHEP01(2016)046}{{\em JHEP} {\bf 01} (2016)
  046},
\href{http://arxiv.org/abs/1507.06017}{{\tt arXiv:1507.06017 [hep-th]}}.
%%CITATION = ARXIV:1507.06017;%%.

\bibitem{Hertog:2015nia}
T.~Hertog and E.~van~der Woerd, ``{Primordial fluctuations from complex AdS
  saddle points},'' \href{http://dx.doi.org/10.1088/1475-7516/2016/02/010}{{\em
  JCAP} {\bf 1602} (2016) no.~02, 010},
\href{http://arxiv.org/abs/1509.03291}{{\tt arXiv:1509.03291 [hep-th]}}.
%%CITATION = ARXIV:1509.03291;%%.

\bibitem{Garriga:2015tea}
J.~Garriga, Y.~Urakawa, and F.~Vernizzi, ``{$\delta N$ formalism from
  superpotential and holography},''
  \href{http://dx.doi.org/10.1088/1475-7516/2016/02/036}{{\em JCAP} {\bf 1602}
  (2016) no.~02, 036},
\href{http://arxiv.org/abs/1509.07339}{{\tt arXiv:1509.07339 [hep-th]}}.
%%CITATION = ARXIV:1509.07339;%%.

\bibitem{Garriga:2016poh}
J.~Garriga and Y.~Urakawa, ``{Consistency relations and conservation of $\zeta$
  in holographic inflation},''
  \href{http://dx.doi.org/10.1088/1475-7516/2016/10/030}{{\em JCAP} {\bf 1610}
  (2016) no.~10, 030},
\href{http://arxiv.org/abs/1606.04767}{{\tt arXiv:1606.04767 [hep-th]}}.
%%CITATION = ARXIV:1606.04767;%%.

\bibitem{Hawking:2017wrd}
S.~W. Hawking and T.~Hertog, ``{A Smooth Exit from Eternal Inflation?},''
  \href{http://dx.doi.org/10.1007/JHEP04(2018)147}{{\em JHEP} {\bf 04} (2018)
  147},
\href{http://arxiv.org/abs/1707.07702}{{\tt arXiv:1707.07702 [hep-th]}}.
%%CITATION = ARXIV:1707.07702;%%.

\bibitem{Arkani-Hamed:2018kmz}
N.~Arkani-Hamed, D.~Baumann, H.~Lee, and G.~L. Pimentel, ``{The Cosmological
  Bootstrap: Inflationary Correlators from Symmetries and Singularities},''
\href{http://arxiv.org/abs/1811.00024}{{\tt arXiv:1811.00024 [hep-th]}}.
%%CITATION = ARXIV:1811.00024;%%.

\bibitem{McFadden:2010na}
P.~McFadden and K.~Skenderis, ``{The Holographic Universe},''
  \href{http://dx.doi.org/10.1088/1742-6596/222/1/012007}{{\em J. Phys. Conf.
  Ser.} {\bf 222} (2010)  012007},
\href{http://arxiv.org/abs/1001.2007}{{\tt arXiv:1001.2007 [hep-th]}}.
%%CITATION = ARXIV:1001.2007;%%.

\bibitem{McFadden:2010vh}
P.~McFadden and K.~Skenderis, ``{Holographic Non-Gaussianity},''
  \href{http://dx.doi.org/10.1088/1475-7516/2011/05/013}{{\em JCAP} {\bf 1105}
  (2011)  013},
\href{http://arxiv.org/abs/1011.0452}{{\tt arXiv:1011.0452 [hep-th]}}.
%%CITATION = ARXIV:1011.0452;%%.

\bibitem{McFadden:2011kk}
P.~McFadden and K.~Skenderis, ``{Cosmological 3-point correlators from
  holography},'' \href{http://dx.doi.org/10.1088/1475-7516/2011/06/030}{{\em
  JCAP} {\bf 1106} (2011)  030},
\href{http://arxiv.org/abs/1104.3894}{{\tt arXiv:1104.3894 [hep-th]}}.
%%CITATION = ARXIV:1104.3894;%%.

\bibitem{Bzowski:2011ab}
A.~Bzowski, P.~McFadden, and K.~Skenderis, ``{Holographic predictions for
  cosmological 3-point functions},''
  \href{http://dx.doi.org/10.1007/JHEP03(2012)091}{{\em JHEP} {\bf 03} (2012)
  091},
\href{http://arxiv.org/abs/1112.1967}{{\tt arXiv:1112.1967 [hep-th]}}.
%%CITATION = ARXIV:1112.1967;%%.

\bibitem{Coriano:2012hd}
C.~Coriano, L.~Delle~Rose, and M.~Serino, ``{Three and Four Point Functions of
  Stress Energy Tensors in D=3 for the Analysis of Cosmological
  Non-Gaussianities},'' \href{http://dx.doi.org/10.1007/JHEP12(2012)090}{{\em
  JHEP} {\bf 12} (2012)  090},
\href{http://arxiv.org/abs/1210.0136}{{\tt arXiv:1210.0136 [hep-th]}}.
%%CITATION = ARXIV:1210.0136;%%.

\bibitem{Kawai:2014vxa}
S.~Kawai and Y.~Nakayama, ``{Improvement of energy-momentum tensor and
  non-Gaussianities in holographic cosmology},''
  \href{http://dx.doi.org/10.1007/JHEP06(2014)052}{{\em JHEP} {\bf 06} (2014)
  052},
\href{http://arxiv.org/abs/1403.6220}{{\tt arXiv:1403.6220 [hep-th]}}.
%%CITATION = ARXIV:1403.6220;%%.

\bibitem{McFadden:2010jw}
P.~McFadden and K.~Skenderis,
  \href{http://dx.doi.org/10.1142/9789814374552_0468}{``{Observational
  signatures of holographic models of inflation},''} in {\em {On recent
  developments in theoretical and experimental general relativity, astrophysics
  and relativistic field theories. Proceedings, 12th Marcel Grossmann Meeting
  on General Relativity, Paris, France, July 12-18, 2009. Vol. 1-3}},
  pp.~2315--2323.
\newblock 2010.
\newblock
\href{http://arxiv.org/abs/1010.0244}{{\tt arXiv:1010.0244 [hep-th]}}.
\newblock
%%CITATION = ARXIV:1010.0244;%%.

\bibitem{Skenderis:2002wp}
K.~Skenderis, ``{Lecture notes on holographic renormalization},''
  \href{http://dx.doi.org/10.1088/0264-9381/19/22/306}{{\em Class. Quant.
  Grav.} {\bf 19} (2002)  5849--5876},
\href{http://arxiv.org/abs/hep-th/0209067}{{\tt arXiv:hep-th/0209067
  [hep-th]}}.
%%CITATION = HEP-TH/0209067;%%.

\bibitem{Papadimitriou:2004ap}
I.~Papadimitriou and K.~Skenderis, ``{AdS / CFT correspondence and geometry},''
  \href{http://dx.doi.org/10.4171/013-1/4}{{\em IRMA Lect. Math. Theor. Phys.}
  {\bf 8} (2005)  73--101},
\href{http://arxiv.org/abs/hep-th/0404176}{{\tt arXiv:hep-th/0404176
  [hep-th]}}.
%%CITATION = HEP-TH/0404176;%%.

\bibitem{Papadimitriou:2004rz}
I.~Papadimitriou and K.~Skenderis, ``{Correlation functions in holographic RG
  flows},'' \href{http://dx.doi.org/10.1088/1126-6708/2004/10/075}{{\em JHEP}
  {\bf 10} (2004)  075},
\href{http://arxiv.org/abs/hep-th/0407071}{{\tt arXiv:hep-th/0407071
  [hep-th]}}.
%%CITATION = HEP-TH/0407071;%%.

\bibitem{Skenderis:2006jq}
K.~Skenderis and P.~K. Townsend, ``{Hidden supersymmetry of domain walls and
  cosmologies},'' \href{http://dx.doi.org/10.1103/PhysRevLett.96.191301}{{\em
  Phys. Rev. Lett.} {\bf 96} (2006)  191301},
\href{http://arxiv.org/abs/hep-th/0602260}{{\tt arXiv:hep-th/0602260
  [hep-th]}}.
%%CITATION = HEP-TH/0602260;%%.

\bibitem{Skenderis:2000in}
K.~Skenderis, ``{Asymptotically Anti-de Sitter space-times and their stress
  energy tensor},'' \href{http://dx.doi.org/10.1142/S0217751X0100386X}{{\em
  Int. J. Mod. Phys.} {\bf A16} (2001)  740--749},
  \href{http://arxiv.org/abs/hep-th/0010138}{{\tt arXiv:hep-th/0010138
  [hep-th]}}.
[,394(2000)].
%%CITATION = HEP-TH/0010138;%%.

\bibitem{Jevicki:1998yr}
A.~Jevicki and T.~Yoneya, ``{Space-time uncertainty principle and conformal
  symmetry in D particle dynamics},''
  \href{http://dx.doi.org/10.1016/S0550-3213(98)00578-1}{{\em Nucl. Phys.} {\bf
  B535} (1998)  335--348},
\href{http://arxiv.org/abs/hep-th/9805069}{{\tt arXiv:hep-th/9805069
  [hep-th]}}.
%%CITATION = HEP-TH/9805069;%%.

\bibitem{Kanitscheider:2008kd}
I.~Kanitscheider, K.~Skenderis, and M.~Taylor, ``{Precision holography for
  non-conformal branes},''
  \href{http://dx.doi.org/10.1088/1126-6708/2008/09/094}{{\em JHEP} {\bf 09}
  (2008)  094},
\href{http://arxiv.org/abs/0807.3324}{{\tt arXiv:0807.3324 [hep-th]}}.
%%CITATION = ARXIV:0807.3324;%%.

\bibitem{Dias:2011in}
M.~Dias, ``{Cosmology at the boundary of de Sitter using the dS/QFT
  correspondence},'' \href{http://dx.doi.org/10.1103/PhysRevD.84.023512}{{\em
  Phys. Rev.} {\bf D84} (2011)  023512},
\href{http://arxiv.org/abs/1104.0625}{{\tt arXiv:1104.0625 [astro-ph.CO]}}.
%%CITATION = ARXIV:1104.0625;%%.

\bibitem{Easther:2011wh}
R.~Easther, R.~Flauger, P.~McFadden, and K.~Skenderis, ``{Constraining
  holographic inflation with WMAP},''
  \href{http://dx.doi.org/10.1088/1475-7516/2011/09/030}{{\em JCAP} {\bf 1109}
  (2011)  030},
\href{http://arxiv.org/abs/1104.2040}{{\tt arXiv:1104.2040 [astro-ph.CO]}}.
%%CITATION = ARXIV:1104.2040;%%.

\bibitem{Bernardo:2018cow}
H.~Bernardo and H.~Nastase, ``{Holographic cosmology from "dimensional
  reduction" of $\mathcal N =4$ SYM vs. AdS$_{5}\times$S$^{5}$},''
  \href{http://dx.doi.org/10.1007/JHEP12(2019)025}{{\em JHEP} {\bf 12} (2019)
  025}, \href{http://arxiv.org/abs/1812.07586}{{\tt arXiv:1812.07586
  [hep-th]}}.

\bibitem{Nastase:2018cbf}
H.~Nastase, ``{Solution of the cosmological constant problem within holographic
  cosmology},'' \href{http://dx.doi.org/10.1016/j.physletb.2019.135168}{{\em
  Phys. Lett. B} {\bf 801} (2020)  135168},
  \href{http://arxiv.org/abs/1812.07597}{{\tt arXiv:1812.07597 [hep-th]}}.

\bibitem{Zeldovich:1978wj}
Y.~Zeldovich and M.~Khlopov, ``{On the Concentration of Relic Magnetic
  Monopoles in the Universe},''
  \href{http://dx.doi.org/10.1016/0370-2693(78)90232-0}{{\em Phys. Lett. B}
  {\bf 79} (1978)  239--241}.

\bibitem{Weinberg:2008zzc}
S.~Weinberg, {\em {Cosmology}}.
\newblock
2008.
\newblock
%%CITATION = INSPIRE-794379;%%.

\bibitem{Kolb:1990vq}
E.~W. Kolb and M.~S. Turner, ``{The Early Universe},''
{\em Front. Phys.} {\bf 69} (1990)  1--547.
%%CITATION = FRPHA,69,1;%%.

\bibitem{Murugan:2014sfa}
J.~Murugan, H.~Nastase, N.~Rughoonauth, and J.~P. Shock, ``{Particle-vortex and
  Maxwell duality in the $AdS_4\times \mathbb{CP}^3$/ABJM correspondence},''
  \href{http://dx.doi.org/10.1007/JHEP10(2014)051}{{\em JHEP} {\bf 10} (2014)
  51},
\href{http://arxiv.org/abs/1404.5926}{{\tt arXiv:1404.5926 [hep-th]}}.
%%CITATION = ARXIV:1404.5926;%%.

\bibitem{Witten:2003ya}
E.~Witten, ``{SL(2,Z) action on three-dimensional conformal field theories with
  Abelian symmetry},''
\href{http://arxiv.org/abs/hep-th/0307041}{{\tt arXiv:hep-th/0307041
  [hep-th]}}.
%%CITATION = HEP-TH/0307041;%%.

\bibitem{Herzog:2007ij}
C.~P. Herzog, P.~Kovtun, S.~Sachdev, and D.~T. Son, ``{Quantum critical
  transport, duality, and M-theory},''
  \href{http://dx.doi.org/10.1103/PhysRevD.75.085020}{{\em Phys. Rev.} {\bf
  D75} (2007)  085020},
\href{http://arxiv.org/abs/hep-th/0701036}{{\tt arXiv:hep-th/0701036
  [hep-th]}}.
%%CITATION = HEP-TH/0701036;%%.

\bibitem{Itzhaki:1998dd}
N.~Itzhaki, J.~M. Maldacena, J.~Sonnenschein, and S.~Yankielowicz,
  ``{Supergravity and the large N limit of theories with sixteen
  supercharges},'' \href{http://dx.doi.org/10.1103/PhysRevD.58.046004}{{\em
  Phys. Rev. D} {\bf 58} (1998)  046004},
  \href{http://arxiv.org/abs/hep-th/9802042}{{\tt arXiv:hep-th/9802042}}.

\bibitem{Graham:1999pm}
C.~R. Graham and E.~Witten, ``{Conformal anomaly of submanifold observables in
  AdS / CFT correspondence},''
  \href{http://dx.doi.org/10.1016/S0550-3213(99)00055-3}{{\em Nucl. Phys.} {\bf
  B546} (1999)  52--64},
\href{http://arxiv.org/abs/hep-th/9901021}{{\tt arXiv:hep-th/9901021
  [hep-th]}}.
%%CITATION = HEP-TH/9901021;%%.

\bibitem{Maldacena:1998im}
J.~M. Maldacena, ``{Wilson loops in large N field theories},''
  \href{http://dx.doi.org/10.1103/PhysRevLett.80.4859}{{\em Phys. Rev. Lett.}
  {\bf 80} (1998)  4859--4862},
\href{http://arxiv.org/abs/hep-th/9803002}{{\tt arXiv:hep-th/9803002
  [hep-th]}}.
%%CITATION = HEP-TH/9803002;%%.

\bibitem{Rey:1998ik}
S.-J. Rey and J.-T. Yee, ``{Macroscopic strings as heavy quarks in large N
  gauge theory and anti-de Sitter supergravity},''
  \href{http://dx.doi.org/10.1007/s100520100799}{{\em Eur. Phys. J.} {\bf C22}
  (2001)  379--394},
\href{http://arxiv.org/abs/hep-th/9803001}{{\tt arXiv:hep-th/9803001
  [hep-th]}}.
%%CITATION = HEP-TH/9803001;%%.

\bibitem{Penrose:1900mp}
R.~Penrose, ``{Singularities and time-asymmetry, Published in: General
  Relativity: An Einstein Centenary Survey. S.W.Hawking, W. Israel, eds.
  Cambridge Univ. Press, },'' pp.~581--638.
\newblock 1979.

\bibitem{Carroll:2004pn}
S.~M. Carroll and J.~Chen, ``{Spontaneous inflation and the origin of the arrow
  of time},'' \href{http://arxiv.org/abs/hep-th/0410270}{{\tt
  arXiv:hep-th/0410270}}.

\bibitem{Wald:2005cb}
R.~M. Wald, ``{The Arrow of time and the initial conditions of the universe},''
  \href{http://dx.doi.org/10.1016/j.shpsb.2006.03.005}{{\em Stud. Hist. Phil.
  Mod. Phys.} {\bf 37} (2006)  394--398},
  \href{http://arxiv.org/abs/gr-qc/0507094}{{\tt arXiv:gr-qc/0507094}}.

\bibitem{KostasSoon}
C.~Corian\`{o}, L.~Delle~Rose, and K.~Skenderis, ``{Two-point function of the
  stress-energy tensor and generalized conformal structure},''
  \href{http://arxiv.org/abs/2008.05346}{{\tt arXiv:2008.05346 [hep-th]}}.

\bibitem{Grozin:2005yg}
A.~Grozin, ``{Lectures on QED and QCD},'' in {\em {3rd Dubna International
  Advanced School of Theoretical Physics Dubna, Russia, January 29-February 6,
  2005}}, pp.~1--156.
\newblock 2005.
\newblock
\href{http://arxiv.org/abs/hep-ph/0508242}{{\tt arXiv:hep-ph/0508242
  [hep-ph]}}.
\newblock
%%CITATION = HEP-PH/0508242;%%.

\bibitem{Burgess:2001sy}
C.~Burgess and B.~P. Dolan, ``{Duality and nonlinear response for quantum Hall
  systems},'' \href{http://dx.doi.org/10.1103/PhysRevB.65.155323}{{\em Phys.
  Rev. B} {\bf 65} (2002)  155323},
  \href{http://arxiv.org/abs/cond-mat/0105621}{{\tt arXiv:cond-mat/0105621}}.

\end{thebibliography}\endgroup
\bibliographystyle{utphys}
%%%%%%%%%%%%%%%%%%%%%%%%%%%%%%%%%%%%%%%%%%%%%%%%%%%%%%%%%%%%%%%%%%%%%%%%%%%%%%%%%%%%%%%%

\end{document}